\documentclass[aip,jcp,amsmath,amssymb,reprint]{revtex4-1}

\usepackage{graphicx}
\usepackage{dcolumn}
\usepackage[tight]{subfigure}
\usepackage{amsmath}
\usepackage{verbatim}
\usepackage[usenames,dvipsnames]{color}
\usepackage{bm} 
\usepackage{bbm}
\usepackage{natbib}
\usepackage{xspace}
\usepackage{marginnote}
\usepackage{mathtools}
\usepackage{dsfont}
\usepackage{hyperref} \hypersetup{colorlinks=true,linktoc=all,linkcolor=blue,breaklinks=true,citecolor=blue,urlcolor=blue}

\usepackage{etoolbox} 
\robustify{\uparrow}
\robustify{\downarrow}
\robustify{\sum}
\robustify{\int}
\robustify{\nonumber}
\robustify{\cite}
\robustify{\footnote}

\renewrobustcmd{\Re}{{\text{Re}}}
\renewrobustcmd{\Im}{{\text{Im}}}

\newrobustcmd{\tot}{\text{tot}} 
\newrobustcmd{\res}{\text{R}}   

\newrobustcmd{\deltah}{\bar{\delta}} 

\renewrobustcmd{\P}{\text{P}} 
\renewrobustcmd{\c}{\text{c}} 
\renewrobustcmd{\r}{\text{r}} 
\newrobustcmd{\e}{\text{e}} 
\newrobustcmd{\p}{\text{p}} 

\newrobustcmd{\sign}{\text{sgn}} 

\newrobustcmd{\g}{h} %
\newrobustcmd{\h}{h} %
\newrobustcmd{\gam}{\gamma} %

\newrobustcmd{\TCL}{\text{TCL}}   

\newrobustcmd{\one}{\mathds{1}}   
\newrobustcmd{\ones}{\mathcal{I}}

\newrobustcmd{\ket}[1]{|#1\rangle}
\newrobustcmd{\bra}[1]{\langle#1|}
\newrobustcmd{\brkt}[1]{\langle #1 \rangle}
\newrobustcmd{\braket}[2]{\langle #1 | #2 \rangle}

\newrobustcmd{\Ket}[1]{\bm{|}#1\bm{)}}
\newrobustcmd{\Bra}[1]{\bm{(}#1\bm{|}}
\newrobustcmd{\Braket}[2]{\bm{(}#1\bm{|}#2\bm{)}}
\newrobustcmd{\Brkt}[1]{\bm{(} #1 \bm{)}}

\newrobustcmd{\op}[1]{\hat{#1}}
\newrobustcmd{\sop}[1]{\mathcal{#1}}

\DeclareMathOperator{\Tr}{Tr}
\newrobustcmd{\tr}[1]{\underset{#1}{\Tr}}
\newrobustcmd{\tri}[1]{{\Tr}_{#1}}

\newrobustcmd{\E}{\text{E}}   
\renewrobustcmd{\S}{\text{S}} 
\newrobustcmd{\R}{\text{R}}   

\newrobustcmd{\suppmat}{\cite{Schulenborg15Suppmat}}

\newrobustcmd{\Eq}[1]{Eq.~(\ref{#1})}
\newrobustcmd{\Eqs}[1]{Eqs.~(\ref{#1})}
\newrobustcmd{\eq}[1]{(\ref{#1})}
\newrobustcmd{\Fig}[1]{Fig.~\ref{#1}}
\newrobustcmd{\fig}[1]{\ref{#1}}
\newrobustcmd{\Figs}[1]{Figs.~\ref{#1}}
\newrobustcmd{\Sec}[1]{Sec.~\ref{#1}}
\newrobustcmd{\App}[1]{App.~\ref{#1}}
\newrobustcmd{\app}[1]{\ref{#1}}
\newrobustcmd{\Ref}[1]{Ref.~[\onlinecite{#1}]}
\newrobustcmd{\Refs}[1]{Refs.~[\onlinecite{#1}]}

\definecolor{grey}{rgb}{0.75,0.75,0.75}
\definecolor{orange}{rgb}{1.0,0.5,0.5}
\definecolor{brown}{rgb}{0.5,0.25,0.0}
\definecolor{pink}{rgb}{1.0,0.4,0.0}
\definecolor{green}{rgb}{0.0,0.75,0.0}
\definecolor{darkblue}{rgb}{0.0,0.0,0.75}
\definecolor{red}{rgb}{1.0,0.0,0.0}
\definecolor{darkred}{rgb}{1.0,0.0,0.0}

\definecolor{myred}{rgb}{0.9,0,0}
\definecolor{myblue}{rgb}{0,0,0.5}
\definecolor{mygreen}{rgb}{0,0.6,0}
\newrobustcmd{\key}[1]{#1}
\renewrobustcmd{\key}[1]{\textcolor{blue}{#1}} 

\newrobustcmd{\refa}[1]{{\textcolor{black}{#1}}}
\newrobustcmd{\refb}[1]{{\textcolor{black}{#1}}} 
\newrobustcmd{\refc}[1]{{\textcolor{black}{#1}}}
\newrobustcmd{\refabc}[1]{{\textcolor{black}{#1}}}

\usepackage{hyperref} \hypersetup{colorlinks=true,linktoc=all,linkcolor=blue,breaklinks=false,citecolor=blue,urlcolor=blue}

\begin{document}

\title{
Five approaches to exact open-system dynamics:\\
Complete positivity, divisibility
and time-dependent observables
}
\author{V. Reimer}
\email{reimer@physik.rwth-aachen.de}
\affiliation{Institute for Theory of Statistical Physics, RWTH Aachen, Germany}
\author{M. R. Wegewijs}
\affiliation{Institute for Theory of Statistical Physics, RWTH Aachen, Germany}
\affiliation{Peter Gr{\"u}nberg Institut, Forschungszentrum J{\"u}lich, Germany}
\affiliation{JARA-FIT, Germany}
\author{K. Nestmann}
\affiliation{Institute for Theory of Statistical Physics, RWTH Aachen, Germany}
\author{M. Pletyukhov}
\affiliation{Institute for Theory of Statistical Physics, RWTH Aachen, Germany}
\begin{abstract}
To extend the classical concept of Markovianity to an open quantum system, different notions of the divisibility of its dynamics have been introduced.
Here we analyze this issue by five complementary approaches:
equations of motion, real-time diagrammatics, Kraus-operator sums,
as well as time-local (TCL) and nonlocal (Nakajima-Zwanzig) quantum master equations.
As a case study featuring several types of divisible dynamics,
we examine in detail an exactly solvable noninteracting fermionic resonant level
coupled arbitrarily strongly to a fermionic bath at arbitrary temperature in the wideband limit.
In particular, the impact of divisibility on the time-dependence of the observable level occupation is investigated
and compared with typical Markovian approximations.
We find that the loss of semigroup-divisibility is accompanied by a prominent reentrant behavior:
Counter to intuition, the level occupation may temporarily \emph{increase} significantly in order to reach a stationary state with \emph{smaller} occupation,
implying a reversal of the measurable transport current.
In contrast, the loss of \refb{the so-called completely-positive} divisibility is more subtly signaled by the \emph{prohibition} of such current reversals in specific time-intervals.
\refabc{Experimentally, it} can be detected in the family of transient currents obtained by varying the initial occupation.
To quantify the nonzero footprint left by the system in its effective environment, we determine the exact time-dependent state of the latter
as well as related information measures such as entropy, exchange entropy and coherent information.\end{abstract}

\maketitle
\section{Introduction\label{sec:intro}}

\refc{%
The continued experimental progress in nanostructuring has increasingly brought electron transport / transfer, chemical dynamics, and  quantum optics in ever closer contact.
An important point of common interest of these fields is the description of nanostructures which are \emph{open quantum systems}
and a large variety of advanced approaches to this problem continue to be developed
\cite{Rebentrost09,
Timm11,Gasbarri18,
Daley14,
Cohen11,Cohen13a,
Tanimura06,Haertle13,
Wilner13,
Koenig96a,Koenig97,Kubala06,Leijnse08,Koller10,Saptsov14,Ansari15,
Koenig96b,Utsumi05,Kern13,
Pedersen05,Pedersen07,Karlstrom13,
Timm08,Modi12,
Schoeller09,Andergassen11a,Pletyukhov12a,Saptsov12,Kashuba13,Reininghaus14,Lindner18,Schoeller18}.
Of particular interest are approaches based on the reduced density operator
that describes the state of a system without access to or information about its environment.
Formulated like this, it underscores that this quantity is also of key interest in quantum information theory.
The rapid advances in the latter field over the last two decades
have spilled over into the study of open system dynamics
and significantly deepened the understanding, e.g., of `memory effects' and Markovian approximations that neglect them, see \Ref{Rivas14} for a review.
}%

\begin{figure}[t]
	\includegraphics[width=\linewidth]{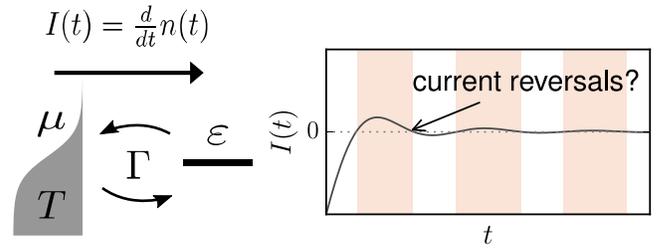}
	\caption{
	\refc{%
	Electron transport involves the tunneling of an electron to and from an electronic reservoir.
	Although a reversal of the transport current $I(t)$ seems like a `non-Markovian' effect,
	there are multiple well-defined notions of `Markovianity' of an open quantum system.
	We investigate in detail their impact on basic transport observables in the noninteracting resonant level model.}
	}%
	\label{fig:intro}
\end{figure}

\refc{%
Anyone interested in such new insights into open systems will immediately notice a formidable gap
between the questions, formulations, and methods used in quantum information 
as compared to traditional approaches known from quantum transport, for example.
The goal of our contribution to this special issue
is to strike a bridge between \emph{basic} techniques and results from quantum information
and the \emph{simplest} possible model of electron transport.
The latter is sketched in \Fig{fig:intro} and
one is interested, for example, in the measurable time-dependent current $I(t)$
to study the time-evolution of the electron level-occupation $n(t)$.
It may at first seem unclear that quantum information concepts such as
`complete positivity' or `coherent information'
have \emph{anything} to do with this problem.
We hope that by the end of this paper it will be evident
how these provide new, interesting and even some surprising insights into transport problems.
}

\refc{%
Throughout the paper, we focus on developing a clear picture of the \emph{dynamics} of such open quantum systems.
Although equivalent general approaches to open-system dynamics
have long ago been formulated in both quantum information~\cite{Sudarshan61,Kraus71} and statistical physics~\cite{Nakajima58,Zwanzig60},
their explicit relation has only recently been pinned down~\cite{Chruscinski13,vanWonderen13,vanWonderen18a,vanWonderen18b,Reimer19a}.
The latter works do not rely on strongly simplifying assumptions
such as weak coupling, high-temperature or off-resonant transport (`superexchange' tunneling)
that have previously enabled information-theoretic discussions of transport models.
Instead, they focus on general parameter regimes relevant to, for example, molecular quantum dot experiments~\cite{Gaudenzi17a}.
For the simple model of \Fig{fig:intro} we consider the combination
of strong tunnel coupling $\Gamma$, finite temperature $T$ and electrically controllable level position ($\varepsilon-\mu$).
Notably, this model has been long known to be exactly solvable~\cite{Caroli71}
and features in textbooks on quantum transport and electron transfer~\cite{ScheerCuevasbook}.
However, we will see that this does not~\cite{Wilner11,Wilner12} imply this problem is fully understood,
even though we do not address the role of Coulomb interactions responsible for various many-body effects in transport measurements~\cite{Gaudenzi17a}.
These aspects are certainly important but beyond the present scope.
}

\refabc{%
The above mentioned recent progress addressed the \emph{microscopic derivation} of the reduced state dynamics $\rho(t)$
of a system $\S$ evolving unitarily [$U(t)$] with an environment $\E$ prepared in a state $\rho^\E$,
\begin{align}
\rho(t) := \Pi(t) \rho(0) = \tr{\E} \, U(t) [ \rho(0) \otimes \rho^\E ] U(t)^\dag
.
\label{eq:pi}
\end{align}
In particular, the so-called complete positivity (CP) of the map $\Pi(t)$ was accounted for.
As we will explain in more detail later on, this property ensures that the reduced state evolves correctly even in the presence of \emph{entanglement}.
}%
\Ref{Chruscinski13} put forward a projection approach for computing the square root $\sqrt{\rho(t)}$ of the state in order to guarantee CP of the dynamics for all $t$.
\Refs{vanWonderen13,vanWonderen18a,vanWonderen18b} instead extended earlier work~\cite{vanWonderen05,vanWonderen06}
\refc{
to microscopically derive the dynamics in form of a Kraus operator-sum~\cite{Sudarshan61,Jordan61,Hellwig69,Kraus71} that explicitly encodes the CP property:
}
\begin{align}
\rho(t) = \sum_m K_m(t) \rho(0) K_m(t)^{\dag}
\label{eq:kraus}
.
\end{align}
The scope of the latter approach has been extended~\cite{Reimer19a} by combining standard techniques of quantum information (purification) and statistical physics (normal-ordering of field operators).
It was shown how the Keldysh diagrammatic series can be reorganized into groups of diagrams which represent well-defined \emph{physical processes}
$\Pi_m(t)\rho(0)=K_m(t) \rho(0) K_m(t)^{\dag}$
and describe an evolution \emph{conditioned} on a specific outcome $m$ of a possible measurement on the environment.
\refc{%
Importantly, the Kraus operator-sum is advantageous beyond guaranteeing CP.
The set of \emph{individual} Kraus operators --not just their operator-sum-- provides further information
about the \refb{coupling of the system to} its environment~\cite{Addis14,Chruscinski11,Chruscinski17,Rivas14,Utsumi15}
in a way that is impossible to achieve otherwise.
As we will explain, they allow the open system's \emph{effective environment} to be quantitatively analyzed
by determining various information measures such as the entropy produced in this environment (exchange entropy).
}%

\refc{
Another point of recent
}%
interest~\cite{Breuer16rev} is whether the dynamics is divisible,
\begin{align}
\Pi(t) = \Pi(t,t') \, \Pi(t') \quad \text{for all } t \geq t' \geq 0
,
\label{eq:pi-cpdiv}
\end{align}
in analogy to the classical notion of Markovianity.
In this context, the CP property also plays a crucial role:
It defines \emph{CP-divisibility}~\cite{Rivas10,Benatti17} which has various clear characterizations in terms of information measures~\cite{Chruscinski16},
but whether it can be considered an extension of Markovianity to the quantum case is still under debate~\cite{Modi19}.
\refc{%
Here we are chiefly interested in understanding its physical implications in the simple setting of \Fig{fig:intro}:
How does CP-divisibility show up in the measurable current $I(t)$
and how does it compare to the classically-inspired notion of semigroup-divisibility?
In addressing this issue
}%
one encounters the problem that
these distinctions typically become pronounced
for systems strongly coupled to continuous environments.
As a result, rigorously investigating their measurable implications is often hindered
by the practical necessity of making approximations in the microscopic derivation of the dynamics,
in particular for strongly interacting open systems with finite-temperature environments.
Whereas weak-coupling approximations typically lead to semigroup dynamics $\Pi(t,t')=\Pi(t-t')$
governed by \refabc{a `Lindblad' quantum master equation
due to Gorini, Kossakowski, Sudarshan~\cite{GKS76} and Lindblad~\cite{Lindblad76} (GKSL)},
microscopic derivations~\cite{Vacchini10,Ferialdi16,Bhattacharya17} may also give rise to more general \emph{time-dependent} GKSL equations.
The latter are known to generate CP-divisible dynamics if the corresponding time-dependent decay rates are nonnegative for all times~\cite{Rivas10}.
\refabc{%
Importantly,
}%
the recent generalizations~\cite{Chruscinski13,vanWonderen13,Reimer19a} that we mentioned
\refabc{%
above
}%
apply beyond these cases
but it is unclear how they reflect divisibility properties.
Therefore, exact solutions of microscopic models where these issues can be understood analytically are of interest.

In search of solvable examples with interesting dynamics, much attention has been given to systems coupled to bosonic environments,
such as the spin-boson and dissipative Jaynes-Cummings~\cite{Leggett87,Scala07,Ferialdi2017a,Ferialdi2017a-erratum,vanWonderen18a} model.
\refc{
Here, we are interested in fermionic models appearing in quantum transport problems, e.g., through strongly coupled quantum dots,
which offer} rich non-equilibrium dynamics due to a variety of many-body effects~\cite{Buot,Andergassen10,Wilner14a,Gaudenzi17a}.
\refc{%
Somewhat surprisingly, the \emph{single} fermionic mode in \Fig{fig:intro}
coupled arbitrarily strongly ($\Gamma$) to a noninteracting fermionic reservoir at arbitrary temperature ($T$)
already features several types of divisible dynamics in different parameter regimes.
}%
It therefore serves as an ideal playground to investigate the above mentioned questions.

\refc{%
A full characterization of the effective environment and the divisibility properties of the dynamics
}%
calls for explicit construction of both the dynamical map $\Pi(t)$ as well as the divisor $\Pi(t,t')$
which seems not to have been discussed for this model despite its long-known exact solvability~\cite{Caroli71}.
\refc{%
To achieve this, the exact solution is approached from several angles in a consistent way.
The importance of this is appreciated when we first
pinpoint the limitations of more traditional approaches (Green's functions, quantum master equations)
before applying less familiar methods (super-fermions in Liouville space, Kraus operators)
to exhaustively study the dynamics of the resonant level in \Fig{fig:intro}.
This will also provide new insights into possible approximations and their implications for generic parameter regimes.
We believe that the benefits of such an in-depth exchange of insights from different research fields outweigh the considerable effort.
}%

The paper is organized as follows.
\refc{%
In \Sec{sec:model} we introduce the transport model of \Fig{fig:intro}
and discuss the simplifying features that enable its exact solution.
Then, in \Sec{sec:open-system} we summarize some general concepts of open system dynamics from an information perspective,
in particular complete positivity and different notions of divisibility.
Doing so, we hope to provide an accessible introduction to some \emph{relevant} jargon of quantum information theory
and motivate the connections to transport problems considered here.
Together with the in-depth discussion in \App{app:entropies}, this should provide sufficient motivation and guidance to consult
standard quantum information literature~\cite{NielsenChuang}
where the connection to open-system dynamics is harder to see.
}%
In \Sec{sec:eom} we first obtain the dynamical map $\Pi(t)$ using Heisenberg equations of motion~\cite{Lehmberg70,Lalumiere13,deVega17} (EOM) for observables.
\refc{This closely relates to Green's function techniques used in electron transport / transfer theory~\cite{Ryndykbook,ScheerCuevasbook}.}
\Sec{sec:superfermion} derives this result from the complementary perspective of state evolution
using the real-time~\cite{Koenig96a,Koenig96b,Schoeller09,Schoeller18} superfermion approach~\cite{Saptsov12,Saptsov14,Schulenborg16}
revealing further properties and physical insights.
In \Sec{sec:kraus} we then \refabc{employ established quantum information techniques and} construct the Kraus operator-sum from these solutions
in order to compute information measures quantifying the system-environment backaction.
In \Sec{sec:qme} the three representations are then used to construct two exact quantum master equations \refabc{for the transport problem},
one time-local or time-convolutionless (TCL)~\cite{Shibata77,Hashitsume77,Chaturvedi79,Shibata80}, and one time-nonlocal (Nakajima~\cite{Nakajima58}-Zwanzig~\cite{Zwanzig60}).
\refc{%
These are extensively used in electron transport, chemical kinetics and quantum optics
since they are more suitable for setting up approximations.
}%
Finally, in \Sec{sec:modes} we construct the exact eigenvectors of the dynamical map $\Pi(t)$ as function of time
and bring the insights from all discussed approaches together.
\refabc{%
We analyze the detailed time-evolution of the level occupation in dependence
of the parameters that control the transport properties.
}%
Those readers particularly interested in \refabc{the model per se} may first traverse \Sec{sec:model} and \ref{sec:eom}
and skip to \Sec{sec:modes}.
Since we expect the reader is not familiar with all five approaches, detailed derivations are collected in the appendices.
Throughout the paper we set $\hbar=k_\text{B}=1$
and denote $\lim_{t \to \infty}f(t)=f(\infty)$.

\section{Resonant level model\label{sec:model}}

\refc{%
The model depicted in \Fig{fig:intro}
}%
describes a single, noninteracting fermionic mode (field operator $d$ at energy $\varepsilon$)
in tunneling contact with a continuous reservoir of fermions (field operator $b_\omega$ at energy $\omega$), all without spin:
\begin{align}
H^\tot = \varepsilon d^\dagger d + \int d\omega \, \omega \, b^\dagger_\omega b_\omega + \sqrt{\frac{\Gamma}{2\pi}}\int d\omega \big(d^\dagger b_\omega + b^\dagger_\omega d \big).
\label{eq:full-hamiltonian} 
\end{align}
\refa{%
As often, each mode of the environment at energy $\omega$ is coupled with a constant tunneling rate $\Gamma > 0$.
}
The environment constitutes a reservoir in thermal equilibrium at temperature $T$ and chemical potential $\mu$
\refa{with the} density operator $\rho^{\E} = \exp \big( - (H^\res-\mu N^\res)/ T  \big) / Z$.
\refa{%
Here $H^\res$ is the second, reservoir-energy term in \Eq{eq:full-hamiltonian}
and the reservoir number operator reads $N^\res := \int d\omega \, b^\dagger_\omega b_\omega$.
The following analysis is much simplified by labeling
}%
field operators with a particle/hole index $\eta= \pm$,
\begin{align}
	d_{\eta}
	=
	\begin{cases}
		d^\dag & \eta = +\\
		d      &  \eta = -
	\end{cases}
	,
	\quad
	\qquad
	b_{\eta \omega}
	=
	\begin{cases}
		b_\omega^\dag & \eta = +\\
		b_\omega      &  \eta = -
	\end{cases}
	,
	\label{eq:eta}
\end{align}
and denoting $\bar{\eta}=-\eta$ such that $d_{\eta}^\dag = d_{\bar{\eta}}$.
\refa{%
Furthermore, we do not use the common convention of setting
}%
$\mu=0$ as this may later on lead to
a confusion of $\varepsilon$ with the level \emph{detuning} $\epsilon := \varepsilon - \mu$, cf. \Eq{eq:krausfinal}.

\refc{%
This model and its extensions to multiple reservoirs have been extensively studied in the context of transport through molecular junctions
where in particular the strong coupling $\Gamma$ to the environment is important~\cite{ScheerCuevasbook}.
Ongoing experimental realizations have revealed interaction effects to be also important for the current-voltage characteristics, see~\Ref{Gaudenzi17a} for a recent review.
Similar trends are observed in models for quantum optics where nanocavities or on-chip waveguides for microwave photons~\cite{Berry09,Jalali10}
not only enable access to the strong-coupling regime but also pave the way to strong, controllable local interactions of photons,
e.g., in Kerr nonlinearities.
In the following we will employ several methods that are in principle capable of
treating these much more difficult models
but refrain from addressing interaction effects
for the sake of a clear comparison with the quantum-information approach. 
Our study thus provides both guidance and a benchmark for these technically more challenging problems.
}%

\refa{%
One of the insights provided by these general methods is that the key observable of a fermonic mode is not so much its occupation $n=d^\dag d$, but rather its \emph{fermion-parity} $(-\one)^n:=e^{i \pi n}=\one-2n$.
Throughout the paper we will thus formulate and discuss the problem of \emph{transient} transport of electrons in terms of the parity operator. \Sec{sec:superfermion} most clearly explains why this is more than a trivial change of variable.
Roughly speaking,%
}
we expect that non-infinite temperature $T<\infty$ leads to memory effects%
		\footnote{
		Additional memory effects appear when lifting the assumption of
		infinite reservoir bandwidth (unrestricted $\omega$ integral) and $\omega$-independent coupling $\Gamma$,
		see page 8 of \Ref{Saptsov12} and also, e.g., \Refs{Roszak12,Ubbelohde12}.
		}
since excitations created in the environment propagate on a finite time-scale $T^{-1}$ before coupling back to the system.
Explicitly following this fact in the derivations of the dynamical map $\Pi(t)$ is both instructive and technically advantageous,
also for more complicated systems~\cite{Saptsov12,Saptsov14,Schulenborg16},
\refa{%
and leads one to consider the fermion-parity.
Also, }%
the infinite-temperature limit of the model plays a central role as the dynamics becomes semigroup-divisible.

It is useful to briefly consider the translation of our model into
\refa{%
the perhaps more familiar language of spin
}
with 
$S^{\pm} = d_{\pm}$, $S^z = d^\dag d -\one/2 = - (-\one)^n/2$ and analogous definitions for the environment.
In this picture one may identify the average parity $\brkt{(-\one)^n}$ and field $\brkt{d}$
with the longitudinal and transverse components of the spin (Bloch vector) describing diagonal and off-diagonal density-matrix elements, respectively.
In fermionic systems, the field amplitude $\brkt{d(t)}$ vanishes at all times due to the fermion-parity superselection~\cite{Wick52,Aharonov67}
and we will
\refabc{%
use this fact
}%
throughout.
However, in the spin formulation it makes sense
to consider $\brkt{d(t)} \neq 0$ and we will comment on this.
Considered as a spin-model, \Eq{eq:full-hamiltonian} is somewhat unconventional:
\begin{align}
H^\tot = \varepsilon S^z + \int d\omega \, \omega \, s^z_\omega
+ \sqrt{\frac{\Gamma}{2\pi}}\int d\omega
\big(S^{+} s^{-}_\omega + s^{+}_\omega S^{-}  \big).
\label{eq:full-hamiltonian-spin} 
\end{align}
It corresponds to purely transverse exchange (spin-flip) \refb{coupling} between a local spin in a magnetic field of strength $\varepsilon$
and a reservoir of spins in magnetic fields of varying strength and direction $\omega$.

In this language, the model can be compared with the spin-boson model~\cite{Leggett87}
of a two-level system exchanging energy with a bosonic thermal bath.
This \refa{latter} model has been solved exactly in a rotating-wave-approximation (RWA) and at zero temperature of the bath
leading to the semigroup Wigner-Weisskopf theory~\cite{WignerWeisskopf30}
but recent extensions included also finite-temperature effects going beyond the RWA~\cite{Ferialdi2017a,Ferialdi2017a-erratum}.
The same holds~\cite{Scala07} for the more complicated setup of the dissipative Jaynes-Cummings model~\cite{JaynesCummings63}
where the two-level system exchanges energy with a radiation field via a single bosonic cavity mode.
An exact treatment is quite complicated in both cases and often restricted to a description in terms of quantum master equations,
see however the Kraus operator-sum treatment of \Ref{vanWonderen18a}.

\begin{figure}[t]
	\includegraphics[width=\linewidth]{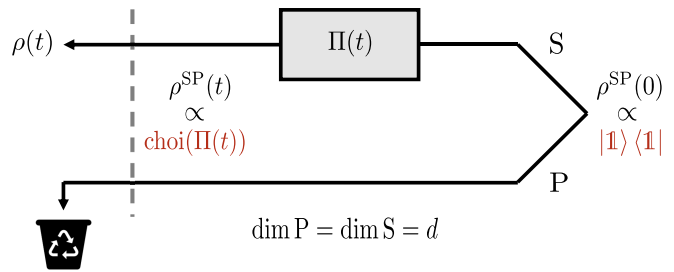}
	\caption{
		Complete positivity.
		The evolution of an initially mixed state $\rho(0)$ of the system $\S$ (upper line) is equivalent to that of a pure state $\rho^{\S\P}(0)$
		entangled (wedge) with a preparing system $\P$ (lower line) of the same dimension $d$ which is eventually traced out (trashcan).
		Evolving $\S$ in time by $\Pi(t)$ (from right to left) without affecting $\P$ results in a positive operator
		$\rho^{\S\P}(t) = (\Pi(t) \otimes \ones) \rho^{\S\P}(0)$ if and only if $\Pi(t)$ is completely positive.
		It is sufficient to check this condition for the Choi operator [\Eq{eq:choi-def}]
		which is obtained when starting from a \emph{maximally} entangled state (`worst case') as indicated in red.
		}
	\label{fig:cp}
\end{figure}

\section{Open system dynamics\label{sec:open-system}}

\refc{%
Even for the above simple model it is of interest to consider
several different representations of the exact dynamical map~\eq{eq:pi}
since these provide access to complementary properties.
In \Sec{sec:eom}-\ref{sec:modes} we will find that
a particular property which seems trivial in one representation
can be very difficult to scrutinize in another.
}%
\refabc{
Here we first outline the properties and representations of interest
and highlight the importance of entanglement creation and distribution between the system and its environment,
i.e., their encoding of information in quantum correlations.
}%

\subsection{Complementary forms of $\Pi(t)$ -- CP \emph{versus} TP}
A primary distinction between the different approaches
derives from two fundamental properties of the dynamical map $\Pi(t)$.
First, trace preservation (TP) requires that the trace over the system $\S$
can be passed through the action of the dynamical map, $\Tr_{\S} \Pi(t)= \Tr_{\S}$,
such that the normalization of the initial state $\Tr_\S \rho(t)=\Tr_\S \rho(0)$ is unaffected for all times $t$.
The second property of $\Pi(t)$ is complete positivity (CP) which not only ensures that the state $\rho(t)=\Pi(t)\rho(0) \geq 0$ remains positive
for every positive initial state $\rho(0)$.
It more strongly guarantees that this still holds true if $\Pi(t)$ acts on half of a positive joint state $\rho^{\S\P}(0)$
of any extension of the system $\S$ by $\P$, $(\Pi(t) \otimes \ones) \rho^{\S\P}(0) \geq 0$, 
see \Fig{fig:cp} for a review.
\refabc{Put differently, understanding and keeping track of CP requires one to consider 
the open system dynamics in combination with its entanglement with other systems.}

These properties derive from the fact that the joint evolution of system and environment in \Eq{eq:pi} is unitary
and that the initial system-environment state factorizes into $\rho(0) \otimes \rho^\E$.
\emph{After} having traced out the environment,
the unitary $U(t)$ and the environment state $\rho^\E$ are however no longer explicitly available
and it becomes difficult to keep track of \emph{both} the CP and TP property~\cite{Chruscinski16,Reimer19a}.
There are two \emph{forms}~\cite{Watrous,Milz17,NielsenChuang,BreuerPetruccione} of the dynamical map $\Pi(t)$
which are tailored to highlight one of these properties.
However, the complementary property then becomes more difficult to check in practice.

\emph{Superoperator form.}
In general, both the equations of motion [\Sec{sec:eom}] and real-time approach [\Sec{sec:superfermion}] prove to be advantageous for obtaining
the exact dynamical map in a Liouville-space superoperator form.
There, the action of $\Pi(t)$ --represented by a $d^2 \times d^2$ matrix with $d=\dim \S$--
on the reduced state $\rho(0)$ --represented as a supervector of dimension $d^2$--
is given by a mere matrix-vector multiplication, in close analogy to operators and vectors in the usual Hilbert space.

This form is suitable for studying the eigenmodes of $\Pi(t)$ as presented in \Sec{sec:modes}.
In fact, the TP property can be naturally considered as a left super-eigenvector equation
$\Bra{a_1} \Pi(t)=1 \cdot \Bra{a_1}$, where $\Bra{a_1}:= \text{Tr}_{\S}$ denotes the trace operation.
Therefore, the TP property may be easily verified even without explicitly diagonalizing $\Pi(t)$.
On the other hand, the CP property is in general not directly related to the eigenspectrum of $\Pi(t)$ itself.
Instead, a necessary and sufficient~\cite{dePillis67,Jamiolkowski72,Choi75} condition for CP is 
that the spectrum of the associated \emph{Choi operator}
\begin{align}
	\text{choi}(\Pi(t)) := (\Pi(t) \otimes \ones ) \ket{\one}\bra{\one} \geq 0
\label{eq:choi-def}
\end{align}
is nonnegative.
It represents a mixed state obtained from the action of $\Pi(t)$ on half of a maximally entangled pure state
$\ket{\one}=\sum_k \ket{k}\otimes\ket{k}$ of the extended system $\S \otimes \P$,
see \Fig{fig:cp}.
The mapping $\Pi \mapsto \text{choi}(\Pi)$ nontrivially mixes up both the eigenvectors \emph{and} eigenvalues of $\Pi(t)$
such that the TP property can no longer be related to a \emph{single} eigenvector of the Choi operator, as we show below.

\emph{Operator-sum form.}
Diagonalization of the Choi operator \eq{eq:choi-def} leads to the dynamical map in the operator-sum form \eq{eq:kraus}:
Normalizing the eigenvectors to their nonnegative eigenvalues $\braket{K_m}{K_{m'}}=\lambda_m \delta_{m m'}$, we have
\begin{align}
\text{choi}(\Pi(t)) = \sum_m \ket{K_m(t)} \bra{K_m(t)} 
\label{eq:choi-kraus}
\end{align}
and using $\ket{K_m(t)} = K_m(t) \otimes \one \ket{\one}$ one obtains a finite set of Kraus operators.
The size of this set, $\text{rank} [ \text{choi}(\Pi(t)) ] \leq d^2=\text{dim} \, \S \otimes \P$,
is minimal relative to any other operator-sum~\footnote{%
	Any two Kraus-operator sets are unitarily [$V(t)$] related~\cite{NielsenChuang,Watrous} as $K_{m}(t)=\sum_{m'} K'_{m'}(t) V_{m'm}(t)$.
	We note that methods which directly compute the Kraus operators from a microscopic expansion~\cite{vanWonderen13,vanWonderen18a,vanWonderen18b,Reimer19a}
	start from a representation with \emph{infinitely} many operators $K_m(t)$ and finding the unitary $V(t)$ is difficult.} 
and depends on the physical parameters of the model as well as on \emph{time}.

The CP property of \Eq{eq:kraus} is now apparent from the quadratic form of each term:
Any joint state $\rho^{\S \P}(t)$ remains positive under the action of $(K_m(t) \bullet K_m(t)^\dag) \otimes \ones
= K_m(t) \otimes \one \bullet K_m(t)^\dag \otimes \one$.
In contrast, the TP property has become complicated to check because of the mixing of eigenvectors and eigenvalues
involved with the Choi mapping pointed out above.
It now requires a nontrivial relation between \emph{all} Kraus operators to hold at all times:
\begin{align}
\sum_m K_m(t)^\dagger K_m(t) = \one
\label{eq:kraus-tp}
.
\end{align}
More generally, the structure of the spectral decomposition of $\Pi(t)$ remains hidden in nonlinear relations of this type,
even for our simple model, see \Sec{sec:modes} and \App{app:kraus-sumrule}.

\begin{figure*}[t]
	\includegraphics[width=\linewidth]{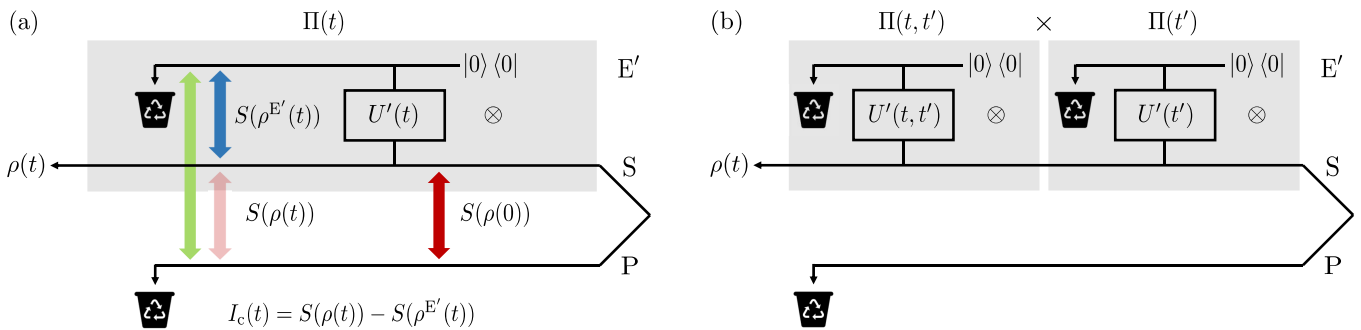}
	\caption{
		(a) Coherent information:
		Equivalent representation of \Fig{fig:cp} indicating the effective environment $\E'$ that is initially pure  and uncorrelated ($\otimes$) with the system $\S$.
		During the unitary joint evolution $U'(t)$, the entanglement of the system $\S$ with its initial preparation system $\P$ (red) is transformed
		into entanglement of $\E'$ with both the system (blue) and the preparation (green) until it is fully consumed in the stationary limit, cf. \Eq{eq:entropy-env-stat}.
		The coherent information $I_\c(t)$ describes the competition between the first two through the difference of their entanglement entropies.
		A strictly zero (positive) mismatch $S(\rho(0))-I_\c(t)$ is known~\cite{NielsenSchumacher96,Preskill} to quantify the (in)ability
		to recover the state of the preparing system $\P$ by processing only the output $\rho(t)$.
		(b) Physical divisibility:
		Interrupting the evolution by a complete measurement of the effective environment at $t'$ and subsequently discarding the outcomes disrupts the system.
		If the evolution is CP-divisible, it is nevertheless possible to reach the uninterrupted final state $\rho(t)$
		by \refb{joint unitary evolution} with a new effective environment which is again discarded. 
		The evolution cannot be the same as the original unless the evolution has the stronger property of being semigroup-divisible.
		}
	\label{fig:div}
\end{figure*}

\subsection{Effective environment and information measures\label{sec:information}}

With the \emph{individual} Kraus operators $K_m(t)$ available,
one can quantify the backaction on the environment due to its coupling to the system.
\refabc{This however requires a `clean count' of the dynamical correlations originating from this coupling,
i.e., we need to distinguish them from the thermal correlations of the environment.}
To see this, one makes use of the converse statement of Kraus' theorem~\cite{Stinespring55,Kraus71}:
Any CP-TP map $\Pi(t)$ can be considered to arise from a joint unitary evolution with an initially uncorrelated \emph{pure}%
	\footnote{Representations with \emph{mixed} effective environments may exist, but only if certain additional conditions hold,
	see, e.g., the appendix `Mixed-state measurement models and extreme operations' in \Ref{Caves-lecture}.}
environment which is eventually discarded.
This allows one to replace in \Eq{eq:pi}
the initial reservoir state $\rho^{\E}(0) \to \rho^{\E'}(0):=\ket{0} \bra{0}$ by an \emph{effective} environment $\E'$ in a pure state,
and the corresponding isometry\footnote
	{
	One can in general extend $U'(t) \to \sum_{m} K_m(t) \otimes \ket{m}\bra{0} + X(t)$ to be unitary~\cite{NielsenChuang,Schumacher96b}.
	However, the additional term obeys the property $X(t)\ket{0}=0$ and thus drops out of all physical quantities
	because the effective environment starts out in a pure state $\ket{0}$.	
	}
$U(t) \to U'(t) = \sum_{m} K_m(t) \otimes \ket{m}\bra{0}$ by a tensor product of the system Kraus operators and environment projectors.
The label of the state $\ket{0}$ corresponds to the index of the single initial Kraus operator $K_m(0) = \delta_{m,0} \one$
which is the identity operator because of $\text{choi}(\Pi(0))=\ket{\one}\bra{\one}$, cf. \Eq{eq:choi-def}.
The resulting reduced dynamics $\Pi(t)$ after tracing out the effective environment $\E'$ remains unaltered.
However, the advantage of the above choice is that the state of the effective environment obtained by instead tracing out the \emph{system}%
\footnote{%
The effective environment state $\rho^{\E'}(t)$ is the result of the action of the so-called \emph{complementary} quantum channel associated with $\Pi(t)$.
It plays a key role in transmission of quantum information~\cite{Preskill,Wilde13} and measurement-information tradeoff~\cite{Holevo12}.%
}%
,
\begin{align}
\rho^{\E'}(t) = \sum_{mm'} \ket{m} \, \Big( \tr{\S} K_m(t) \rho(0) K_{m'}(t)^\dag \Big) \, \bra{m'}
,
\label{eq:rho-E'}
\end{align}
is now represented by a matrix which only features the Kraus operators of $\Pi(t)$ and the initial system state $\rho(0)$.

Knowledge of the density \emph{matrices} for both system $\rho(t)$ and effective environment $\rho^{\E'}(t)$ allows
the computation of interesting information measures based on the quantum entropy $S( x ) := - \text{Tr} \, x \log_2 x$
which are reviewed in more detail in \App{app:entropies-coh}.
The competition between the entropy of the system and the effective environment determines the \emph{coherent information}:
\begin{align}
	I_\c(t):= S( \rho(t) ) - S(\rho^{\E'}(t))
	.
	\label{eq:I}
\end{align}
For pure initial states $\rho(0)$, the entropies can be shown to coincide such that $I_\c(t)=0$ for all $t$.
More strongly, the entire \emph{spectra} of $\rho(t)$ and $\rho^{\E'}(t)$ coincide in this case.
If, however, $\rho(0)$ is \emph{mixed}, the coherent information $I_\c(t) \neq 0$ provides additional insight
into how well entanglement is preserved during the time-evolution as illustrated in \Fig{fig:div}(a):
A mixed state of the system $\S$ can only be prepared with the help of an auxiliary system $\P$,
and it is the entanglement with this system (red) that the coherent information keeps track of.
During the evolution towards a unique stationary state
it is continuously being converted into entanglement between the effective environment $\E'$ and $\S$ (blue), respectively $\P$ (green),
until it is entirely broken.
In the stationary limit the dynamical map then takes the form
of an entanglement-breaking map [\App{app:entropies-factor-stationary}] 
which implies that the stationary effective environment factorizes
\begin{align}
\rho^{\E'}(\infty) = \rho(\infty) \otimes \rho(0)
\label{eq:entropy-env-stat}
\end{align}
into the initial and stationary \emph{system} density matrix.
From a physical perspective, a state tomography of the effective environment in principle allows one to extract the initial and final data of the system evolution:
The initial state is not lost but can be found explicitly in the effective environment.
What matters here is that only the effective environment keeps a `clean' count of the \refb{coupling} of the system with the outside world.
\refabc{This is overlooked in traditional open-system approaches which do not provide access to this effective environment through the Kraus operator-sum.}

Finally, the mismatch of the coherent information with the initial system entropy,
\begin{align}
	S(\rho(0))- I_\c(t)
	=  
	S(\rho^{\E'}(t)) - [S(\rho(t))-S(\rho(0))]
	\geq 0
	,
	\label{eq:mismatch}
\end{align}
is the difference between entropy \emph{productions} in the system and effective environment with $S(\rho^{\E'}(0))=0$.
This quantity is always nonnegative due to the fundamental triangle inequality for quantum entropies~\cite{LiebAraki}.
\refa{%
The inequality \eq{eq:mismatch}
implies that the entropy produced in the effective environment is always larger than the entropy produced in the system if the latter is positive.
}
Due to \Eq{eq:entropy-env-stat}, it reaches the maximal value $2S(\rho(0))$ in the stationary limit
indicating that the entanglement between $\S$ and $\P$ has been completely broken in favor of entanglement between $\S$ and $\E$ and, separately, between $\P$ and $\E$.

\subsection{Complete positivity and divisibility}

The nontrivial constraint of complete positivity is also important when inquiring whether the dynamics is divisible [\Eq{eq:pi-cpdiv}].
This question arises when aiming to extend the notion of Markovianity to the quantum realm where it is not as clear-cut as in the classical case.
Analogous to the classical case, one may first define Markovianity as a factorization of the dynamics at \emph{any} intermediate time $0 \leq t' \leq t$
into a repeated action $\Pi(t)=\Pi(t-t') \Pi(t')$ of the \emph{same} CP-TP evolution over different time-intervals.
This \emph{semigroup} property implies that the dynamics is insensitive to a re-initialization of the environment at any time $t'$,
and the splitting is always physically meaningful because $\Pi(t-t')$ is CP-TP for all $t\geq t'$.

Evolutions that fail to be a semigroup, may however \emph{still} be considered Markovian in a weaker sense:
A factorization $\Pi(t)=\Pi(t,t') \Pi(t')$ of the CP-TP map $\Pi(t)$ into two \emph{different} evolutions at any time $0 \leq t' \leq t$
is still physically meaningful if we explicitly require the \emph{divisor} $\Pi(t,t')$ to be CP-TP as well.
This \emph{CP-divisibility} then implies according to Kraus' theorem [\Sec{sec:information}] that the dynamics is also insensitive to a re-initialization of the environment,
but now with a different unitary evolution $U'(t,t')$ which depends parametrically on the time $t'$, see \Fig{fig:div}(b).
If the divisor $\Pi(t,t')$ is not CP-TP, this argument breaks down and divisibility becomes meaningless:
Whenever $\Pi(t)$ is invertible, it is also divisible without further constraining the divisor.
As we shall see, this is indeed true for our model [\Eq{eq:g2-h}]
but the dynamics fails to fulfill both the semigroup-divisibility and the CP-divisibility criterion in broad parameter regimes.
In \Sec{sec:modes} we will, however, see that whereas the loss of the semigroup-divisibility is clearly reflected in observables,
the loss of CP-divisibility is more subtle to detect.

Formally, semigroup- and CP-divisibility can most easily be assessed from the quantum master equation (QME) generating the evolution
\refa{and this motivates our considerations in \Sec{sec:qme}.}
Both properties require that it has the special GKSL form
\begin{gather}
\tfrac{d}{dt} \rho(t) =  -i [H(t),\rho(t)]
\label{eq:gksl} \\
+ \sum_m \kappa_m(t) \Big( L_m(t)\rho(t) L_m(t)^\dag - \tfrac{1}{2}[L_m(t)^\dag L_m(t),\rho(t)]_{+} \Big)
,
\notag    	
\end{gather}
where the coefficients $\kappa_m(t)$ of the jump operators $L_m(t)$ should be nonnegative~\cite{Rivas10} for all times $t$ (CP-divisibility),
or both the nonnegative coefficients and the jump operators are time-independent~\cite{GKS76,Lindblad76} (semigroup-divisibility).
In either case, the exponential solution allows one to construct the Kraus operators~\cite{Andersson07,Prosen08,Prosen10}.

It should be stressed what a failure of these conditions implies:
While a time-independent GKSL form with negative coefficients $\kappa_m < 0$ indicates a breakdown of the CP property,
\refa{%
one cannot conclude that CP fails if the time-dependent coefficients are temporarily~\cite{Megier17} negative, $\kappa_m(t) < 0$.
}%
This situation is encountered even in our simple model.
The dynamics may then \emph{still} be CP but in order to explicitly see this,
one either has to rely on model-specific forms which reveal this property [\Sec{sec:superfermion}]
or construct the operator-sum [\Sec{sec:kraus}] for the dynamical map $\Pi(t)$.
Explicit construction of the divisor $\Pi(t,t')$ allows one to investigate what happens
\refa{%
in such cases where CP-divisibility \emph{fails}.
}%

\section{Equation of motion approach\label{sec:eom}}

We start the discussion of the exact solution from the Heisenberg equations of motion (EOM) for a set of observable operators.
\refabc{%
This is familiar, e.g., from Green's function techniques~\cite{MeirWingreen,Caroli71,Rammer,Rabani15,StefanuccivanLeeuwen,Ryndykbook}
used extensively in electron transport theory
or input-output formalisms in quantum optics~\cite{Lehmberg70,Lalumiere13,GardinerZoller,deVega17}.
}%
Often, the approach is used in conjunction with various approximations and/or limits~\cite{Clerk10,Lalumiere13,Reimer18a}.
However, in this case we can follow it through exactly due to the wideband-limit~\cite{Caroli71}.

\subsection{Equations of motion and solution for $\Pi(t)$}
As explained in \App{app:eom-general}, the superoperator \eq{eq:pi} can be expanded as
\begin{align}
\Pi(t)
& = \sum_{A} A \, \tr{\S} \{ \brkt{A(t)^\dag}^\E \bullet \}
\label{eq:pi-op-expansion}
\end{align}
in terms of a complete orthogonal set of system operators $A$ normalized as $\Tr_{\S} A^\dag A=1$.
Here, $A(t):=U(t)^\dag A U(t)$ is the Heisenberg-picture operator which for $t>0$ also acts on the environment.
After performing the partial average over the environment, $\brkt{A(t)}^\E := \Tr_{\E} A(t) \rho^\E$, it is reduced to a system-only operator,
and expectation values $\brkt{A(t)} := \text{Tr}_{\S} \brkt{A(t)}^\E \rho(0)$ of such operators
are obtained by additionally averaging with respect to the initial system state.
To compute the partial averages $\left\langle A(t) \right\rangle^\E$ one first sets up the Heisenberg EOM for $A(t)$,
\begin{align}
\tfrac{d}{dt} A(t) = i[H^{\text{tot}}(t),A(t)]
,
\end{align}
and complements these by equations of motion for additional environment operators until the system of equations closes.
Taking the partial averages \emph{before} integrating these equations simplifies matters
since by Wick's theorem terms containing an odd number of environment fields vanish, $\brkt{b_{\eta\omega}}^\E = 0$,
and pairs of environment fields correspond to Fermi-distribution functions
\begin{align}
\brkt{b_{\eta'\omega'} b_{\eta\omega} }^\E
&= \delta_{\eta',\bar{\eta}}
\, 
\delta(\omega'-\omega) \left[ \frac{1}{2} + \frac{\eta}{2} \tanh \left( \frac{\omega-\mu}{2T} \right) \right]
\label{eq:fermi-distrib}
.
\end{align}
The splitting into parts symmetric and anti-symmetric in $\omega$
is convenient because it corresponds to time-local and time-nonlocal contributions to the evolution,
cf. \Sec{sec:superfermion}.

We choose the orthogonal system operator basis $\{d_{+},d_{-},\one,(-\one)^n \}$ which simplifies matters in two ways:
First, the dynamics $\brkt{\one(t)}^\E = \one $ of the identity is trivial and
second, the parity plays a crucial role in fermionic systems which becomes explicit in the superfermion method of \Sec{sec:superfermion}.
There, the fermion-parity superselection is exploited and allows an application even to interacting multi-level models~\cite{Schulenborg14,Schulenborg16,Vanherck17,Schulenborg18a}.
To close the equations of motion, it suffices to add the environment fields $b_{\eta \omega}$
which are conveniently chosen~\cite{Saptsov14} to \emph{anticommute} with system fields: $[d_{\eta'}, b_{\eta \omega}]_{+}=0$.
Wick-averaging the EOM over the environment then yields [\App{app:eom-parity}]:
\begin{subequations}
\begin{align}
 \frac{d}{dt}  \brkt{d_\eta(t)}^\E
&= i\eta\left(\varepsilon+i\eta\tfrac{1}{2} \Gamma \right) \brkt{ d_\eta(t) }^\E
\label{eq:eom-averaged-field}
\\
\frac{d}{dt} \brkt{ (-\one)^n (t)}^\E
&= - \Gamma \brkt{(-\one)^n (t)}^\E + \Gamma h(t) \, \one
.
\label{eq:eom-averaged-parity}
\end{align}
\label{eq:eom-averaged}%
\end{subequations}%
In \Eq{eq:eom-averaged-parity}, the inhomogeneous term is determined by
\begin{align}
h(t):=\int_0^t ds e^{-\frac{\Gamma}{2}s} \gamma(s) 
,
\label{eq:h-def}
\end{align}%
which involves the Keldysh \emph{correlation function}
\refb{%
of the fields $b(t):= \int d \omega e^{i \omega t} b_{\omega} / \sqrt{2\pi}$.
}%
\begin{subequations}
\begin{align}
\gamma(s)
&:=
\refb{%
2 \, \Re \left( e^{-i \varepsilon (t-t')} 
 \big\langle \, 
 	\big[
 	b(t),b(t')^\dagger
 	\big]_{-}
 \big\rangle^\E
  \right)
}%
\label{eq:fermi-contraction-keldysh}
\\
&= \frac{1}{\pi} \int d\omega \cos \left[(\omega-\varepsilon)s \right] \tanh \left(\frac{\omega-\mu}{2T} \right)
\label{eq:fermi-contraction-a}
\\
&= 2T \frac{\sin \left[\epsilon s \right]}{\sinh \left[\pi T s \right]}
\label{eq:fermi-contraction-b}
\end{align}%
\label{eq:fermi-contraction}%
\end{subequations}%
depending on the relative time $s=t-t'>0$.
It stems from the anti-symmetric part of the Fermi function \eq{eq:fermi-distrib}
and accounts for the time-nonlocal effects due to propagation in the thermal environment at the energy of the resonant level.
Thus, all temperature dependence is incorporated in this function and only the level \emph{detuning} $\epsilon=\varepsilon-\mu$
with respect to the chemical potential enters.

\begin{figure}[t]
	\includegraphics[width=\linewidth]{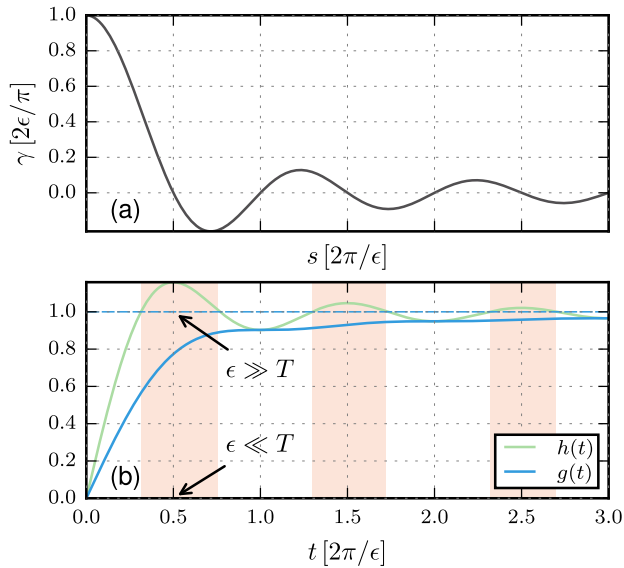}
	\caption{
		Time-dependent functions \refabc{determining} the dynamics for $2\epsilon / \pi  = 20 \Gamma$ and  $\pi T = 10^{-3} \cdot \Gamma/2$.
		(a) Reservoir correlation function $\gamma(s)$ [\Eq{eq:fermi-contraction}] describing the time-nonlocal `memory' effects.
		(b) The oscillatory behavior of $\gamma(s)$, while still present in the function $h(t)$ [\Eq{eq:h-def}, green]
		appearing in the \emph{time-local} EOM \eq{eq:eom-averaged},
		is smoothed out in the function $g(t)$ [\Eq{eq:g-h}, blue] which describes the cumulative nontrivial time-dependence of the EOM solution \eq{eq:eom-sol}.
		In the semigroup limits of small [\Eq{eq:eps-zero}] and large [\Eq{eq:eps-infty}] detuning,
		both these functions instantly reduce to constant values of $0$ and $1$, respectively, as indicated by arrows.
	}
	\label{fig:functions}
\end{figure}

Integrating the equations of motion \eq{eq:eom-averaged} with initial values $\brkt{d_\eta(0)}^\E=d_\eta$ and $\brkt{ (-\one)^n (0)}^\E =(-\one)^n$ gives
\begin{subequations}%
\begin{align}%
\brkt{ d_\eta (t)}^\E
& = e^{i\eta(\varepsilon + i\eta\frac{\Gamma}{2})t} \, d_\eta
\label{eq:eom-sol-a}
\\
	\brkt{(-\one)^n(t)}^\E  & = e^{-\Gamma t} (-\one)^n + (1-e^{-\Gamma t})g(t)  \one
	,
	\label{eq:eom-parity}
\end{align}
\label{eq:eom-sol}%
\end{subequations}%
where we extract the conventional factor $( 1-e^{-\Gamma t})$ in \Eq{eq:eom-parity} 
to isolate the dimensionless function
\begin{align}
g(t) & := 
\frac{\Gamma}{1-e^{-\Gamma t} }
\int_0^t d\tau e^{- \Gamma(t-\tau) } h(\tau)
\label{eq:g-h}
.
\end{align}
This is physically motivated since it reduces to the expectation value of the parity in the stationary limit:
\begin{align}
	\brkt{(-\one)^n(\infty)} = g(\infty)
	.
	\label{eq:g-infty0}
\end{align}%
The nontrivial time-dependence of the solution is thus fully determined by the functions \eq{eq:h-def} and \eq{eq:g-h}
deriving from the Keldysh correlation function \eq{eq:fermi-contraction}.
In \Fig{fig:functions}, we show how the oscillatory behavior of the correlation function is translated into both $h(t)$ and $g(t)$.
The detuning sets the sign $\sign \, \epsilon = \sign \, h(t)= \sign \, g(t)$ for all $t>0$.

Substituting \Eq{eq:eom-sol} into \eq{eq:pi-op-expansion} and normalizing the operators gives the dynamical map in superoperator form:
\begin{align}
	\Pi(t) 
	&=
\sum_{\eta} d^\dag_\eta \, \tr{\S}
 \Big[ e^{(i\eta\varepsilon -\frac{\Gamma}{2})t} d_\eta \bullet \Big]
+ \tfrac{1}{2} \one \, \tr{\S} \Big[ \bullet \Big]
\label{eq:pi-eom} \\
&\phantom{=} \quad 
+ \tfrac{1}{2} (-\one)^n  \tr{\S} \Big[ (1-e^{-\Gamma t})g(t) \one \bullet  + e^{-\Gamma t} (-\one)^n \bullet  \Big]
.
\notag
\end{align}
The TP property $\Tr_\S \Pi(t) = \Tr_\S $ is explicit through the second term as all other terms produce traceless operators.
Indeed, all further eigen-supervectors of the dynamics are easily derived from this form, cf. \Sec{sec:modes} and \Eq{eq:app-pi-diag}.
In contrast, the CP property is unclear from this representation
and this motivates the complementary treatments in Sec.~\ref{sec:superfermion}-\ref{sec:kraus}.

Postponing the issue of CP to \Sec{sec:superfermion}, we may nevertheless inquire about the divisibility of the evolution into two factors as in \Eq{eq:pi-cpdiv}.
To this end, we integrate \Eq{eq:eom-averaged} from $t'$ onward.
The result for the divisor $\Pi(t,t')$ can be obtained from \Eq{eq:pi-eom} by replacing
\begin{subequations}%
\begin{align}
	(1-e^{-\Gamma t})g(t) \to
	(1-e^{-\Gamma [t-t']})g(t,t')
	\label{eq:div-subst}
\end{align}
and adjusting the time-interval to $[t,t']$ in all other terms.
The function
\begin{align}
g(t,t') & := 
\frac{\Gamma}{1-e^{-\Gamma [t-t']} }
\int_{t'}^t d\tau e^{- \Gamma(t-\tau) } h(\tau)
\label{eq:g2-h}
\end{align}%
\label{eq:div-subst-final}%
\end{subequations}%
interpolates between $g(t,0)=g(t)$ and $g(t,t)=h(t)$ and therefore shares many of their features, see \Fig{fig:g-divisor}.
The evolution $\Pi(t)$ is thus \emph{always} divisible and because the divisor $\Pi(t,t')$ has the same structure \eq{eq:pi-eom} it is always a TP map.
To distinguish meaningful types divisibility one has to investigate the properties of $g(t,t')$.

\subsection{Physical properties of time-dependent functions\label{sec:functions-physical}}
The important functions $\gamma(s)$, $h(t)$, $g(t)$ and $g(t,t')$ incorporate the nontrivial time-dependence of the evolution and its divisor.
Each function can in principle be expressed in terms of Digamma and Lerch functions reported before, see \Ref{Saptsov14} and references therein.
Their physical properties however rely on different representations which are derived in \App{app:functions}
and summarized below.

\begin{figure}[t]
	\includegraphics[width=\linewidth]{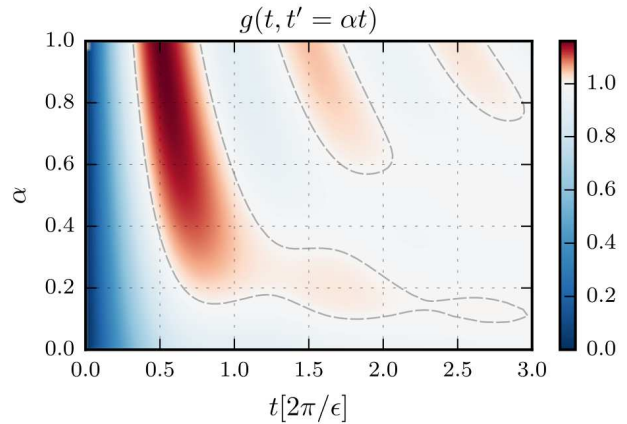}
	\caption{
		Time-dependent function $g(t,t')$ [\Eq{eq:g2-h}] determining the dynamics of the divisor $\Pi(t,t')$ for the same parameters as in \Fig{fig:functions}.
		As a function of the division ratio $\alpha = t'/t$, it interpolates between the functions $g(t)$ at $\alpha=0$ and $h(t)$ at $\alpha=1$.		
		In contrast to $g(t)$, it may exceed the value of $1$ indicated by the contour, cf. the discussion in \Sec{sec:functions-physical}.
	}
	\label{fig:g-divisor}
\end{figure}

\emph{Semigroup-divisible limits}.
When the system is \emph{thermally} close to resonance, $|\epsilon| \ll T$,
the correlation function $\gamma(s)$ shows exponential decay with memory time $T^{-1}$
and thus vanishes in the limit
\begin{align}
\lim_{T/|\epsilon| \to \infty} \gamma(s) =
\lim_{T/|\epsilon| \to \infty} h(t)=
\lim_{T/|\epsilon| \to \infty} g(t) = 0
,
\label{eq:eps-zero}
\end{align}%
suppressing both $h(t)$ and $g(t)$ as a consequence.
Far off resonance, $|\epsilon| \gg T$, the correlation function instead shows oscillating power-law decay with time-scale $|\epsilon|^{-1}$
and becomes time-local in the limit
\begin{subequations}
\begin{align}
\lim_{|\epsilon|/T \to \infty} \gamma(t-t')
= \deltah(t-t'),
\label{eq:factor2}
\end{align}
noting that the relevant $\delta$-function is normalized on the real half-line $\int_0^t dt' \deltah(t-t') = 1$ since we solve an initial-value problem.
In this limit, $h(t)$ and $g(t)$ still coincide,
\begin{align}
\lim_{|\epsilon|/T \to \infty} h(t)
=
\lim_{|\epsilon|/T \to \infty} g(t)
=
\sign (\epsilon) 
\,
\theta(t)
,
\label{eq:theta}
\end{align}%
\label{eq:eps-infty}%
\end{subequations}%
and reduce independently of $T$ and $\Gamma$ to a step function with $\theta(0):=0$. 
Although both limits lack memory for different physical reasons,
they share the semigroup property that $g(t)$ --but not the parity $\brkt{ (-\one)^n (t)}$--
is instantly stationary.

\emph{Generic behavior.}
In all other cases of finite temperatures or level detunings, the functions differ as shown in \Fig{fig:functions}.
The deviations are already pronounced at small times [\App{app:functions-extra}] as $h(t)$ starts out with double the slope:
\begin{align}
	h(t) \approx 2 g(t) 
	= \frac{2\epsilon}{\pi} \, t
	\label{eq:hg-linear}
	.	
\end{align}
Whereas $h(t)$ is always nonmonotonic, the function $g(t)$ is nonmonotonic only for $\Gamma/2 > \pi T$, cf. \App{app:functions-g}.
This is due to the fact that the latter is obtained by an integration [\Eq{eq:g-h}] which smoothens the oscillations of $h(t)$.
However, both approach the same stationary limit
\begin{align}
	g(\infty)
	=
	h(\infty)
	=
	\frac{2}{\pi} \, \Im \, \psi \left( \frac{1}{2} + \frac{\tfrac{\Gamma}{2}+i\epsilon}{2\pi T} \right)
	\label{eq:g-infty}
\end{align}
expressed in terms of the imaginary part of the Digamma function $\psi$, cf. \Eq{eq:ginfty-app}.

\emph{Physical bounds.}
The function $g(t)$ is bounded for any time and any set of parameters as
\begin{align}
|g(t)|
\leq 1
\label{eq:g-cp}.
\end{align}
That this condition holds for our model can be seen in two steps.
First, the function never exceeds its asymptotic value $|g(t)| \leq |g(\infty)|$.
This fact is not obvious and seems to rely on model-specific details [\App{app:functions-g}].
Second, due to its relation \eq{eq:g-infty0} to the stationary expectation value of the parity,
it is physically clear that $g(\infty)$ becomes maximal in the off-resonant semigroup limit $|\epsilon| \gg T$
where the stationary state is pure:
\begin{align}
	|g(\infty)| \leq \lim_{|\epsilon|/T \to \infty} |\brkt{(-\one)^n(\infty)}| = 1
	.
	\label{eq:g-infty-parity}
\end{align}
A physical reason why the bound \eq{eq:g-cp} \emph{must necessarily} hold,
may be sought in the preservation of the positivity (PP) of the state under evolution,
i.e., the property
$\Pi(t) \rho(0) \geq 0$ for every $\rho(0) \geq 0$.
In \App{app:pp}, we verify that indeed the condition \eq{eq:g-cp} is equivalent to PP.

\emph{Divisibility.}
In contrast, $h(t)$ is not bounded by its asymptote:
It may significantly overshoot it and more strongly $|h(t)| > 1$ for certain parameter regimes [\App{app:functions-h}]
as seen in \Fig{fig:functions}.
This behavior directly translates to the function $g(t,t')$.
Specifically, the equivalence
\begin{align}
|g(t,t')| \leq 1 \text{ for $t \geq t' \geq 0$ } \iff |h(t)| \leq 1 \text{ for $t \geq 0	$}
\label{eq:h-cpdiv}
\end{align}
holds as one verifies as follows:
Assuming first the left-hand side, this particularly includes $|g(t,t)|=|h(t)| \leq 1$ for all $t \geq 0$ by \Eq{eq:g2-h}.
Conversely, the right-hand side implies in \Eq{eq:g2-h} that $|g(t,t')| \leq 1$ for any $t \geq t' \geq 0$ as the integral is normalized to one.
In contrast to $g(t)$, there is no physical reason why values $g(t,t') > 1$ should not occur and in \Fig{fig:g-divisor} they indeed occur (red).
The observable implications of the criterion
\begin{align}
		|g(t,t')| > 1 \iff |h(t)| > 1
		\label{eq:cpdiv-loss}
\end{align}
marked red in \Fig{fig:functions} are discussed in \Sec{sec:modes}.

\section{Real-time  superfermion approach \label{sec:superfermion}}

Although the EOM approach has a straightforward derivation and identifies the relevant time-dependent functions,
the form \eq{eq:pi-eom} no longer reflects that the evolution \eq{eq:pi} is actually CP-- a stronger condition than the PP property verified in the EOM approach.
In general PP does \emph{not} imply CP nor any of its useful consequences that we need in the remainder of the paper.
It is a coincidence that in the present model, both PP and CP \emph{happen} to be equivalent to the bound \eq{eq:g-cp},
and some other models also exhibit this special property, see \Refs{Hall08,Whitney2008a} and \App{app:pp}.
This equivalence does however \emph{not} become clear in the EOM approach,
nor that the bound \eq{eq:h-cpdiv} \emph{happens} to be equivalent to CP-divisibility.

To see this, we turn to an approach that also facilitates the exact solution but
naturally leads to a representation of the dynamics $\Pi(t)$ which avoids these issues.
\refabc{%
This real-time superfermion approach~\cite{Saptsov12,Saptsov14,Schulenborg16,Schulenborg18a}
has been developed~\cite{Schoeller09,Schoeller18} for transport through strongly correlated quantum dots, where higher-order calculations are mandatory due to, e.g., Kondo processes~\cite{Pletyukhov12a}.
In contrast to the previous section, it focuses on the evolution of the state rather than observables.
It provides a physically motivated connection to the quantum information approach
by inherently separating finite-temperature corrections from the infinite-temperature limit
in which the system is \emph{maximally entangled} with its environment, cf. \Eq{eq:choi-def}.
}

\subsection{Renormalized perturbation theory for $\Pi(t)$}
A naive expansion of $U(t)=e^{-i H^\tot t}$ in \Eq{eq:pi}
in the coupling $H^V=\sqrt{{\Gamma}/{2\pi}} \int d\omega \sum_{\eta} d_\eta b_{\bar{\eta} \omega}$
does not lead to an obviously summable series for the state evolution $\Pi(t)$, even for quadratic fermionic models.
Such a summable series can be obtained, however, when abandoning the Hamiltonian formulation and using an \emph{intrinsically dissipative} reference for the expansion.
This emerges naturally when using Liouville-space methods developed in \Ref{Schoeller09}
and can be formulated elegantly~\cite{Saptsov14} in terms of fermionic superfields outlined below.
This formulation importantly reveals the existence of \emph{two} fundamental expansion `parameters' in the wideband limit:
the retarded reservoir correlation function capturing the time-local effects of the physical \emph{infinite-temperature} limit,
and the Keldysh correlation function $\gamma(s)$ [\Eq{eq:fermi-contraction}] incorporating the time-nonlocal \emph{finite}-temperature corrections.
Exploiting this structure results in a renormalized perturbation theory that remarkably stops at finite order for quadratic models~\cite{Saptsov14}.

\emph{Superfermions.}
We express the time-evolution \eq{eq:pi} directly as $\Pi(t) = \Tr_\E \{ e^{-iL^{\text{tot}}t} (\bullet \otimes \rho^\E) \}$
in terms of the total Liouvillian superoperator $L^{\text{tot}}:= [H^{\text{tot}},\bullet]_-$.
It acts on the fermionic Liouville-Fock space and allows for a `second quantization' representation
\begin{align}
L^{\text{tot}}
= & \sum_{\eta}
\eta \Big[ \varepsilon G^+_\eta G^-_{\bar{\eta}} + {\tiny \int d\omega} \omega J^+_{\eta\omega} J^-_{\bar{\eta} \omega} \Big]
\notag
\\
&
 + \sum_{\eta,q} \sqrt{\frac{\Gamma}{2\pi}} {\tiny \int d\omega} J^q_{\eta\omega} G^{\bar{q}}_{\bar{\eta}}
\end{align}
that is analogous to that in Hilbert-Fock space but also takes into account \emph{mixed} states and superoperators~\cite{Schmutz78,Prosen08,Kosov09,Mukamel08,Kosov11,Dzhioev12}.
The fermionic creation ($q=+$) and annihilation ($q=-$) superfields
\begin{subequations}
\begin{gather}
G_\eta^q = \tfrac{1}{\sqrt{2}} \big(d_\eta \bullet \one + q (-\one)^n \bullet (-\one)^n d_\eta \big)
\\
J_{\eta\omega}^q = \tfrac{1}{\sqrt{2}} \big( b_{\eta\omega} \bullet \one + q (-\one)^{n^\text{R}} \bullet (-\one)^{n^\text{R}} b_{\eta\omega} \big)
\end{gather}
\label{eq:def-superfields}%
\end{subequations}
for system and reservoir, respectively, obey the anti-commutation relations
$[G^{q'}_{\eta'}, G^q_\eta  ]_+ = \delta_{\eta',\bar{\eta}} 
\, \delta_{q',\bar{q}}$
and
$[J^{q'}_{\eta'\omega'}, J^q_{\eta\omega} ]_+ = \delta_{\eta',\bar{\eta}} \,  \delta_{q',\bar{q}} \,  \delta(\omega'-\omega)$.
Contrary to the EOM approach, maximal simplicity is achieved with the convention~\cite{Schoeller09}
of \emph{commuting}~\footnote{
\refb{%
It is often not realized that fermionic particle statistics only requires fields acting on the \emph{same mode} to anticommute, see paragraph 65 of Ref.~[\onlinecite{LandauLifshitz}].
Fields acting on different modes can be chosen to either commute or anticommute, see \Ref{Saptsov14} for a detailed discussion.
We stress that this is not specific to noninteracting models
and allows for drastic simplifications, especially in open-system dynamics.%
}%
}
system and environment fields, $[d_{\eta'},b_{\eta \omega}]_{-}=0$,
which translates to $[G^{q'}_{\eta'},J^{q}_{\eta\omega} ]_- = 0$.

Through the parity-check in the definition \eq{eq:def-superfields},
both the fermion-parity superselection as well as the \emph{Pauli principle} are incorporated~\cite{Saptsov14}.
The latter is expressed by the superfermion identity
\begin{align}
[G^+_{\eta}]^2 = 0
\label{eq:pauli}
\end{align}
and indicates that the particle/hole index $\eta$ [\Eq{eq:eta}] takes on the new formal role of a quantum number which cannot be `doubly filled'.
A second key feature of this method was overlooked in its previous applications:
Bilinear products of superfields
\begin{subequations}
\begin{gather}
\sum_\eta G^+_{\eta} G^-_{\bar{\eta}} = - \sum_\eta \mathcal{L}_\eta, \quad
\sum_\eta \eta G^+_{\eta} G^+_{\bar{\eta}} =  \sum_\eta \eta \mathcal{L}_\eta
\label{eq:Leta-g}
\end{gather}
introduced by pair contractions correspond to superoperators of GKSL-form
\begin{align}
\mathcal{L}_\eta := d_{\eta} \bullet d_{\bar{\eta}} - \tfrac{1}{2} \big[ d_{\bar{\eta}} d_\eta , \bullet \big]_+
\label{eq:Leta}
.
\end{align}%
\label{eq:Leta-superfermion}%
\end{subequations}%
The Pauli principle \eq{eq:pauli} therefore implies a nonlinear relation between these generators: $[ \sum_\eta \eta \mathcal{L}_\eta ]^2 = 0$.

For the reservoir, the parity-check reveals that only two (out of four) types of pair-contractions contribute:
\begin{subequations}
\begin{gather}
\left\langle  J^-_{\eta'\omega'} J^+_{\eta\omega} \right\rangle^\E = \delta_{\eta',\bar{\eta}}\delta(\omega'-\omega)
\label{eq:gf-res-retarded}
\\
\left\langle  J^-_{\eta'\omega'} J^-_{\eta\omega} \right\rangle^\E = \eta \delta_{\eta',\bar{\eta}}\delta(\omega'-\omega) \tanh\left(\frac{\omega-\mu}{2T}\right)
.
\label{eq:gf-res-keldysh}
\end{gather}
\label{eq:gf-res}%
\end{subequations}
These respectively correspond to the symmetric and anti-symmetric part of the Fermi function \eq{eq:fermi-distrib}.
The algebraic structure thus \emph{automatically} separates time-local (memoryless) contributions \eq{eq:gf-res-retarded} that remain at infinite temperature
from the time-nonlocal (memory) contributions \eq{eq:gf-res-keldysh} accounting for the finite-temperature corrections.

\emph{Renormalized perturbation theory.}
With these insights, we can identify two stages in the naive expansion of $\Pi(t)$ in powers of the coupling Liouvillian $L^{V} = \sum_{\eta,q} \sqrt{{\Gamma}/{2\pi}} {\tiny \int d\omega} J^q_{\eta\omega} G^{\bar{q}}_{\bar{\eta}} $,
see \App{app:superfermion} for details.
Using Wick's theorem, each term in the series is decomposed~\cite{Schoeller09,Saptsov14,Schoeller18} into products of reservoir pair-contractions
\begin{align}
\brkt{ L^V(\tau) L^V(\tau')}^\E
& = - \tfrac{\Gamma}{2} \sum_\eta \left[\deltah(\tau-\tau') - \eta \gamma(\tau-\tau')\right] \mathcal{L}_\eta
,
\label{eq:lv-contraction}
\end{align}
where the time-dependence of $L^{V}(\tau) = e^{iL^0\tau} L^{V} e^{-iL^0\tau}$ denotes the interaction-picture with respect to the free evolution
$L^0 = \sum_{\eta} \eta [ \varepsilon G^+_\eta G^-_{\bar{\eta}} + {\tiny \int d\omega} \omega J^+_{\eta\omega} J^-_{\bar{\eta} \omega}]$.
As anticipated in the EOM approach, each pair-contraction has a time-local [$\deltah$, \Eq{eq:factor2}]
and time-nonlocal contribution [$\gamma$, \Eq{eq:fermi-contraction}].
Importantly, these contributions are now distinguished by their \emph{algebraic} superoperator structure \Eq{eq:Leta-g}:
In contrast to the the time-local term $\sum_\eta \mathcal{L}_\eta \sim G^+ G^-$ combining creation \emph{and} annihilation superfields,
the time-nonlocal term $\sum_\eta \eta \mathcal{L}_\eta \sim G^+ G^+$ \emph{exclusively} features creation superfields
such that any higher power of this term vanishes by \Eq{eq:pauli}.

The perturbation series can thus be summed exactly in a two-stage procedure:
First, the time-local infinite-temperature contributions from \Eq{eq:lv-contraction} form a geometric series which sums to the exponential
\begin{align}
\Pi_\infty(t) = e^{-iL t + \frac{\Gamma}{2}t \sum_\eta \mathcal{L}_\eta }
.
\label{eq:pi-inf-temp}
\end{align}
For the second stage, this semigroup limit is used as the reference of a renormalized perturbation series featuring \emph{only} creation superfields $G^+$
implying that it exactly terminates after a \emph{single} correction due to \Eq{eq:pauli}:
\begin{subequations}
\begin{align}
\Pi(t)
&= \Pi_\infty(t) - \tfrac{1}{2}\left(1 - e^{-\Gamma t}\right)g(t) \sum_\eta \eta \mathcal{L}_\eta
\label{eq:pi-superfermion-a}
\\
&= e^{-i L t + \frac{\Gamma}{2}t \sum_\eta \left[1-\eta g(t) \right]\mathcal{L}_\eta}
.
\label{eq:pi-superfermion-b}
\end{align}
\label{eq:pi-superfermion}%
\end{subequations}%
It is also the Pauli-exclusion principle that
suggests to re-exponentiate the correction to obtain the elegant exponential form \eq{eq:pi-superfermion-b}.
This can be done using the Baker-Campbell-Hausdorff formula since the algebra of operators is closed, cf. \App{app:superfermion}.
The appearance of the function $g(t)$ [\Eq{eq:g-h}] already indicates that both expressions coincide with the EOM solution
as one verifies using \Eq{eq:app-Leta}.

\subsection{Complete positivity and divisibility \label{sec:pi-discuss}}

The exponential form \eq{eq:pi-superfermion-b} explicitly reveals the CP and divisibility properties of the dynamical map $\Pi(t)$
without taking recourse to operator-sums [\Sec{sec:kraus}] or quantum master equations [\Sec{sec:qme}].

\emph{Complete positivity.}
The exponential form suggests to consider an auxiliary dynamics $\frac{d}{d \lambda} X (\lambda) = \Lambda X(\lambda)$ with $X(\lambda)|_{\lambda=0} = \ones$:
At a \emph{fixed} time $t$, the physical map \eq{eq:pi-superfermion-b}
is equivalent to the formal solution $\Pi(t)=X(\lambda)|_{\lambda=1}$
of the auxiliary dynamics evaluated at the flow parameter $\lambda=1$.
This allows for the application of the GKSL theorem~\cite{GKS76,Lindblad76} because the auxiliary generator given by
$\Lambda = -i Lt + \frac{\Gamma}{2}t \sum_\eta \left[1-\eta g(t) \right]\mathcal{L}_\eta$
has a $\lambda$-independent GKSL form%
		\footnote{
		One cannot apply the GKSL semigroup or CP-divisibility theorem \eq{eq:gksl} to the \emph{physical} evolution
		because neither theorem is conclusive for many model parameters, see \Sec{sec:qme}.
		}.
Thus, $X(\lambda)$ is CP-TP if and only if the coefficients of the jump operators are positive, i.e., $|g(t)| \leq 1$,
which clarifies that the bound \eq{eq:g-cp} on the function $g(t)$ found in the EOM approach 
is actually the condition for CP which \emph{must} hold for each $t$.
The PP property thus happens to coincide with CP.
We can extend this result to resolve the question of divisibility.

\emph{Divisibility.}
First, in the two semigroup limits where $g(t)=g(\infty)$ for $t > 0$ [\Eqs{eq:eps-zero} and \eq{eq:eps-infty}],
the time-linear exponent in \Eq{eq:pi-superfermion-b} indeed implies a Markovian semigroup $\Pi(t)=\Pi(t-t')\Pi(t')$.
In all other cases, the dynamics is not a Markovian semigroup but we can explicitly construct the \emph{divisor} $\Pi(t,t')$ to analyze the CP-divisibility condition.
Analogous to the EOM approach, it is obtained by setting $g(t) \to g(t,t')$ and adjusting the time interval to $[t,t']$ in $\Pi(t)$ as given by \Eq{eq:pi-superfermion-b},
\begin{align}
	\Pi(t,t')= e^{-iL(t-t') +\frac{\Gamma}{2}t \sum_\eta \left[1-\eta g(t,t') \right]\mathcal{L}_\eta}
	,
	\label{eq:pi-divisor}
\end{align}
see \App{app:superfermion-divisor} for the justification.
Applying the same argument as in the discussion of the complete positivity of $\Pi(t)$ establishes
that the divisor is CP-TP if and only if $|g(t,t')| \leq 1$ for all $t \geq t' \geq 0$,
which we showed earlier [\Eq{eq:h-cpdiv}] to be equivalent to the bound $|h(t)| \leq 1$.

If this condition fails to hold [\Eq{eq:cpdiv-loss}] --as it does for a wide range of parameters in our model--
the division of the evolution is no longer physically meaningful in the sense of \Sec{sec:open-system}.
The dynamics then exhibits `truly' non-Markovian behavior which is neither semigroup- nor CP-divisible.
As a criterion, this is equivalent to \Eq{eq:gksl} based on the quantum master equations as we show in \Sec{sec:qme}.
However, the single function $g(t,t')$ completely characterizes the \emph{explicit} divisor $\Pi(t,t')$ and its behavior is explored in \Sec{sec:modes}.

\section{Operator-sum representation\label{sec:kraus}}

From the exact solution $\Pi(t)$ in the form \eq{eq:pi-eom} or \eq{eq:pi-superfermion}, we can construct the Kraus operator-sum.
This also reveals the CP restriction $|g(t)| \leq 1$ just found in \Sec{sec:pi-discuss},
but it furthermore gives access to information measures quantifying
\refb{%
the effects of the system-environment coupling as explained in \Sec{sec:information}.
By constructing the Kraus operators we gain access to the effective environment state
which is clearly beyond the previously discussed approaches.
This has been intensely studied in quantum information in the context of so-called `complementary quantum channels'
but has so far attracted little attention in quantum transport problems.
Our exact expressions for the Kraus operators demonstrate their relevance for this problem.
They also provide a benchmark for new} approaches that aim at \emph{directly} deriving them~\cite{vanWonderen13,Chruscinski13,vanWonderen18a,vanWonderen18b,Reimer19a}
in order to produce CP maps even under approximations.
For instance, the $T\to\infty$ limit of the result reported below has been previously obtained within such an approach~\cite{Reimer19a},
but it \refabc{is still} unclear how finite-temperature contributions can be summed up.

As explained in \Sec{sec:open-system}, the Kraus operator-sum
\begin{subequations}
\begin{align}
\Pi(t) = \sum_{k=0,1} \sum_{\eta=\pm} K_{\eta}^{k}(t) \bullet K_{\eta}^{k}(t)^\dag
\label{eq:pi-kraus}
\end{align}
follows from diagonalizing the Choi operator \eq{eq:choi-kraus} which is explicitly carried out in \App{app:kraus-choi}.
For generic parameters $\varepsilon$, $\mu$, $\Gamma$ and $T$, the Choi operator is full rank and we get $d^2=4$ nonzero Kraus operators.
Because of the block-diagonal structure of $\text{choi}(\Pi(t))$, these four Kraus operators separate into two groups labeled by $k=\{0,1\}$
where $(-1)^k$ is their fermion parity, see \App{app:kraus-sumrule}.

For $k=0$, we have
\begin{gather}
K_\eta^0(t) = \sqrt{\lambda^0_\eta(t)}
\,
\frac{\eta \sqrt{r(t)}^{\eta} d d^\dagger + \frac{1}{\sqrt{r(t)}^{\eta}} e^{-i \varepsilon t} d^\dagger d }
{\sqrt{r(t)+\frac{1}{r(t)}} }
,
\label{eq:kraus0}
\end{gather}
where both the Choi eigenvalues $\lambda^0_\eta(t)$ as well as the Choi eigenvectors determined by the single function $r(t)$ enter:
\begin{align}
\lambda^0_\eta(t) &= 
\label{eq:lambda0} \\
\notag
&\tfrac{1}{2} (1+e^{-\Gamma t}) + \eta \sqrt{e^{-\Gamma t}+[\tfrac{1}{2}(1-e^{-\Gamma t})g(t) ]^2 } \\
r(t) &= e^{\frac{\Gamma}{2}t} \times
\label{eq:r-t} \\
\notag
& \Big[ \tfrac{1}{2}(1-e^{-\Gamma t})g(t) +  \sqrt{e^{-\Gamma t}+[\tfrac{1}{2}(1-e^{-\Gamma t})g(t) ]^2 } \, \Big]
.
\end{align}
Note the separate dependence on the level position $\varepsilon$: It drops out only in the spectrum of $\rho(t)$
which depends on the level \emph{detuning} $\epsilon=\varepsilon -\mu$ through $g(t)$ [\Eq{eq:g-h}].
The operators $K^0_\eta(t)$ evolve the system conditional on no \emph{net} fermion transfer having occurred with
the $\eta$-index reflecting a nontrivial relative phase between an occupied and an unoccupied system.
As pointed out in \Sec{sec:information}, only a single Kraus operator $K^0_\eta(0)=\delta_{\eta,+} \one$ proportional to the identity operator remains at the initial time.

For $k=1$, the time-dependence of the Kraus operators enters only through the Choi eigenvalues $\lambda^1_\eta(t)$:
\begin{gather}
K_\eta^1(t) = \sqrt{\lambda^1_\eta(t)} \, d_{\eta}
\label{eq:kraus1} 
\\
\lambda^1_\eta(t)  = \tfrac{1}{2}(1-e^{-\Gamma t})[1-\eta g(t)]
.
\label{eq:lambda1}
\end{gather}
\label{eq:krausfinal}%
\end{subequations}%
These operators correspond to a conditional evolution
in which a fermion has been effectively added to $(\eta=+)$ or removed from $(\eta=-)$ the system up to the finite time $t$.
In accordance with this, both these Kraus operators initially vanish, $K^1_\eta(0)=0$.
Similarly, in the limit of large detuning ($|\epsilon| \gg  T$) where $|g(t)|=|g(\infty)|=1$ [\Eq{eq:eps-infty}],
tunneling into $(\epsilon> 0)$ or out of $(\epsilon < 0)$  the system becomes impossible for all times.
This is reflected by the vanishing of two of the four Choi eigenvalues and their corresponding Kraus operators.

It is admittedly a disadvantage of the operator-sum \eq{eq:krausfinal} that
\refabc{%
it
}%
represents the dynamics
as an intricate competition between exponentially decaying terms depending only on $\Gamma$
and the nontrivial evolution of $g(t)$ depending on all parameters $\Gamma$, $T$ and $\epsilon$.
This is easier analyzed using the spectral decomposition of $\Pi(t)$ as obtained from the EOM in \Sec{sec:modes}.
Here, we merely note that the eigenvectors of $\Pi(t)$ are fully determined by sum-rules for the Kraus operators
--similar to the TP condition \eq{eq:kraus-tp}-- given in \App{app:kraus-sumrule}.
Clearly, the CP property of the dynamical map $\Pi(t)$ is easily inferred from the Choi eigenvalues $\lambda_\eta^k(t) \geq 0$
and is equivalent to $|g(t)| \leq 1$ [\Eq{eq:g-cp}], independently confirming the result of \Sec{sec:pi-discuss}. 

\emph{System density matrix.}
\refabc{%
The key advantage of the Kraus operators is that they
}%
give access to the spectra of both the system and effective environment state
required for computing the information measures introduced in \Sec{sec:information}.
In case of the system, the spectrum is however also easily expressed
in terms of the parity evolution which fully determines%
		\footnote{
		Within the spin formulation \eq{eq:full-hamiltonian-spin}, the lack of parity-superselection
		complicates the structure of both system and effective environment density matrices
		but leaves their positivity properties unaffected, see \App{app:kraus-spin} for a detailed discussion.
		}
the system state
\begin{align}
\rho(t) = \tfrac{1}{2} \Big[ \one + \brkt{(-\one)^n(t)} \, (-\one)^n \Big]
.
\label{eq:rho-t}
\end{align}
The expectation value $\brkt{(-\one)^n(t)}=\Tr_\S (-\one)^n \rho(t)=\Tr_\S \brkt{(-\one)^n(t)}^\E \rho(0)$ 
is the same as that obtained from the EOM [\Eq{eq:eom-sol}]:
\begin{align}
\brkt{(-\one)^n(t)} &= 
e^{-\Gamma t} \brkt{(-\one)^n(0)} + (1-e^{-\Gamma t})g(t)
.
\label{eq:parity-sol}
\end{align}
The eigenvalues of the system state,
\begin{align}
	\Lambda_\eta(t)
	& = \tfrac{1}{2} \big[ 1 + \eta \brkt{(-\one)^n(t)}  \big]
	,
	\label{eq:spectra-relations}
\end{align}
are positive due to the CP condition $|g(t)|\leq1$ [cf. \App{app:pp}]
and enter into the binary entropy $S(\rho(t)) = - \sum_{\eta} \Lambda_{\eta}(t) \log_2 \Lambda_{\eta}(t)$ of the system.

\emph{Effective environment density matrix.}
The density matrix of the effective environment can be constructed from \Eq{eq:rho-E'}:
$\rho^{\E'}(t)^{k k'}_{\eta \eta'} = \Tr_\S K_{\eta}^k(t) \rho(0) K_{\eta'}^{k'}(t)^\dag$, see \App{app:kraus-env}.
Due to parity-superselection [\Sec{sec:model}], the state decomposes into parity blocks
and can therefore be considered as the mixed state of a \emph{two-fermion}
\footnote{%
Since the effective environment has finite dimension $4$, it can always be considered as consisting of two qubits~\cite{NielsenChuang}.
However, because the joint state is block-diagonal by the superselection property of the dynamics,
it is consistent to consider the environment as two fermions%
}
environment $\E'$:
\begin{align}
	\rho^{\E'}(t) =
	\begin{bmatrix}
		\rho^{\E'}(t)^{00} & 0 \\
		0                  & \rho^{\E'}(t)^{11}
	\end{bmatrix}
	.
	\label{eq:effective-environment-state}
\end{align}
These \emph{blocks} are not independent: One finds that the effective environment 
factorizes for all times $t\geq0$ into two \emph{independent} fermion modes, $\rho^{\E'}(t)=\rho^{\E'+}(t)\otimes\rho^{\E'-}(t)$.
The eigenvalues of their states $\rho^{\E'\lambda}(t)$ for $\lambda=\pm$ read
\begin{widetext}
\begin{align}
	\Lambda^{\E'\lambda}_{\eta}(t) =
	\frac{1}{2}
	\Bigg[
	1 + \eta \frac{
		\sqrt{
			[\coth(\Gamma t/2)+ \brkt{(-\one)^n(0)} g(t)]^2 - [1-g(t)^2] [1-\brkt{(-\one)^n(0)}^2]
		}
		+ \lambda [\brkt{(-\one)^n(0)} - g(t)]
	}{\coth(\Gamma t/2)+1}
	\Bigg]
	.
	\label{eq:factors}
\end{align}%
\end{widetext}
These reveal the generic short- and long-time behavior discussed in \Sec{sec:information}:
Initially $\Lambda^{\E'\lambda}_\eta(0)=\tfrac{1}{2}[1+\eta]=\delta_{\eta,+}$, i.e., each mode is pure as it should be by construction.
In the stationary limit, the factorization \eq{eq:entropy-env-stat} is recovered
as one effective mode acquires the spectrum of $\rho(0)$ and the other that of $\rho(\infty)$, see \App{app:kraus-factor} for details.

\emph{Information measures.}
Due to the factorization of the effective environment state, its entropy is a sum
$S(\rho^{\E'}(t)) = \sum_{\lambda} S(\rho^{\E'\lambda}(t))=- \sum_{\eta,\lambda} \Lambda^{\E'\lambda}_{\eta}(t) \log_2 \Lambda^{\E'\lambda}_{\eta}(t)$
of binary entropies of the two effective modes.
This fact simplifies the coherent information \eq{eq:I} in two interesting cases:
For a pure \emph{initial} system state [$\brkt{(-\one)^n(0)} = \pm 1$],
the spectra of the effective environment and the system generally coincide for \emph{all} times, see \App{app:entropies-factor-stationary}.
One of the modes $\Lambda^{\E'\lambda}_\eta(t)$ is then fixed to the pure initial system state $\rho(0)$ with zero entropy 
whereas the other has the spectrum of $\rho(t)$.
Consequently, the entropies of the effective environment and the system match exactly, causing $I_\c(t) = 0$ as it should. 
This is different for the pure \emph{stationary} system state [$g(\infty) = \sigma = \pm 1$]
obtained in the off-resonant semigroup limit \eq{eq:eps-infty} with $g(t)=\theta(t)\sigma$,
where one mode instantly acquires the pure spectrum of $\rho(\infty)$.
The coherent information does not vanish in this case but reduces to the difference
$I_\c(t)=S(\rho(t))- S(\rho^{\E'\sigma}(t))$
of just two binary entropies because the other mode evolves different from $\rho(t)$.
Remarkably, this other mode starts in the \emph{stationary} system state and converges to the \emph{initial} system state $\rho(0)$ as $t\to \infty$
to reproduce the stationary factorization \eq{eq:entropy-env-stat}.

Finally, we note that the positivity of the coherent information mismatch \eq{eq:mismatch} cannot be understood
by a cancellation of $S(\rho(t))$ with the entropy of one of these effective modes, see \Eq{eq:app-no-explanation} and \Fig{fig:entropies}.
Its correct long-time limit $2S(\rho(0))$ [\Eq{eq:app-mismatch-bounds}] is however evident.

\section{Exact quantum master equations\label{sec:qme}}

The exact dynamics in the form of \Eqs{eq:pi-eom}, \eq{eq:pi-superfermion}, or \eq{eq:krausfinal} is the solution of
\refabc{%
two \emph{exact} quantum master equations (QME) which are extensively used in transport, chemical dynamics, and quantum optics.
Here we discuss them for two reasons:
First of all, the time-local form of the QME allows to connect to \Eq{eq:gksl}
which more easily reveals the divisibility properties.
Furthermore, approximations are often formulated on the level of QMEs
such that their impact on divisibility can be discussed~\cite{Kidon18}.
}

\subsection{Time-nonlocal QME (Nakajima-Zwanzig)\label{sec:qme-nonlocal}}
In the real-time approach of \Sec{sec:superfermion}, the natural quantum master equation to consider is time-nonlocal:
\begin{subequations}
\begin{align}
	\frac{d}{dt} \Pi(t) = -i L \Pi(t) + \int_0^t dt' \Sigma(t-t') \Pi(t').
	\label{eq:kineq2}
\end{align}
Here, the Liouvillian $L = [\varepsilon d^\dag d, \bullet]_-$ accounts for the uncoupled system whereas the (Nakajima-Zwanzig) \emph{memory-kernel}
\begin{align}
\Sigma(s) 
&= \tfrac{\Gamma}{2} \sum_\eta \Big[ \deltah(s) - \eta e^{-\frac{\Gamma}{2}s} \gamma(s)  \Big] \mathcal{L}_\eta
\label{eq:kineq-kernel}
\end{align}
\label{eq:kineq}%
\end{subequations}
describes its \refb{coupling} with the environment and depends on the relative time $s=t-t'$ only.
The QME \eq{eq:kineq} is obtained similarly to \Eq{eq:pi-superfermion}
by considering irreducible contributions to the perturbation series, see \App{app:superfermion-kernel}.
It likewise clearly separates the time-local infinite-temperature result from the time-nonlocal corrections
in terms of the correlation function $\gamma$ [\Eq{eq:lv-contraction}].

Although clear from its solution \eq{eq:pi-superfermion}, it is difficult to analyze the CP property of this time-nonlocal QME
because the GKSL theorem \eq{eq:gksl} only applies to the \emph{time-local} form discussed later on.
In fact, the explicit structure of the memory kernel $\Sigma(s)$ ensuring CP has only recently been determined~\cite{Reimer19a}.
However, this structure is quite complicated and even here not easily identified due to the finite-temperature effects introduced by $\gamma(s)$.

In contrast, semigroup-divisibility is readily identified and corresponds to a time-local Keldysh correlation function $\gamma(s) \to \deltah(s)$,
which is only the case for the on-resonant [\Eq{eq:eps-zero}] and off-resonant [\Eq{eq:eps-infty}] limits.
A further distinction of CP-divisibility is not obvious outside these semigroup-limits ($0 < \epsilon < \infty$):
The frequency-dependence of the Laplace-transformed kernel [\App{app:superfermion-kernel}]
\begin{align}
\Sigma(z) &= \int_0^\infty dt e^{i z t} \Sigma(t) \label{eq:sigma-z} \\ \notag
&= \tfrac{\Gamma}{2} \sum_{\eta} \left[ 1 + i\frac{\eta}{\pi}\sum_{\chi = \pm}
\chi \psi \left(\frac{1}{2} + \frac{\tfrac{\Gamma}{2} - i(z-\chi \epsilon)}{2\pi T}\right) \right] \mathcal{L}_\eta
\end{align}
does not seem to suggest any further qualitative difference, in particular, concerning the CP-divisibility property [\Eq{eq:h-cpdiv}].
Thus, the time-nonlocal QME is less suited for discussing the CP and CP-divisibility properties of the dynamics.
Nevertheless, the frequency-representation is a crucial starting point for advanced approximations~\cite{Andergassen10,Schoeller09,Pletyukhov12a,Schoeller18}
because the Laplace variable $z$ represents the physical energy.

\subsection{Time-local QME (TCL)\label{sec:qme-local}}

The evolution is equivalently described by an exact \emph{time-local} or time-convolutionless (TCL) quantum master equation~\cite{Shibata77,Hashitsume77,Chaturvedi79,Shibata80,Rabani15}
\begin{subequations}
\begin{align}
\frac{d}{dt} \Pi(t)
&= [-i L+ \Sigma^\text{TCL}(t)] \,  \Pi(t)
.
\label{eq:tcl2}
\end{align}
In our case, we can directly construct the generator~\footnote
	{\refa{The time-local generator~\eq{eq:tcl2} is obtained either
	by taking the time-derivative of the exponential form \eq{eq:pi-superfermion-b},
	by superfermion considerations [\App{app:superfermion-tcl}],
	or by explicitly inverting the EOM result \eq{eq:pi-eom} to calculate $\big[ \tfrac{d}{dt} \Pi(t) \big] \Pi(t)^{-1}$ [\App{app:eom-tcl}].
	}}
\begin{align}
	\Sigma^\text{TCL}(t) = \tfrac{\Gamma}{2} \sum_\eta \left[ 1 - \eta h(t) \right] \mathcal{L}_\eta
\label{eq:tcl-kernel}
\end{align}
\label{eq:tcl}%
\end{subequations}%
\refa{and find that it} is composed of constant GKSL superoperators
with time-dependent coefficients involving the function $h(t)$ discussed earlier [\Eq{eq:h-def}].

Different from the time-nonlocal QME, the structure of \eq{eq:tcl} clearly distinguishes both types of Markovian evolution occurring in our model:
In the semigroup-divisible limits where $h(t)=h(\infty)$ is constant, the GKSL version of theorem \eq{eq:gksl} applies
and we obtain the necessary and sufficient condition $|h(\infty)|=|g(\infty)| \leq 1$ for complete positivity
in agreement with the real-time approach [\Eq{eq:pi-superfermion}] and the time-nonlocal QME discussed above.
If $h(t)$ is time-dependent, the CP-divisibility version of
theorem \eq{eq:gksl} requires $|h(t)| \leq 1$ for all $t \geq 0$ for the evolution to be CP-divisible,
recovering the result from the real-time approach [\Sec{sec:pi-discuss}].
If $|h(t)| \geq 1$ for some $t$, the GKSL coefficients are temporarily negative and we conclude that the evolution is not CP-divisible.
In contrast to the real-time solution \eq{eq:pi-superfermion}, the latter does \emph{not} make any general statement about CP.

Finally, we note that despite the simplicity of the model,
solving the QME \eq{eq:tcl2} to obtain the solution [\Eqs{eq:pi-eom}, \eq{eq:pi-superfermion}, or \eq{eq:krausfinal}] requires time-ordering
since the time-local generator \eq{eq:tcl-kernel} does not commute $[ \Sigma^\text{TCL}(t), \Sigma^\text{TCL}(t')]_- \neq 0$
with itself at different times~\cite{Chruscinski16}.
Nevertheless, the relation between the time-local generator and the time-nonlocal memory-kernel involves \emph{no} time-ordering
but due to \Eq{eq:h-def} a simple integration,
\begin{subequations}
\begin{align}
\Sigma^\text{TCL}(t)
 = \int_{0}^{t} d s \,  \Sigma(s)
 .
\label{eq:miracle}
\end{align}
This is remarkable in the light of the general relation
\begin{align}
\Sigma^{\text{TCL}}(t) := \int_0^t dt' \Sigma(t-t')\Pi(t')\Pi^{-1}(t)
.
\label{eq:miracle-b}
\end{align}%
\label{eq:miracle-tcl}%
\end{subequations}%
The reason that both relations are valid for this model is \emph{not} that $\Sigma(s)$ is time-local --which it is not--
or that we approximated $\Pi(t) \approx \Pi(t')$ --which we did not.
That \Eq{eq:miracle-tcl} indeed holds for this model, is easy to see in the real-time approach [\App{app:superfermion-tcl}],
where it can be tied to the absence of interparticle interaction in the model,
and is also confirmed by an explicit calculation [\App{app:eom-generators}] revealing that the superoperator identity
$\Sigma(t-t')\Pi(t')\Pi^{-1}(t) = \Sigma(t-t')$ holds for all $t$, $t'$
even though $\Pi(t')\Pi^{-1}(t) \neq \mathcal{I}$.

\begin{table*}
	\caption{\label{tab:spectrum}
		Time-dependent eigenvectors and eigenvalues of the evolution $\Pi(t)$ and its time-local generator $\Sigma^\text{TCL}(t)$.
		Note that the separation of eigenvectors into subsets $k=1,4$ and $k=2,3$ does \emph{not} correspond
		to subsets of Kraus operators in \Eq{eq:krausfinal}.
		The operators $m_1(t)$ and $m_1^\TCL(t)$ are trace-normalized and positive if and only if $|g(t)| \leq 1$ and $|h(t)|\leq 1$, respectively.
		}.
	\begin{ruledtabular}
		\begin{tabular}{lrcl rcl}
			& & $\Pi(t)$ & & 
			& $\Sigma^\text{TCL}(t) $ &
			\\
			$k$ & Amplitude & Eigenvalue & Mode &
			Amplitude & Eigenvalue & Mode
			\\
			$1$ & $\Bra{\one}$ &
			$1$ &
			$\tfrac{1}{2} \big[ \Ket{\one}+g(t) \Ket{(-\one)^n} \big]$ &
			$\Bra{\one}$ &
			$0$ &
			$\tfrac{1}{2} \big[ \Ket{\one}+h(t) \Ket{(-\one)^n} \big]$ 
			\\
			$_3^2$ & $\Bra{d_\eta^\dag}$ &
			$e^{(i \eta \varepsilon -\frac{1}{2}\Gamma)t}$ &
			$\Ket{d_\eta^\dag}$ &
			$\Bra{d_\eta^\dag}$ &
			$ i \eta \varepsilon -\frac{1}{2}\Gamma $ &
			$\Ket{d_\eta^\dag}$ 
			\\
			$4$ & $\tfrac{1}{2} \big[ \Bra{(-\one)^n}-g(t) \Bra{\one} \big]$ &
			$e^{-\Gamma t}$ &
			$\Ket{(-\one)^n}$ &
			$\tfrac{1}{2} \big[ \Bra{(-\one)^n}-h(t) \Bra{\one} \big]$ &
			$-\Gamma$ &
			$\Ket{(-\one)^n}$ 
		\end{tabular}
	\end{ruledtabular}
\end{table*}

\subsection{`Markov-only' approximate QME}
We can now investigate the impact of some approximations formulated on the level of the QMEs~\cite{Whitney2008a}.
Based on the semigroup-divisible limits discussed in \Sec{sec:functions-physical}, we first discuss an obvious approximation%
	\footnote%
	{In the `initial slip' approach~\cite{Geigenmuller83,Gnutzmann96} one also replaces the time-local QME
	by another time-constant QME which has the same stationary state as the original equation.
	To further improve this approximation one additionally modifies the \emph{initial condition}, see \Ref{Whitney2008a}.
	Importantly, a constant initial slip will never recover the interesting reentrant behavior
	discussed in \Sec{sec:modes} because the dynamical map remains a semigroup.
	}
obtained by setting $h(t) \approx h(\infty)$ [\Eq{eq:g-infty}]
which amounts to replacing the generator in the time-local QME \eq{eq:tcl} by its \emph{exact stationary value}:
\begin{align}
	\Sigma^\text{TCL}(t) & \approx
	\Sigma^\text{TCL}(\infty) 
	=\tfrac{\Gamma}{2} \sum_\eta \left[1 - \eta h(\infty) \right] \mathcal{L}_\eta
	=\Sigma(i0^+)
	.
	\label{eq:tcl-markov}
\end{align}
The resulting approximate dynamics is governed by a GKSL equation with constant operators and coefficients,
the latter being automatically positive since $|h(\infty)| = |g(\infty)| \leq 1$ [\Eq{eq:g-infty-parity}].
The approximation \eq{eq:tcl-markov} thus strictly respects both CP and TP.

By construction, the approximation reproduces the \emph{exact} stationary state [\Sec{sec:modes}]
which is possible because \eq{eq:tcl-markov} is nonperturbative in the parameters $\Gamma$, $T$, and $\epsilon$,
in contrast to the $\Gamma$-linear Born-Markov approximation discussed below.
Despite this, it does not capture the interesting reentrant behavior discussed in \Sec{sec:modes}
since the nontrivial time-dependence of the functions $\gamma(s)$, $h(t)$ and $g(t)$ has been dropped.
Only in the semigroup limits, close to resonance ($|\epsilon| \ll T$) and far off resonance ($|\epsilon|  \gg T$),
also the transient dynamics is exact.
Whereas the first case forms the starting point for the renormalized perturbation theory [\Sec{sec:superfermion}] around the infinite-temperature limit,
the second case relates to infinite-bias approximations~\cite{Nazarov89,Gurvitz96,Gurvitz97,Oguri02,Wunsch05,Pedersen05,Pedersen07}.

The approximation \eq{eq:tcl-markov} may be called `Markov-only'
since we obtain the same approximation in the time-\emph{nonlocal} QME
from an `exact coarse-graining' procedure $\Sigma(t-t') \approx \Sigma(i0^+) \, \deltah(t-t')$
where only the zero Laplace-frequency component $z \to i0^+$ of the \emph{exact} memory kernel \eq{eq:sigma-z} is retained.
Note carefully that the zero-frequency and long-time approximations relate to \emph{different} objects
which are not simply related by a time-energy uncertainty relation:
$\Sigma(z)$ is \emph{not} the Laplace transform of $\Sigma^\TCL(t)$, note the upper integration limit \Eq{eq:miracle}.
\refabc{%
In fact, the equality $\Sigma^\text{TCL}(\infty)=\Sigma(i0^+)$ is remarkable
since the two procedures in general do \emph{not} give the same `Markov-only' approximation~\cite{Timm11,Contreras12,Nestmann19a}.
}%

Furthermore, the approximation can also be implemented on the level of individual Kraus operators by replacing $g(t) \to g(\infty)$.
An interesting open question is how this can be effected
in a direct microscopic calculation of the Kraus operators~\cite{vanWonderen13,vanWonderen18a,vanWonderen18b} at finite $T$.

\subsection{Born-Markov approximation\label{sec:qme-approx2}}

Finally, we note that the combined Born-Markov approximation recovers the well-known Golden-Rule result
$\Sigma^\TCL(\infty)=\Sigma(i0^+)\approx\Gamma \sum_\eta f({\eta \epsilon}/{T})\mathcal{L}_\eta$
in GKSL form where $f(x)=[e^{x}+1]^{-1}$ is the Fermi function.
For this, one additionally expands the time-local generator up to linear order in $\Gamma$ (Born)
which implies a zeroth-order approximation for $h(\infty)$ directly obtained from \Eq{eq:g-infty}:
\begin{gather}
	h(\infty)
	\approx
	 \tfrac{2}{\pi} \, \Im \, \psi \left(\tfrac{1}{2} + \tfrac{i\epsilon}{2\pi T} \right) 
	= 1-2f(\epsilon/T).
\end{gather}%
This approximation still respects TP as well as CP since $|f( \epsilon / T )| \leq 1$ \emph{independent} of $\Gamma$,
i.e., \emph{even} for $\Gamma$ values where the approximation is inapplicable.
In this respect, Born-Markov GKSL equations may be deceptive as the CP-TP property does not indicate a good approximation.

This is different when including higher order corrections.
Then, too large values of $\Gamma$ definitely lead to loss of CP signaling an inconsistency in the calculation, which is a good thing.
The presented expressions can be used as a benchmark
to study how systematic higher-order approximations~\cite{Koenig95,Leijnse08,Koller10,Timm08} in $\Gamma$
improve upon the Born-Markov results with respect to the CP,  semigroup- and CP-divisibility properties.
We note that the real-time renormalization group~\cite{Schoeller09,Schoeller18} 
already obtains the exact solution for this model in the \emph{one-loop} approximation~\cite{Schoeller00,Saptsov12}
and is in fact equivalent~\cite{Saptsov14} to our derivation in \Sec{sec:superfermion}.
\refc{%
This approach allows for a simultaneous treatment of strong coupling and (the here neglected) strong interaction effects
most relevant for molecular quantum dots, see~\Ref{Gaudenzi17a} and the references therein.
}

\section{Impact on observable dynamics\label{sec:modes}}

\refabc{
We now combine the insights of all approaches to address the question raised in \Fig{fig:intro}:
How are \emph{measurable} current reversals related to failure of the divisibility criteria for non-Markovian dynamics?
We analyze in detail the time-evolution of the level occupation
-- through the parity $\brkt{(-\one)^n(t)}= 1-2\brkt{n(t)}$ --
as well as the time-dependent information measures introduced in \Sec{sec:information}.
} 

\subsection{Spectral decomposition}
\refc{%
As usual, time-dependent solutions of a linear evolution problem are best analyzed in terms its eigenvectors.
Therefore we
start from the result \eq{eq:pi-eom} obtained in the EOM approach
which allows
}%
for an easy diagonalization [\Eq{eq:app-pi-diag}] of the superoperator $\Pi(t)$ by rearranging terms:
With the notation $\Bra{A}\bullet := \Tr_{\S}A^\dag \bullet$ and $\Ket{B}:=B$
to denote the Hilbert-Schmidt scalar product $\Braket{A}{B}=\Tr_{\S}A^\dag B$,
we have in the parity-basis:
\begin{align}
\Pi(t)  = \sum_{k=1}^4  \Pi_k(t) \, \Ket{m_k(t)} \Bra{a_k(t)}
.
\label{eq:pi-diag}%
\end{align}
Since $\Pi(t)$ is a non-normal superoperator,
it has distinct right eigenvectors $\Ket{m_k(t)}$ (modes) and left eigenvectors $\Bra{a_k(t)}$ (amplitudes) to the same eigenvalue $\Pi_k(t)$
which are listed in Table \ref{tab:spectrum}.
Similarly, the time-local generator \eq{eq:tcl-kernel} is diagonalized [\App{app:eom-tcl}] after converting to braket-form:
\begin{align}
\Sigma^\text{TCL}(t) 
& = \sum_{k=1}^4  \Sigma^\text{TCL}_k \, \Ket{m_k^\TCL(t)} \Bra{a_k^\TCL(t)}
.
\label{eq:tcl-diag}
\end{align}
Remarkably, the eigenvalues of the time-local generator are \emph{time-constant}
and related to the evolution eigenvalues as $\Pi_k(t)= e^{\Sigma_k^\TCL t}$, a form suggestive of a Markovian semigroup
where time-ordering is not an issue, cf. \Eq{eq:miracle}.
Also, the evolution eigenvalues always \emph{decay} in time, irrespective of the restrictions imposed by CP [\Eq{eq:g-cp}] or CP-divisibility [\Eq{eq:h-cpdiv}].
These restrictions are thus entirely incorporated in the \emph{eigenvectors} in Table~\ref{tab:spectrum}.
In particular, only the first mode vector ($k=1$) and last amplitude covector ($k=4$) feature a nontrivial time-dependence in terms of the functions $g(t)$ and $h(t)$,
which prevents the generator $\Sigma^{\text{TCL}}(t)$ from commuting with itself at different times.
As noted in \Sec{sec:qme-local} the latter complicates the solution of the time-local QME and the corresponding summation of the perturbation expansion.
For the following discussion, we therefore focus on the nontrivial dynamics of these modes.

\subsection{Fixed point of $\Pi(t)$ -- Reentrant states}

\refa{%
The time-independent eigenvalue $\Pi_1(t)=1$ corresponds to a \emph{fixed point} at finite time $t$,
$\Pi(t) \Ket{m_1(t)} = \Ket{m_1(t)}$,
whose existence is guaranteed by the trace preservation of the evolution.
This \emph{time-dependent} fixed point of the dynamical map is often not discussed
and should be clearly distinguished from the \emph{stationary state}
}%
\begin{align}
\rho(\infty)= \Pi(\infty) \rho(0) = m_1(\infty),
\label{eq:rho-infty-def}
\end{align}
which is reached \emph{independently} of the initial state $\rho(0)$.
In agreement with general fixed-point theorems [\App{app:pp}],
the fixed point $m_1(t)$ is guaranteed to be a physical state if the condition $|g(t)|\leq 1$ holds, see the caption of Table~\ref{tab:spectrum}.
As we show below, preparing the system in such a state, $\rho(0)=m_1(t^\r)$ with some parameter $t^\r$,
will force the evolution $\Pi(t)$ to exactly recover it at least once at time $t=t^\r$.
This distinct \emph{reentrant} behavior cannot arise within semigroup limits or approximations
and is therefore an indicator for the loss of semigroup-divisible dynamics.
In terms of particle transport, it implies a \emph{reversal} of the time-dependent \emph{particle current} to the environment
--a pronounced effect that can equivalently be seen in the parity of the fermionic level given by \Eq{eq:parity-sol} and repeated here for convenience:
\begin{align}
	\brkt{(-\one)^n(t)} = 
	e^{-\Gamma t} \brkt{(-\one)^n(0)} + (1-e^{-\Gamma t})g(t)
	.
	\label{eq:parity-sol2}
\end{align}

\begin{figure}[t]
	\includegraphics[width=\linewidth]{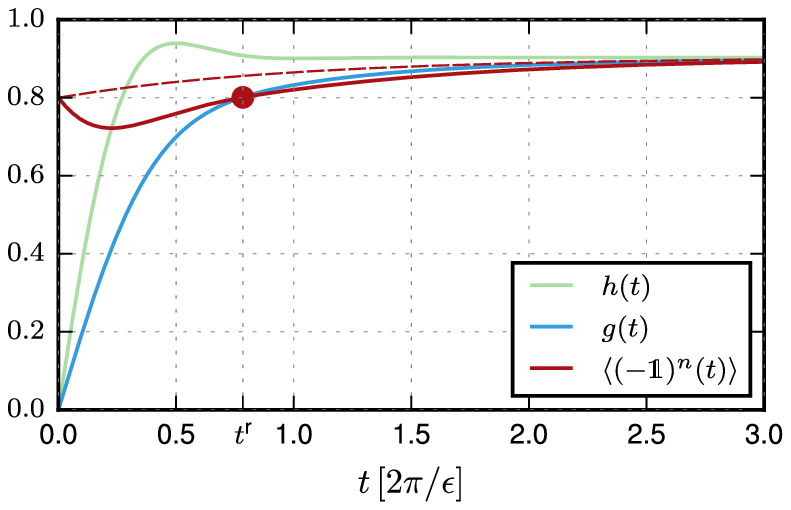}
	\caption{
		\refabc{Reentrance at time $t^\r$ (dot) of the level occupation expressed in terms of the parity $\brkt{(-\one)^n(t)}$ (red)
		for $2\epsilon/\pi = 4\Gamma$ and $\pi T = 10 \cdot \Gamma/2$.}
		Asymptotically the parity approaches the function $g(t)$ (blue).
		Note that the minimum implied by the reentrance occurs precisely when the parity crosses the function $h(t)$ (green), cf. \Eq{eq:extremum}.
		The reentrant behavior is not captured by the non-perturbative `Markov-only' approximation \eq{eq:tcl-markov} to the parity (red dashed).
		}
	\label{fig:reentrant}
\end{figure}

\begin{figure}[t]
	\includegraphics[width=\linewidth]{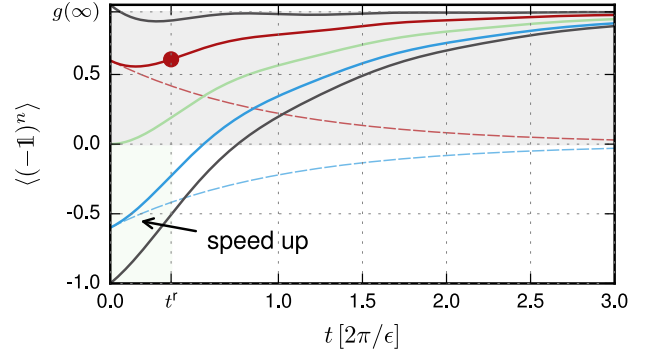}
	\caption{
		\refabc{Evolution of the level occupation expressed in terms of the parity}
		for different initial conditions and $2\epsilon/\pi = 4\Gamma$ and $\pi T = 10^{-3} \cdot \Gamma/2 $.
		When initialized in the range~$[0,g(\infty))$ (gray shade) the parity shows reentrant behavior (red curve, dot).
		For a fixed initial value, the reentrant time (green shade) scales as $t^\r \sim \pi / \epsilon$ with the level detuning $\epsilon=\varepsilon-\mu$.
		In this window, the deviation from the leading exponential behavior $e^{-\Gamma t}$ (dashed lines) sets in.
		For reentrant dynamics, this causes the parity to reverse while for non-reentrant dynamics it speeds up the approach to the stationary state from below.
		Neither effect is captured by the `Markov-only' semigroup approximation \eq{eq:tcl-markov} (not shown).
	}
	\label{fig:reentrant2}
\end{figure}

\Fig{fig:reentrant} illustrates that the parity (red curve) is dominated at short times $t \sim \Gamma^{-1}$ by the exponentially decaying first term.
This decay is however soon counteracted with the nonlinear time-dependence introduced by $g(t)$ (blue curve) in the second term
which sets in on an independent time-scale $t \sim \pi/\epsilon$ determined by the level detuning $\epsilon$, cf. \Eq{eq:hg-linear}.
As shown in \Fig{fig:reentrant2} for $\epsilon \geq 0$,
their competition causes the parity evolution to display two qualitatively different types of behavior depending on the initial parity value $\brkt{(-\one)^n(0)}$.

(a) When the initial parity lies within the range of $g(t)$, i.e., within $[0,g(\infty))$ for $\epsilon > 0$ and $(g(\infty),0]$ for $\epsilon < 0$,
see the discussion following \Eq{eq:g-h}, one can find a parameter $t^{\r}$ such that $\brkt{(-\one)^n(0)} = g(t^{\r})$.
The evolution then revisits this parity value at time $t=t^\r$ (red curve):
\begin{align}
\brkt{(-\one)^n(t^{\r})} = \brkt{(-\one)^n(0)} = g(t^\r)
.
\label{eq:reentrant-point}
\end{align}
In cases where $g(t)$ is nonmonotonic, the parity may revisit the initial value several times whenever $g(t)$ revisits it (not shown).
It is remarkable that the initial decay with $e^{-\Gamma t}$ (red dashed line in \Fig{fig:reentrant2}) goes in the `wrong' direction, \emph{away} from the stationary value.
Thus, in these cases where the level is initially prepared as being `too empty' relative to the stationary value,
it first becomes significantly \emph{more empty} before starting to fill up again.

This behavior can be controlled with the level detuning $\epsilon=\varepsilon -\mu$ and the effects are most pronounced in between the two Markovian semigroup limits:
For $|\epsilon| \ll T$, the range of reentrant initial values (gray vertical range in \Fig{fig:reentrant2}) shrinks because $g(t) \to 0$ [\Eq{eq:eps-zero}].
For $|\epsilon| \gg T$, instead the time-scale for revisiting the initial value (green horizontal range) shrinks because $g(t) \to \theta(t)$ [\Eq{eq:eps-infty}].
The reentrant behavior at low temperatures is most pronounced when $|\epsilon| \sim \Gamma \gg T$,
i.e., when $\mu$ is positioned on the \emph{flank} of the $\Gamma$-broadened resonance at $\varepsilon$.
Note that even with all these parameters fixed, the time-scale $t^\r$ for reentrance still depends on the initial condition in the range of $g(t)$
and can thus be chosen independently.

(b) No reentrance occurs when the initial value of $\brkt{(-\one)^n(0)}$ lies outside the range of $g(t)$.
The parity then decays to its stationary value $g(\infty)$ either from above or below (gray curves in \Fig{fig:reentrant2}).
The approach from above may still be strongly nonmonotonic, in particular when $|\epsilon| \gg T$,
i.e. when the stationary state is nearly pure, i.e., the stationary parity is large  $|g(\infty)| \to 1$.

\subsection{Fixed point of $\Sigma^\TCL(t)$ -- Local stationary states\label{sec:fixedpoint-tcl}}

The reentrant behavior of the parity implies by continuity that it must have gone through an extremum (minimum) at some earlier time $t^\e \leq t^\r$.
We thus turn to analyzing its time-\emph{local} behavior reflected by the derivative
\begin{align}
	\tfrac{d}{dt} \brkt{ (-\one)^n(t)}
	 = - \Gamma \Big[ \brkt{(-\one)^n(t)} - h(t) \Big]
	 ,
\label{eq:dpdt}
\end{align}
and more generally by the eigenvectors of the time-local generator $\Sigma^\text{TCL}(t)$
which differ from those of $\Pi(t)$ only by the replacement $g(t) \to h(t)$, see Table~\ref{tab:spectrum}.
The existence of a time-dependent zero-mode $\Ket{m^\TCL_1(t)}$ is implied by the trace-preservation.
In contrast to $\Ket{m_1(t)}$, this mode is \emph{not} a physical state unless $|h(t)| \leq 1$ holds at a \emph{fixed} time $t$.
As a consequence, the parity cannot have any kind of extremum%
		\footnote{
		This argument fails for more complicated systems where a single scalar observable is not sufficient to describe the system density operator [\Eq{eq:rho-t}]:
		The physical zero-mode is only visited if all observables spanning an orthonormal operator-basis are extremal at the \emph{same} time.
		}
at instants where $|h(t)| > 1$, no matter how the initial parity is chosen.

\begin{figure}[t]
\includegraphics[width=\linewidth]{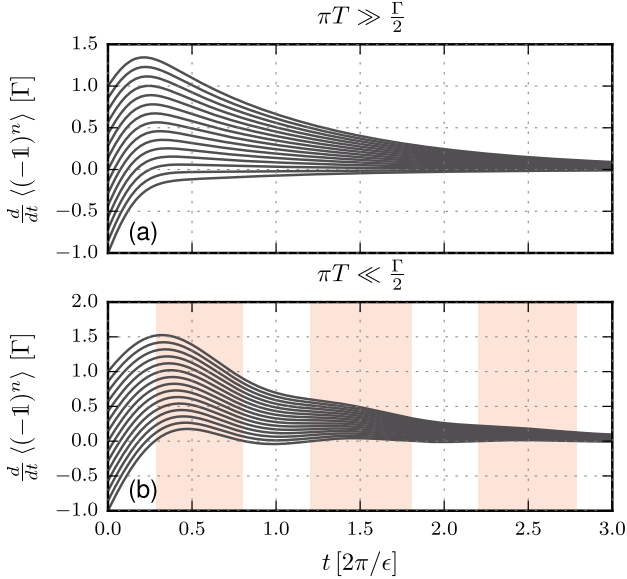}
	\caption{
		A family of parity evolutions is obtained by varying the initial conditions $\brkt{(-\one)^n(0)} \in [-1,1]$.
		Plotted are the corresponding \emph{transport currents} for $2\epsilon/\pi = 4\Gamma$.
		(a) At high temperature $\pi T = 20 \cdot \Gamma /2$,
		we have CP-divisible evolution characterized by the occurrence of current-reversals at any time.
		(b) At low temperature $\pi T = 10^{-3} \cdot \Gamma /2$,
		we have non-CP-divisible evolution, as a current-reversal is prohibited in broad time-intervals (red shade), irrespective of the initial condition.
		In the complementary intervals, the dynamics is still \emph{locally} CP-divisible, see discussion in the text.
	}
	\label{fig:cpdiv}
\end{figure}

We stress that the condition for an extremum at time $t$ is weaker than the CP-divisibility criterion \eq{eq:h-cpdiv}
demanding $|h(t)|\leq 1$ for \emph{all} times $t>0$ of the evolution.
In this simple system, CP-divisibility can be studied by measuring families of occupation evolutions with varying initial conditions:
The evolution is CP-divisible, if this family reveals the existence of extrema in \emph{any} time-interval as is illustrated in \Fig{fig:cpdiv}(a)
by plotting the corresponding transport current.
Conversely, the loss of CP-divisibility is accompanied by the emergence of time-intervals \emph{without} extrema
as illustrated in \Fig{fig:cpdiv}(b) for a lower temperature.
Within the time-intervals where extrema do occur, $|h(t')|\leq 1$, and the dynamical map still factorizes as $\Pi(t)=\Pi(t,t')\Pi(t')$ into two CP-TP maps [\Eq{eq:pi-divisor}].
It thus makes sense to denote the dynamics as \emph{locally} CP-divisible at time $t'$.
Physically, this means that the dynamics is still insensitive to reinitializing the environment but only in restricted time-intervals
which are observable quantifiers of non-semigroup dynamics.

\emph{Nature of extrema.}
Let us now focus on the nature of extremal points $t^\e$ which occur when the physical zero-mode is visited, $\rho(t^\e) = m_1^\TCL(t^\e)$ during some evolution.
At such points, the evolution temporarily comes to a complete halt, $\tfrac{d}{d t} \rho(t^{\e}) = 0$, before speeding up again towards the stationary state.
As noticed above, this halting is equivalent to an extremum of the parity and by \Eq{eq:dpdt} requires
\begin{align}
\brkt{(-\one)^n(t^\e)}=h(t^\e)
\label{eq:extremum}
.
\end{align}
In \App{app:functions-parity}, we show that  for $\Gamma/2 \geq \pi T$,
\emph{any} fixed initial condition $\brkt{(-\one)^n(0)} \in [-1,1]$ satisfies \Eq{eq:extremum} for infinitely many times $t^\e$. This means that the parity keeps on oscillating although with ever decreasing amplitude.
For $\Gamma/2 \leq \pi T$, there exist initial parity conditions which \emph{never} satisfy the condition for any $t^\e$ and their evolutions show no extrema at all.
However, there is always a range of initial parities that does satisfy it for some $t^\e$
and it is in this range that the reentrant behavior occurs.

\begin{figure}[t]
	\includegraphics[width=\linewidth]{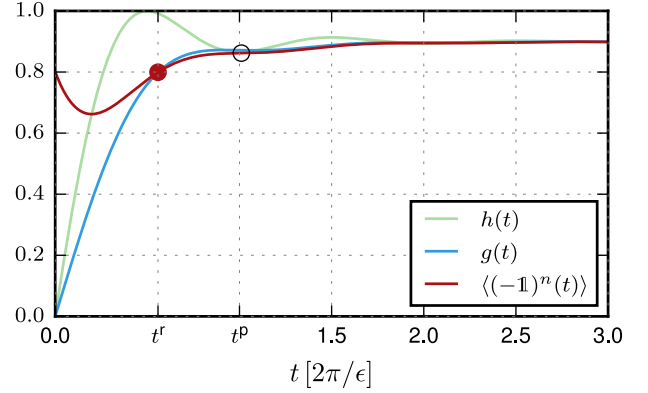}
	\caption{
		Local stationary state:
		The \refabc{level occupation} parity (red) can develop a plateau (circle) when it crosses with $h(t)$ (green) at $t^\p=\ell \pi/\epsilon$ [\Eq{eq:tstar}] for any $\ell = 1,2,\ldots$.
		Here $\ell = 2$ and
		we have taken low temperature $\pi T = 10^{-3} \cdot \Gamma / 2$ for which the curvature \eq{eq:stationary-local3} is negligibly small. 
		The time-width of the plateau scales as $\epsilon^{-1}$ with the level-detuning since $\epsilon=\varepsilon-\mu$
		determines the inverse slope of $\gamma(t^\p)$ [\Eq{eq:fermi-contraction}], here $2\epsilon/\pi = 2\Gamma$.
		We note that the value \eq{eq:stationary-local3} is only exactly zero for a specific, much higher temperature where the dynamics is very close to a featureless semigroup.
	}
	\label{fig:halting}
\end{figure}

The halting $\tfrac{d}{d t} \rho(t^{\e})=0$ does not imply that higher time-derivatives are zero at $t^\e$ as it does in the stationary limit.
The curvature at an extremal point is determined due to \Eq{eq:miracle} by the \emph{memory-kernel} of the time-\emph{non}local QME,
\begin{align}
\tfrac{d^2}{d t^2} \rho(t^{\e}) = [\tfrac{d}{dt}\Sigma^\TCL(t^\e)] \,  \rho(t^\e)
=\Sigma(t^\e) \rho(t^\e)
.
\label{eq:stationary-local2}
\end{align}
Although $\rho(t^\e)=m_1^\TCL(t^\e)$ is never a zero eigenvector of $\Sigma(t^\e)$, the curvature may still vanish.
This happens at times $t^\p$ where the memory-kernel \eq{eq:kineq-kernel} itself vanishes
and coincides with the zeros
\begin{align}
\gamma(t^\p) = 0 \quad \Leftrightarrow \quad 
t^\p = \frac{\pi}{\epsilon} \ell
\quad
\ell=1,2,\ldots
\label{eq:tstar}
\end{align}
of the correlation function \eq{eq:fermi-contraction}.
Since $\tfrac{d}{dt}h(t) =e^{-\frac{\Gamma}{2}t} \gamma(t)$ [\Eq{eq:h-def}], the function $h(t)$ equivalently becomes extremal at the time $t^\p$.
Thus, if the parity $\brkt{ (-\one)^n(t)}$ crosses $h(t)$ at one of its extremal points $t^\p$, its curvature
\begin{align}
	\tfrac{d^2}{dt^2} \brkt{ (-\one)^n(t)}  &= -\Gamma \left[\tfrac{d}{dt} \brkt{ (-\one)^n(t) }-\tfrac{d}{dt}h(t) \right]
	\label{eq:stationary-local3}
\end{align}
vanishes exactly. 
Moreover, around points where $t^\e \approx t^\p$ the parity can also develop a pronounced \emph{approximate} plateau as illustrated in \Fig{fig:halting}.
The state may be called \emph{locally stationary} close to such points.

\begin{figure}[t]
	\includegraphics[width=\linewidth]{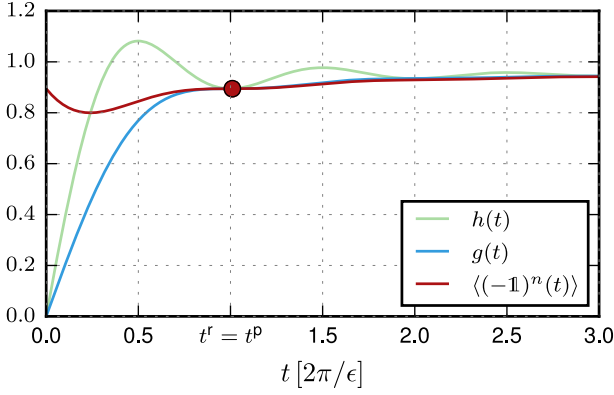}
	\caption{
		Local stationarity at reentrance:
		The reentrant point (dot) of the \refabc{level occupation} parity (red) can coincide with a plateau (circle)
		only at one of the \emph{touching points} of $g(t)$ (blue) and $h(t)$ (green) located at the minima $t^\p =\ell \pi/\epsilon$ for $\ell = 0,2,4,\ldots$.
		These curves first touch exactly for $\Gamma/2=\pi T$ and are generic also for lower temperatures:
		It requires very large level-detunings $\epsilon$ to significantly pull the green curve below the blue one.
		Here $2\epsilon/\pi = 4\Gamma$ and $\ell = 2$.
	}
	\label{fig:halting-reentrant}
\end{figure}

\emph{Local stationarity at reentrance.}
As seen in Table~\ref{tab:spectrum}, the eigenvectors of the evolution and its time-local generator are in general different.
However, there are special cases where they coincide, namely at the crossing points $t^\r$ of the functions $g(t)$ and $h(t)$.
The initial condition can thus be chosen as $\brkt{ (-\one)^n(0)} = g(t^\r) = h(t^\r)= \brkt{ (-\one)^n(t^\r)}$
such that the reentrant point [\Eq{eq:reentrant-point}] is also an extremum ($t^\r=t^\e$).
This is always possible for $\Gamma / 2 > \pi T$, where the reentrant point is either a maximum or a minimum.
For $\Gamma / 2 = \pi T$ it is instead a saddle point [\App{app:functions-extra}] and the parity shows a plateau at the reentrant point ($t^\r = t^\p$).
The system then not only recovers the initial state but even becomes \emph{locally} stationary at reentrance as is illustrated in \Fig{fig:halting-reentrant}.
For $\Gamma / 2 < \pi T$ such a situation can never arise.

\subsection{Information measures}

\refabc{%
Above we made a distinction between the initial state and the reentrant state by following the time-evolution between $t=0$ and $t=t^\r$.
However, if one is limited to only performing measurements on the system at these two times, one cannot distinguish these states.
Measures of information exchange [\Sec{sec:information}] at these two times
}%
do allow for such a distinction and furthermore quantify the backaction of the system on its environment in detail.

\begin{figure}[t]
	\includegraphics[width=\linewidth]{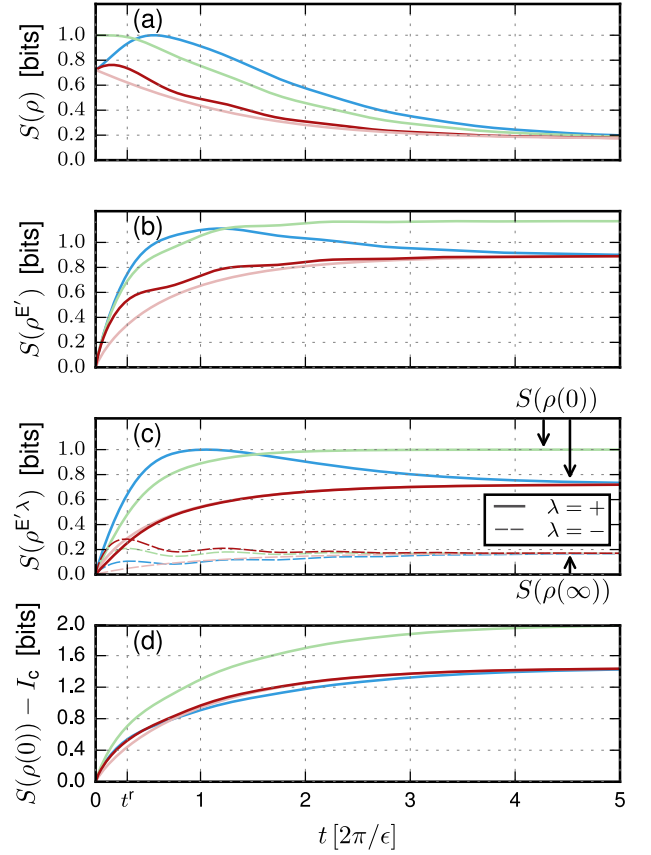}
	\caption{
         (a) System entropy for the middle three parity evolutions of \Fig{fig:reentrant2} (same color coding).
         The system entropies may initially differ (mixture of preparation) but converge to the same stationary value.
         For the reentrant dynamics (red), we compare with the `Markov-only' approximation \eq{eq:tcl-markov} (light red).
         (b) The corresponding effective environment entropies are initially the same (by purification)
         but converge to different stationary values (backaction on the effective environment).
         (c) Decomposition of (b) into entropies of the effective environment modes,
         $S(\rho^{\E'}(t)) = \sum_{\lambda} S(\rho^{\E'\lambda}(t))$, see \Eq{eq:factors}.
         One mode (solid, $\lambda=+$) captures the smooth large variations
         while the other (dashed, $\lambda=-$) adds smaller oscillatory behavior.
         Consistent with \Eq{eq:entropy-env-stat}, the former mode converges to the spectrum of $\rho(\infty)$ and the latter to that of $\rho(0)$.
         (d) Coherent-information mismatch [\Eq{eq:mismatch}].
         In this quantity the `Markov-only' approximation deviates only very little.
     }
	\label{fig:entropies}
\end{figure}

In \Fig{fig:entropies} we plot the respective entropies of the system and the effective environment
for the three different initially mixed states of \Fig{fig:reentrant2}.
For zero initial parity (green), the system entropy in \Fig{fig:entropies}(a) decreases from its maximal value.
For the evolution without reentrance (blue), the system entropy first increases until it reaches the maximal entropy state of one bit before decreasing again.
In this case, the nonmonotonic behavior is solely due to the fact that the system entropy cannot distinguish the sign of the parity
which has to pass through zero to reach the stationary value of opposite sign.
Because of the fermionic superselection, this reversal can only be achieved via the maximal entropy state,
i.e., the corresponding Bloch vector must shrink to zero in order to reverse,
and is also captured by the `Markov-only' semigroup approximation (not shown).
This behavior should be clearly distinguished from the reentrant dynamics (red)
where the system entropy also increases initially but never reaches the maximal value:
The system initially evolves in the 'wrong' direction before turning around towards the stationary state
--a behavior not present in the 'Markov-only' approximation (light red).
	
The effective environment entropy in \Fig{fig:entropies}(b) also depends on the initial system state.
Unlike the system entropy this dependence remains in the stationary limit where it converges to
$S(\rho(0))$ [\Eq{eq:entropy-env-stat}, \Sec{app:entropies-factor-stationary}].
At large times it therefore does not distinguish between the reentrant and non-reentrant evolutions starting from opposite parity signs
(blue and red curves merge at large times).
However, at short times it distinguishes these particularly clearly:
The reentrant and non-reentrant evolutions have a distinct backaction, the former having a smaller --but still sizable-- entropy footprint in the effective environment.
This shows that the state of the environment is definitely different at the reentrant point $t^\r$ although the system state is recovered exactly.
The splitting between the blue and red curves does not arise in the $t$-linear regime
where both evolutions speed towards the maximally mixed state and their effective-environment entropies still coincide.
They only split up when the nonlinearity caused by $g(t)$ kicks in which is precisely what distinguishes the two parity evolutions in \Fig{fig:reentrant2}:
It is responsible for the reversal of the parity evolution in the reentrant case, and its speed-up in the non-reentrant one.

The contributions of the individual effective environment modes $\rho^{\E'\lambda}(t)$ discussed at the end of \Sec{sec:kraus} are shown in \Fig{fig:entropies}(c).
Remarkably, these modes separate the oscillatory behavior of the effective environment entropy ($\rho^{\E'-}$, dashed curves)
from the dominant evolution ($\rho^{\E'+}$, solid curves).
Notably, the `Markov-only' approximation (light red) significantly deviates from the former but nearly perfectly reproduces the latter.

Finally, in \Fig{fig:entropies}(d) we show the corresponding coherent-information mismatch $S(\rho(0)) - I_\c(t)$.
Its overall dominating increase quantitatively underscores that the initial entanglement with the auxiliary system $\P$,
involved in the preparation of the mixture $\rho(0)$ [\Sec{sec:information}],
is continuously being converted into entanglement with the effective environment $\E'$ until all entanglement is broken in the stationary state, see \Fig{fig:div}(a).
This happens irrespective of the reentrant behavior of the level occupation and the corresponding reversals of the particle current.
The curves in \Fig{fig:entropies}(d) evolve in groups split according to whether the magnitude of the initial parity is small (green) or large (blue, red).
The sign of the initial parity --decisive for the occurrance of reentrant behavior-- causes only a small splitting (blue, red).
One further notices that modulations on the timescale $\pi/\epsilon$ --clearly present in the individual entropies--
are absent in the (mismatch of the) coherent information.
This highlights a $\pi$-phase-shift between the modulations of the two entropies which may be easily overlooked in \Fig{fig:entropies}(a) and (b).
In fact, only in the limit of large detuning $\epsilon \gg \Gamma /2 \geq  \pi T$,
one can find nonmonotonic behavior on time-scales where the stationary state is far from reached.
Still, both $S(\rho(t))$ and $S(\rho^{\E'}(t))$ show oscillations in this case long after the oscillations in $I_\c(t)$ have died out.

\section{Summary and outlook\label{sec:discuss}}
\refabc{%
Motivated by the pivotal role of open quantum systems
in a range of disciplines, 
we addressed a number of recent questions regarding their dynamics
by revisiting the simplest possible case of common interest.
Our analysis of the time-dependent transport for
a resonant level with arbitrary coupling $\Gamma$ to a thermal reservoir at arbitrary temperature $T$
relied on several approaches to the exact dynamics providing very different insights.
As anticipated in the introduction,
the application of quantum-information methods required a significant effort,
but has provided interesting insights even though this model
is often declared `solved' in the study of open systems.
}%
In particular, the Kraus operator-sum enabled an exact analysis in terms of a strongly correlated system of one fermion
coupled to an \emph{effective} two-fermion \emph{environment}.
The presented exact expressions may provide useful benchmarks, in particular for general methods
that aim at directly computing Kraus operators for open quantum systems~\cite{vanWonderen13,Chruscinski13,vanWonderen18a,vanWonderen18b,Reimer19a} to enable novel CP approximations.

To evaluate the impact of approximations, it proved useful to also discuss two
\refabc{%
extensively used
}%
exact quantum master equations for the same dynamics, one time-local and one time-nonlocal.
Notably, we found an example of a `Markov-only' semigroup approximation that is non-perturbative in all parameters,
yet simultaneously completely positive (CP) and trace-preserving (TP).
\refabc{%
It should be noted that this approximation relies \emph{only} on the exact value of the \emph{stationary occupation} $\brkt{n(\infty)}=\tfrac{1}{2}[1-g(\infty)]$.
This may be relevant for recently developed time-dependent density functional theory (TD-DFT) approaches to interacting open systems~\cite{Dittmann18,Kurth18}
that are based on a mapping to the \emph{noninteracting} limit studied here.
}%

Despite its simplicity, \refabc{the resonant level model} displays both semigroup- and CP-divisible behavior
as well as dynamics that is neither
\refabc{%
and thus `non-Markovian'.
Experimental
}%
parameters allow one to tune between these three regimes, in particular the level-detuning $\epsilon$
and the competition between thermal ($T$) and quantum fluctuations ($\Gamma$).
We explored how these distinctions are reflected in the transient dynamics of the level occupation when varying the initial condition.
We
\refabc{%
found
}%
that the loss of semigroup-divisibility is the most pronounced distinction~\cite{Modi19}
\refabc{%
as the time-dependent fixed point of the dynamical map becomes markedly distinct from the stationary state.
This reveals that,
counter}
to intuition, the system occupation may temporarily \emph{increase} significantly in order to reach a stationary state with \emph{smaller} occupation.
\refabc{%
Experimentally, this means that the measurable transport current is reversed
for a wide range of parameters and a finite, sharp range of initial level occupations.
}

We also found that the occupation dynamics can come to a halt at several extrema before continuing towards the unique
\refabc{%
stationary state.
For strong coupling
$\Gamma  > 2\pi T$,
}%
this generically happens irrespective of the initial
\refabc{
level occupation,
}%
whereas in the opposite case a definite window of initial
\refabc{
occupations
}%
is required.
Additionally, the extrema can turn into \emph{locally stationary} states where even the \emph{evolution curvature} is strongly suppressed.
\refabc{%
We showed
that in this system
}%
CP-divisibility can be observed by studying families of occupation measurements with varying initial conditions:
Whereas the loss of semigroup-divisibility is accompanied by the \emph{appearance} of extrema in such families,
the loss of CP-divisibility is instead associated with the \emph{loss} of extrema.
The dynamics may however still be \emph{locally} CP-divisible in definite time-windows
--a distinction that allows for a more fine-grained characterization
\refabc{%
of `non-Markovianity'.
Thus, CP-divisibility provides an interesting quantitative guide to the more subtle features of the dynamics
beyond the semigroup-divisibility.
}%

\acknowledgments

We thank R. Saptsov, \'A. Rivas, and S. Maniscalco for useful discussions.
V. R. and K. N. acknowledge support by the Deutsche Forschungsgemeinschaft (RTG 1995).
\appendix

\section{Purifications, information measures and effective environment [\Sec{sec:open-system}]\label{app:entropies}}

In this appendix we collect some basic derivations of the results reviewed in \Sec{sec:open-system}.
It is useful to deviate from the notation of the main text by $\rho \to \rho^{\S}$ and $\rho^\P \to \rho^{\P_\S}$
to clearly distinguish the subsystems introduced below.

\subsection{Purifications and complementary subsystems}

\begin{figure}[t]
	\includegraphics[width=\linewidth]{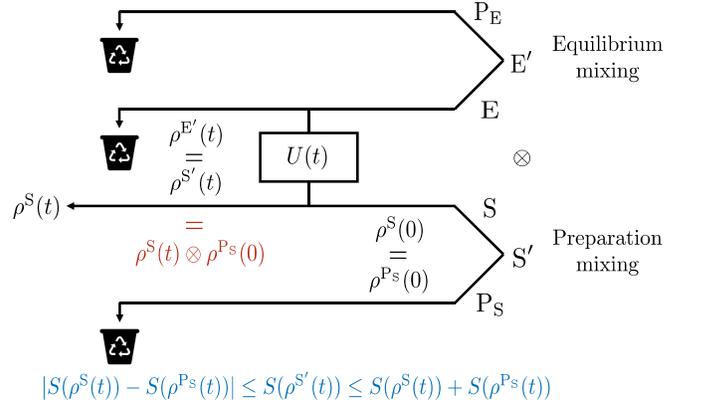}
	\caption{
		Dynamics $\rho^\S(t)= \Tr_{\E} U(t) [ \rho^\S(0) \otimes \rho^{\E}(0) ] U(t)^\dag$
		with \emph{mixed} initial states for both system and environment.
		The extensions by the purifications $\P_\S$ and $\P_\E$ account for this mixing as initial entanglement of pure systems $\S'$ and $\E'$, respectively.
		The trashcan indicates that a system has become inaccessible (is traced out).
		The \emph{red} product state only holds for a \emph{pure} initial state $\rho^\S(0)$.
		An equality such as $\rho^\S(0)=\rho^{\P_\S}(0)$
		indicates that the \emph{spectra} of the reduced \emph{density matrices} of two complementary subsystems are the same,
		while the operators act on different subsystem spaces $\S$ and $\P$.
	}
	\label{fig:app-entropies}
\end{figure}

The information quantities \eq{eq:I}-\eq{eq:mismatch} in the main text and their fundamental properties are best understood
by considering instead of the \emph{mixed} system $\S$ and its thermally \emph{mixed} environment $\E$
their \emph{purifications}~\cite{NielsenChuang} as sketched in \Fig{fig:app-entropies}.
At $t=0$ system and environment are decoupled and can each be considered to as the marginal state of a pure entangled state
of an extended system $\S' = \S \otimes \P_\S$, respectively environment $\E'=\E \otimes \P_\E$:
\begin{subequations}
\begin{align}
	\rho^{\S}(0) &=\tr {\P_\S} \ket{\psi^{\S'}(0)}\bra{\psi^{\S'}(0)}
	\label{eq:rho-S-0}
	\\
	\rho^{\E}(0) & =\tr {\P_\E} \ket{\psi^{\E'}(0)}\bra{\psi^{\E'}(0)}
	\label{eq:rho-E-0}
	.
\end{align}
\end{subequations}
Here, it is sufficient~\cite{NielsenChuang} to consider $\P_\S$ and $\P_\E$ as copies of the original Hilbert spaces $\S$ and $\E$.
Physically, $\P_\S$ can be understood as an auxiliary system involved in the \emph{preparation} of $\S$ in the initial \emph{mixed} state \eq{eq:rho-S-0}
as discussed in Figs.~\fig{fig:cp} and \ref{fig:div} of the main text.
We now similarly represent the mixture of the environment $\E$ --due to thermal and electrochemical equilibrium--
in terms of entanglement with an auxiliary system $\P_\E$.

The joint state of $\S' \otimes \E'$ is initially pure and factorizes as $\ket{\psi^{\S'\E'}(0)}=\ket{\psi^{\S'}(0)} \otimes \ket{\psi^{\E'}(0)}$.
The unitary evolution $U(t)$ in \Eq{eq:pi} only couples $\S$ and $\E$ and thereby generates entanglement between $\S'$ and $\E'$,
causing the initial factorization to break down:
\begin{align}
	\ket{\psi^{\S'\E'} (t)}
	&
	= \one^{\P_\S \P_\E} \otimes U(t)
	\, \,
	\ket{\psi^{\S'\E'}(0)}
	.
	\label{eq:psi-se} 
\end{align}
The key point for the following is illustrated in \Fig{fig:app-entropies}:
Because the joint state stays pure, the \emph{spectra} of marginal states obtained by tracing out any pair of complementary subsystems are all equal,
and as a consequence also the corresponding entropies are equal.
If at time $t$ we trace out all subsystems except $\S$, respectively $\E'$,
we obtain the time-dependent states of the system, respectively \emph{effective} environment that are discussed in the main text:%
\begin{subequations}%
	\begin{align}
	\rho^{\S}(t) &=\tr {\E'\P_\S } \ket{\psi^{\S'\E'} (t)}\bra{\psi^{\S'\E'} (t)}
	\\
	\rho^{\E'}(t) & = \tr{\,\,\, \S \P_\S} \ket{\psi^{\S'\E'} (t)}\bra{\psi^{\S'\E'} (t)}
	.
\end{align}%
Importantly, the above holds for any choice of purifications which produces the same reduced dynamics $\rho^{\S}(t)$ and specifically
includes the \emph{effective environment} introduced in the main text.
There, we denoted the initial purified environment state as $\rho^{\E'}(0):=\ket{0} \bra{0}$
and expressed the evolution on $\S \otimes \E'$ in terms of the system Kraus operators as $U'(t) = \sum_{m} K_m(t) \otimes \ket{m}\bra{0}$, cf. \Eq{eq:rho-E'}. 
Since the \emph{spectrum of} the marginal state $\rho^{\E'}(t)$ is independent of the choice of this particular purification,
it provides \emph{intrinsic} information about the \refb{effect of the coupling} of the system with \emph{any} initially pure effective environment\cite{NielsenChuang}.
The initial purity of the effective environment is important as it allows a clean 'count' of the entanglement generated by \refb{its joint evolution} with the system.
For the 'original' environment $\E$, this is not possible because it is already mixed at $t=0$ due its equilibrium state represented as entanglement with $\P_\E$.

\subsection{Coherent information and mismatch\label{app:entropies-coh}}

It is important to also consider the marginal state
\begin{align}
	\rho^{\S'}(t) & = \tr{\E'} \ket{\psi^{\S'\E'} (t)}\bra{\psi^{\S'\E'} (t)}
\end{align}%
\end{subequations}%
of the system $\S'=\S \otimes \P_\S$ including the auxiliary preparation:
The \emph{negative of its conditional entropy}~\cite{NielsenSchumacher96,Preskill} given the system $\S$ defines the coherent information \eq{eq:I}:
\begin{subequations}
	\begin{align}
I_\c(t) & = - [ S(\rho^{\S'}(t)) - S(\rho^{\S}(t)) ]
\label{eq:app-Ic-a}
\\
	& = S(\rho^{\S}(t)) - S(\rho^{\E'}(t))
	.
	\label{eq:app-Ic-b}
\end{align}
\end{subequations}
From its definition \eq{eq:app-Ic-a} one can show~\cite{Wilde13,Horodecki05,Horodecki07} that $\S$ and $\P_\S$ are entangled if $I_\c(t) > 0$,
i.e., the positive coherent information quantifies the remaining `preparation-entanglement' \emph{after} \refb{joint evolution with the environment and discarding the latter.}
For initially pure states $\rho^\S(0)$, there is no such entanglement and $I_\c(t) = 0$ for all $t \geq 0$  as claimed after \Eq{eq:I} in the main text
and illustrated in red in \Fig{fig:app-entropies}:
In this case, the purification $\rho^{\S'}(0)=\rho^{\S}(0)\otimes \rho^{\P_\S}(0)$ is necessarily a tensor product of pure states (entanglement monogamy~\cite{NielsenChuang}).
Because the extension $\P_\S$ does not evolve in time [\Eq{eq:psi-se}] it
remains in its initial pure state $\rho^{\P_\S}(t)=\rho^{\P_\S}(0)$ with vanishing entropy $S(\rho^{\P_\S}(t))=0$. 
Consequently, all entropy of $\S'$ originates from the system itself, $S(\rho^{\S'}(t)) = S(\rho^{\S}(t))$ for $t\geq 0$.

The other form \eq{eq:app-Ic-b} follows from the fact noted earlier that the spectra of $\rho^{\E'}(t)$ and $\rho^{\S'}(t)$ are always equal, see \Fig{fig:app-entropies}.
This in particular implies for the case of an initially pure system $\rho^{\S}(0)$ discussed above
that due to $\rho^{\S'}(t)=\rho^{\S}(t) \otimes \rho^{\P_\S}(0)$ the nonzero eigenvalues of the $\E'$ and $\S$ spectra coincide for all $t\geq 0$:
\begin{align}
	\text{spec}\, \rho^{\E'}(t) = 	\text{spec}\, \rho^{\S}(t)
	\quad
	\text{ if $\rho^{\S}(0)$ pure.}
	\label{eq:app-same-spectra}
\end{align}
This shows that $I_\c(t)$ quantifies the competition
between the time-evolution's action (entangling $\S$ with the effective environment $\E'$) and its backaction (of $\S$ on $\E'$).
Action and backaction only differ when there is initial entanglement with the \emph{preparing} system $\P_\S$, i.e., when $\rho^\S(0)$ is mixed.

Finally, we discuss the \emph{mismatch} \eq{eq:mismatch} of the coherent information with the initial system entropy:
\begin{subequations}%
\begin{align}
& S(\rho^{\S}(0)) - I_\c(t)
\notag
\\
& = S(\rho^{\S'}(t)) - [ S(\rho^{\S}(t)) - S(\rho^{\P_\S}(t) ]
\\
& =S(\rho^{\E'}(t)) - [ S(\rho^{\S}(t)) - S(\rho^{\S}(0)) ]
.
\label{eq:app-mismatch-b}%
\end{align}%
\label{eq:app-mismatch}%
\end{subequations}%
The form \eq{eq:app-mismatch-b} follows from $\rho^{\P_\S}(t)=\rho^{\P_\S}(0)$ and shows that
first, the mismatch is positive by the Araki-Lieb~\cite{LiebAraki} lower bound on entropies,
$|S(\rho^\text{A}) - S(\rho^\text{B})| \leq S(\rho^\text{AB}) $,
and that second, it is less than twice the initial entropy by the subadditivity upper bound,
$S(\rho^\text{AB}) \leq S(\rho^\text{A}) + S(\rho^\text{B})$:
\begin{align}
	0 \quad \leq \quad S(\rho^{\S}(0)) - I_\c(t) \quad \leq  \quad 2 S(\rho^{\S}(0))
	.
	\label{eq:app-mismatch-bounds}
\end{align}
The second form \eq{eq:app-mismatch-b} is given in the main text [\Eq{eq:mismatch}]
and expresses the mismatch as the competition between entropy \emph{changes} on system and environment.
It again follows from the equivalence of spectra, see \Fig{fig:app-entropies}.

\subsection{Factorization stationary effective environment\label{app:entropies-factor-stationary}}

Returning to the notation $\rho^\S = \rho$ used in the main text, we conclude with a derivation of \Eq{eq:entropy-env-stat}
which applies to dynamics with a unique stationary state $\rho(\infty)$ reached independently of the initial state $\rho(0)$.
This implies that $\Pi(\infty)$ has the form of an entanglement-breaking map~\cite{Wilde13,Preskill},
\begin{subequations}
\begin{align}
	\Pi(\infty)
	& = \Ket{\rho(\infty)} \Bra{\one}
	= \sum_{ij} K_{ij} \bullet K_{ij}
	\\
	K_{ij} &= \sqrt{ \bra{j}\rho(\infty)\ket{j} } \,  \ket{j}\bra{i}
	,
\end{align}
\end{subequations}
which can be conveniently written in terms of a Kraus operator-sum determined by the spectrum of the stationary state $\rho(\infty)$ alone.
Denoting with $\ket{j}$ and $\ket{i}$ the eigenbases of $\rho(\infty)$ and $\rho(0)$, respectively,
we find from \Eq{eq:rho-E'} that the stationary \emph{effective environment} state is the tensor product of the \emph{initial and stationary system} states:
\begin{subequations}
	\begin{align}
\bra{ij}\rho^{\E'}(\infty)\ket{i'j'}
& =  \bra{i}\rho(0)\ket{i} \bra{j}\rho(\infty)\ket{j}
\delta_{ii'} \delta_{jj'}
\\
& = \bra{ij} \rho(0)\otimes \rho(\infty) \ket{i'j'}
.
\end{align}%
\label{eq:app-rho-env-stationary}%
\end{subequations}%
The subadditivity upper bound in \Eq{eq:app-mismatch-bounds} corresponds to $S(\rho^{\E'}(t)) \leq S(\rho(t))+S(\rho(0))$,
and we consequently see from \Eq{eq:app-rho-env-stationary} that for dynamics of this type
this bound is reached in the stationary limit as $S(\rho^{\E'}(\infty)) = S(\rho(0)) + S(\rho(\infty))$.
This indicates that the `preparation entanglement' in \Fig{fig:div} is completely broken for $t \to \infty$.

\section{Heisenberg EOM approach [\Sec{sec:eom}]\label{app:eom}}

\subsection{Evolution superoperator\label{app:eom-general}}

We first discuss expression \eq{eq:pi-op-expansion} for the superoperator form of the dynamical map $\Pi(t)$.
By inserting a complete orthonormal set of operators $\Ket{A}=A$ 
on the left, we get
\begin{align}
\Pi(t)  = \sum_{A} \Ket{A} \Bra{A}\Pi(t)  \bullet
= \sum_{A} \Ket{A} \, \Bra{ \brkt{A(t)^\dag}^{\E} } \, \bullet
.
\label{eq:app-orthog}
\end{align}
The action of $\Bra{A}\bullet =\Tr_{\S} A^\dag \bullet$ on the \emph{system}-operator argument $\bullet$
was expressed in terms of the environment average $\brkt{ \circ }^\E = \Tr_\E \circ (\one^\S \otimes \rho^\E )$
of the \emph{system-environment} Heisenberg-picture operator $A(t) := U^\dag(t) A U(t)$:
\begin{align}
\Bra{A}\Pi(t) \bullet
& =
\tr{\S \E} A^\dag U(t) ( \bullet \otimes \rho^\E) U(t)^\dag
\label{eq:app-pi-exp}\\
& =
\tr{\S \E} U(t)^\dag A U(t) ( \one^\S \otimes \rho^\E ) \bullet
\notag
\\
&=
\tr{\S} \Big[ \tr{\E} A(t)^\dag ( \one^\S \otimes \rho^\E ) \Big] \bullet
:=
\tr{\S} \brkt{A(t)^\dag}^{\E} \bullet
\notag
.
\end{align}
From the start we see that the EOM approach is a method to compute the super-adjoint evolution $\Pi(t)^\dag$ of system-\emph{observables} $A$.
It is thus the Heisenberg-dual to the real-time approach of \Sec{sec:superfermion} which computes the system \emph{state} evolution $\Pi(t)$.

The last form in \Eq{eq:app-pi-exp} provides the connection to the EOM approach
which is exploited in the main text and in the next section.
This approach is similar to familiar Green's function methods.
The fundamental properties of the dynamical map $\Pi$ [\Sec{sec:open-system}] are preserved in terms of EOM quantities as follows:
By unitarity of $U(t)$, trace preservation (TP) $\Bra{\one}\Pi(t) =\Bra{\one}$ corresponds to the trivial evolution of the observable 
$\brkt{\one(t)}^{\E} = \one $ for all $t\geq 0$.
In contrast, complete positivity (CP) or, equivalently, positivity of the Choi-operator \eq{eq:choi-kraus},
\begin{subequations}
	\begin{align}
	\text{choi}(\Pi)
	& = (\Pi \otimes \ones) \ket{\one}\bra{\one}
	\\
	& = \sum_{AB} \brkt{A(t)^\dag B} \cdot A \otimes B^{*}
	\geq 0
	\label{eq:app-constraint}
	,
	\end{align}
\end{subequations}
involves the set of \emph{all} `Green's functions' $\brkt{A(t)^\dag B}$ and is therefore not as easy to check as trace preservation.
In deriving \Eq{eq:app-constraint}, the identity $X \otimes \one  \ket{\one} = \one \otimes X^T \ket{\one}$ has been used.

\subsection{Exact EOMs for parity and field\label{app:eom-parity}}

We next discuss the steps leading to the exact EOM result \eq{eq:eom-averaged} for the dynamical map $\Pi(t)$,
highlighting the correspondence to the real-time superfermion approach of \Sec{sec:superfermion}.
In particular, the specific superoperator structure of the latter approach clearly separates the steps in the derivation
that need to be applied repeatedly in the EOM calculation.

\emph{Field evolution.}
The equation of motion for Heisenberg operators (distinguished only by the time-argument) for the reservoir fields,
\begin{align}
\frac{d}{dt} b_{\eta\omega}(t)
&= i \eta\omega b_{\eta\omega}(t) + i \eta \sqrt{\frac{\Gamma}{2\pi}} d_\eta(t)
,
\label{eq:eom-res}
\end{align}
couples to the system fields via the second inhomogeneous term.
The exact solvability follows from the linearity of this coupling,
in contrast to the real-time approach where this is seen only \emph{at the end} in form of the superfermion property \eq{eq:pauli}.
Integration of \Eq{eq:eom-res} expresses the field dynamics as free, time-local evolution in the system
superposed with time-nonlocal dynamics resulting from tunneling of a fermion at $\tau$ and subsequent free evolution in the reservoir:
\begin{align}
b_{\eta\omega}(t) &= e^{i \eta \omega t} b_{\eta\omega} + i\eta \sqrt{\frac{\Gamma}{2\pi}} \int_0^t d\tau  e^{i \eta \omega (t-\tau)} d_\eta(\tau).
\label{eq:sol-res}
\end{align}
However, inserting this into the EOM for the \emph{system} fields gives a \emph{time-local} equation,
\begin{align}
\frac{d}{dt} d_\eta(t)  
&= i\eta \varepsilon d_\eta(t) + i\eta \sqrt{\frac{\Gamma}{2\pi}} \int d\omega b_{\eta\omega}(t)
\label{eq:eom-sys} \\
&= i\eta\left(\varepsilon+i\eta\frac{\Gamma}{2}\right)d_\eta(t) + i\eta \sqrt{\frac{\Gamma}{2\pi}} \int d\omega e^{i\eta\omega t} b_{\eta\omega},
\notag
\end{align}
since the wide-band limit energy-integration in the last term yielded a factor $\tfrac{1}{2} \times 2\pi \delta(t-\tau)$
noting that the time integrations run over the real \emph{half}-axis, cf. the discussion after \Eq{eq:factor2}.
As before, we can integrate this differential equation to
\begin{align}
d_\eta(t) &= e^{(i\eta\varepsilon -\frac{\Gamma}{2})t} d_\eta
			\label{eq:sol-sys} \\
		  &\qquad + i\eta \sqrt{\frac{\Gamma}{2\pi}} \int d\omega \int_0^t d\tau
		  e^{(i\eta\varepsilon - \frac{\Gamma}{2})(t-\tau)} e^{i\eta\omega t} b_{\eta\omega}
		  \notag
		  .
\end{align}
Relative to the free case, the system fields have acquired a dissipative renormalization $\tfrac{1}{2}\Gamma$ in the exponent
which corresponds to the solution of the $T\to\infty$ limit in the first stage of the real-time approach.
Partial averaging over the initial reservoir state, we get rid of the reservoir part
and obtain the \emph{time-local} EOM \eq{eq:eom-averaged-field} for the field:
\begin{align}
\frac{d}{dt} \brkt{d_\eta(t)}^\E
&= \left(i\eta \varepsilon - \tfrac{1}{2} \Gamma \right)
\brkt{d_\eta(t)}^\E
.
\label{eq:app-eom-field}
\end{align}
Note that no temperature-dependence has arisen yet even though we averaged over the equilibrium reservoir.
Furthermore, the initial superselection, $\brkt{d_\eta(0)}=0$, is preserved by the dynamics.
This does however \emph{not} mean that the field amplitudes can be eliminated from the beginning: 
They are crucial for obtaining the dissipative renormalization factors $e^{-\frac{\Gamma}{2} t}$. 

\emph{Occupations, parity and identity.}
To complete the operator-basis in the decomposition \eq{eq:app-orthog}, we need two additional \emph{system} operators.
Their EOMs \eq{eq:eom-averaged-parity} however do not require new equations
because they follow by applying the product rule to the EOMs \eq{eq:eom-sys} and partially averaging only \emph{afterwards}.
A common choice --when using Green's functions-- is to take the orthogonal system particle- and hole-occupation operators:
\begin{align}
& \frac{d}{dt} \left[d_\eta(t) d_{\bar{\eta}}(t)\right] \notag
=
\frac{d}{dt} \left[ d_\eta(t) \right] d_{\bar{\eta}}(t) + d_\eta(t) \frac{d}{dt}\left[ d_{\bar{\eta}}(t)\right]
\\
&\quad= - \Gamma d_\eta(t) d_{\bar{\eta}}(t)
\label{eq:app-eom-occ}
\\
&\qquad \, + \sqrt{\frac{\Gamma}{2\pi}} \int d\omega \left( i\eta e^{i\eta(\omega-\varepsilon+i\eta\frac{\Gamma}{2})t} b_{\eta\omega} d_{\bar{\eta}} + \text{h.c.} \right)
\notag
\\
&\qquad \, + \frac{\Gamma}{2\pi} \int d\omega \int d\omega' \int_0^t dt' \notag \\
&\qquad \times \left( e^{i\eta(\omega-\varepsilon+i\frac{\Gamma}{2})(t-t')} e^{i\eta(\omega t-\omega't')} b_{\eta\omega} b_{\bar{\eta}\omega'} + \text{h.c.}\right)
\one.
\notag
\end{align}
Partial-averaging causes all terms with a single reservoir field to vanish identically
and due to \Eq{eq:fermi-distrib} only a single energy-integral over $\omega=\omega'$ remains:
\begin{align}
&
\frac{d}{dt}  \brkt{ d_\eta(t) d_{\bar{\eta}}(t) }^\E
= - \Gamma \brkt{d_\eta(t) d_{\bar{\eta}}(t)}^\E
+
\label{eq:eom-av-occ}
\\
& 
\frac{\Gamma}{\pi} \int d\omega \int_0^t dt'
e^{-\frac{\Gamma}{2}(t-t')}
f(\eta\omega) \cos\left[(\omega + \mu - \varepsilon)(t-t')\right]
\one
\notag \\
&= - \Gamma \brkt{ d_\eta(t) d_{\bar{\eta}}(t) }^\E
-\tfrac{\Gamma}{2} \tiny
\Big[
1- \eta \int_0^t d\tau e^{-\frac{\Gamma}{2}(t-t')}
\gamma(t-t')
\Big] \one
.
\notag
\end{align}
Only at this step finite temperature effects have been accounted for through the second term containing the reservoir distribution function
$f(\eta\omega) = (e^{\eta\omega/T}+1)^{-1} = \frac{1}{2} - \frac{\eta}{2}\tanh \left( \frac{\omega}{2T}\right)$
which we decomposed as in \Eq{eq:fermi-distrib}.
This is similar to the second stage of summing the renormalized perturbation series in the real-time approach in \Sec{sec:superfermion}.
Finally, by taking the difference of \Eq{eq:eom-av-occ} for $\eta =\mp$, one obtains the EOM \eq{eq:eom-averaged-parity}
for the parity $(-\one)^n=- \sum_{\eta} \eta d_\eta d_{\bar{\eta}}$ given in the main text.
Summing  \Eq{eq:eom-av-occ} over $\eta =\pm$ verifies that the dynamics is trace-preserving $\brkt{\one(t)}^{\E} = \one$.

\subsection{Positivity preservation in EOM approach\label{app:pp}}

Now we verify which property of the EOM solution is \emph{equivalent} to positivity preservation (PP) of the dynamics,
requiring that $\rho(t)=\Pi(t)\rho(0)\geq 0$ for every initial $\rho(0) \geq 0$, irrespective of whether the fermion-parity superselection is obeyed.
We thus consider \emph{any} initial state and explicitly take into account the field $\brkt{d_\eta(t)}$:
\begin{align}
\rho(t) = \tfrac{1}{2} \Big[ \one + \brkt{(-\one)^n(t)} \, (-\one)^n \Big]
+ \sum_{\eta} \brkt{d_\eta(t)} \, d_\eta
.
\label{eq:app-rho-t}
\end{align}
Expressed in terms of the Bloch-vector magnitude $|b(t)|^2:=|\brkt{(-\one)^n(t)}|^2 + |2\brkt{d(t)}|^2$ of this state,
the PP property for this model requires that $|b(t)|^2 \leq 1$ for all $|b(0)|^2 \leq 1$.
Eliminating $|2\brkt{d(0)}|$ we can consider the parity and the Bloch-vector magnitude
as independent variables up to the constraint $|\brkt{(-\one)^n(0)}| \leq |b(0)| \leq 1$.
Inserting the EOM solutions
\begin{subequations}
	\begin{align}
	\brkt{ d_\eta(t)}
	& = e^{(i\eta \varepsilon - \frac{\Gamma}{2})t} \, \brkt{ d_\eta(0)}
	\label{eq:app-field-sol}
	\\
	\brkt{(-\one)^n(t)} &= 
	e^{-\Gamma t} \brkt{(-\one)^n(0)} + (1-e^{-\Gamma t}) g(t)
	,
	\label{eq:app-parity-sol}
	\end{align}
\end{subequations}
one can rewrite the explicit PP constraint as
\begin{subequations}
	\begin{align}
	0 \leq
	& 1-|b(t)|^2
	\\
	= &
	1 +
	(1-e^{-\Gamma t})e^{-\Gamma t} \Big[ \brkt{(-\one)^n(0)} - g(t)\Big]^2
	\notag\\
	&
	\,\,\, -
	(1-e^{-\Gamma t}) [g(t)]^2 - e^{-\Gamma t} |b(0)|^2
	.
	\label{eq:app-quadratic}
\end{align}
\end{subequations}
We minimize over all states $\rho(0)$ by first setting $|b(0)| = 1$ in the last term and then varying $|\brkt{(-\one)^n(0)}| \leq 1$.
In the case $|g(t)| \leq 1$, the minimum of the quadratic term in \eq{eq:app-quadratic}
is achievable for a physically allowed value $\brkt{(-\one)^n(0)}=g(t)$ and equals $(1-e^{-\Gamma t}) [1-g(t)^2]$. which is positive if and only if $|g(t)| \leq 1$.
Note that the minimum occurs for a mixed state with a nonzero $|\brkt{d(0)}|=\sqrt{1-|g(t)|^2}/2$.
In the other case $|g(t)| > 1$, the minimal achievable value of \eq{eq:app-quadratic}
occurs for $\brkt{(-\one)^n(0)}= \sign \, g(t)$ (and $|\brkt{d(0)}|=0$) and equals
$(1-e^{-\Gamma t}) (1-|g(t)|) [ 1 +e^{-\Gamma t} +(1 -e^{-\Gamma t})|g(t)| ] < 0$ which is always negative.
Thus $\Pi(t)$ is PP if and only if $|g(t)| \leq 1$ as claimed in the main text after \Eq{eq:g-infty-parity}.

This condition is also equivalent to the positivity of the probabilities for states obeying superselection, $|\brkt{d(0)}|=0$.
In this case, the restriction on eigenvalues \eq{eq:spectra-relations}, $\Lambda_\eta = [ 1 + \eta \brkt{(-\one)^n(t)} ]/2 \leq 1$,
requires $\text{max} \, |\brkt{(-\one)^n(t)}| =e^{-\Gamma t}+(1-e^{-\Gamma t}) |g(t)| \leq 1$ at time $t$ which is equivalent to $|g(t)| \leq 1$.
The maximal value of \Eq{eq:app-parity-sol} is achieved for an initial pure state with parity $\brkt{(-1)^n(0)} = \sign \, g(t)$.

Finally, we note that our results for the spectrum of $\Pi(t)$ are consistent with general fixed-point theorems, see Chap. 6. of \Ref{Wolf12}:
These state first that any continuous PP-TP map from the set of finite-dimensional density operators into that set has at least one fixed point density operator.
Second, this set of fixed points is linearly spanned by a finite set of fixed-point density operators.
In our case, the left and right eigen-supervectors for eigenvalue $\Pi_1=1$, are positive operators if $|g(t)| \leq 1$
as mentioned in the caption of Table \ref{tab:spectrum} of the main text.
Consequently, $|g(t)|>1$ excludes the map $\Pi(t)$ to be PP-TP.
For our model, this argument does however not allow one to conclude that $|g(t)|\leq1$ is equivalent to CP and thereby PP as discussed below.

\subsection{Complete positivity in EOM approach}
As mentioned in \Sec{sec:pi-discuss}, the PP condition for our model happens to be equivalent with the CP condition \eq{eq:g-cp}.
Although CP implies PP, the converse does \emph{not} hold in general
since the PP restriction ignores the entanglement with the auxiliary system $\P$ required to prepare initial mixed system states, see \Fig{fig:cp}.
Also, CP is the more relevant condition because of its useful consequences [\Sec{sec:open-system}] which are \emph{not} implied by PP.

To verify the stronger CP property, the EOM result \eq{eq:pi-eom} is not suitable
and requires either conversion to a Kraus operator sum [\Sec{sec:kraus}]
or --for our specific model-- conversion to the exponential form as obtained in the real-time approach [\Sec{sec:superfermion}].
We stress that CP can \emph{not} be inferred from by the spectral decomposition \eq{eq:pi-diag} derived easily from the EOM result for $\Pi(t)$ [\Sec{sec:modes}]
as the positivity of fixed points discussed in the previous section follows from $\Pi(t)$ being PP-TP.
Thus, Table \ref{tab:spectrum} is not in contradiction with our general remark in \Sec{sec:open-system}
that the eigenvectors of $\Pi(t)$ and $\text{choi}(\Pi(t))$ provide complementary information
in the sense that if TP is easy to see, CP is hard infer and \emph{vice versa}.
Complete positivity can also \emph{not} be inferred from the time-local QME for \emph{any} set of model parameters:
Even in the present model its form remains inconclusive in non-CP-divisible parameter regimes where $|h(t)| > 1$,
but nevertheless $|g(t)| \leq 1$, see Eqs.~\eq{eq:gksl} and \eq{eq:tcl}.

\subsection{Time-local quantum master equation\label{app:eom-tcl}}

We noted that for our model the EOM approach naturally leads to the \emph{time-local} equations.
In the above derivation \Eq{eq:app-eom-field} and \eq{eq:eom-av-occ} are already time-local,
in particular because in \Eq{eq:app-eom-occ} and \Eq{eq:eom-av-occ} the operator $\one$ is $t'$-independent.
To derive a corresponding time-local QME \eq{eq:tcl} for $\rho(t)$
we need compute the inverse appearing in $-iL+\Sigma^\TCL(t) = [\tfrac{d}{dt} \Pi(t)] \Pi(t)^{-1}$.
Both the derivative and the inverse are easily constructed from the EOM solution \eqref{eq:pi-eom} and its spectral decomposition \eq{eq:pi-diag}:
	\begin{align}
	-iL+\Sigma^\TCL(t)
	&= \sum_\eta (i \eta \varepsilon -\tfrac{\Gamma}{2}) \Ket{d_\eta^\dag}\Bra{d_\eta^\dag}
	\label{eq:app-tcl-line}\\
	& \notag \quad\,\,\, -\tfrac{\Gamma}{2}
	\Ket{(-\one)^n} \Big[\Bra{(-\one)^n} - h(t) \Bra{\one} \Big]
	.
\end{align}
This is equivalent to the spectral decomposition \eq{eq:tcl-diag} given in the main text,
while the representation \eq{eq:tcl-kernel} of the generator,
$\Sigma^\text{TCL}(t) = \tfrac{\Gamma}{2} \sum_\eta \left[ 1 - \eta h(t) \right] \mathcal{L}_\eta$,
is obtained from the identity
\begin{align}
	\mathcal{L}_\eta
	&=
	d_\eta \bullet d_{\bar{\eta}} -\tfrac{1}{2} [d_{\bar{\eta}}d_\eta ,\bullet]_+
	\label{eq:app-Leta}
	\\
	&= - \tfrac{1}{2}\Big\{
	\sum_{\eta'}\Ket{d_{\eta'}^\dag}\Bra{d_{\eta'}^\dag}
	+ \Ket{(-\one)^n} \Big[ \Bra{(-\one)^n} + \eta \Bra{\one} \Big]
	\Big\}
	\notag
	.
\end{align}

\subsection{Relation time-local and nonlocal generators\label{app:eom-generators}}

Finally, we show how the general expression \eq{eq:miracle-b} for the time-local generator $\Sigma^\TCL(t)$
in terms of the time-nonlocal memory kernel $\Sigma(t-t')$ is consistent with the simple relation \eq{eq:miracle}:
\begin{align}
	\Sigma^{\text{TCL}}(t)
	=
	\int_0^t dt' \Sigma(t-t')\Pi(t')\Pi^{-1}(t)
	\overset{!}{=}
	\int_{0}^{t} d s \Sigma(s)
	.
	\notag
\end{align}
This relation surprisingly holds for our model despite $\Pi(t')\Pi^{-1}(t) \neq \mathcal{I}$.
Using the spectral decomposition \eq{eq:pi-diag} of the EOM result \eq{eq:pi-eom} one instead finds
\begin{align}
	\Pi(t')\Pi^{-1}(t)
	= &
	\sum_{k=1}^4
	\frac{\Pi_{k'}(t')}{\Pi_k(t)} \,
	\Ket{m_{k'}(t')}\Braket{a_{k'}(t')}{m_k(t)} \Bra{a_k(t)}
	\notag \\
	= &
	\sum_{k=1}^4
	\Ket{m_k(t')} \Bra{a_k(t)}
	\label{eq:app-one}
	\\
	& \quad
	+ e^{-\Gamma t}\tfrac{1}{2}[g(t)-g(t')] \Ket{m_4(t')} \Bra{a_1(t)}
	\notag
	.
\end{align}
This superoperator would map the right eigenvectors at time $t$ to those at time $t'$
if the second term of \Eq{eq:app-one} were not present.
However, when the memory kernel \eq{eq:kineq-kernel}, rewritten in super braket form using \Eq{eq:app-Leta}
\begin{align}
	&\Sigma(t-t')
	\\
	&\quad
	=
	-\tfrac{\Gamma}{2}
	\Big\{
	\Big[
	\sum_{\eta }
	\Ket{d_\eta^\dag}\Bra{d_\eta^\dag}+
	\Ket{(-\one)^n} \Bra{(-\one)^n}
	\Big]
	\deltah (t-t')
	\notag
	\\
	&
	\quad\quad\quad\quad
	-e^{-\tfrac{1}{2}\Gamma (t-t')}
	\gamma(t-t') \, \Ket{(-\one)^n} \Bra{\one}
	\Big\},
	\notag
\end{align}
acts on this term from the left, it has no effect:
The first term in $\Sigma(t-t')$ nullifies it due to $\bar{\delta}(t-t')$
and so does the second term because of $\Braket{a_1}{m_k}=\delta_{1,k}$, see Table \ref{tab:spectrum}.
For the same reasons, we can set $t=t'$ also in the first term of \Eq{eq:app-one}, reducing it to the identity since $\sum_{k=1}^4 \Ket{m_k(t)} \Bra{a_k(t)} = \ones$.
We thus find $\Sigma(t-t')\Pi(t')\Pi^{-1}(t) = \Sigma(t-t')$ for all $t$, $t'$ as claimed in the main text, see also \App{app:superfermion-tcl}.

\section{Time-dependent functions [\Sec{sec:functions-physical}] \label{app:functions}}

The complementary approaches in \Sec{sec:eom}-\ref{sec:qme} all rely on three functions whose key properties we derive here.

\subsection{$\gamma(t)$ -- Time-nonlocal memory kernel\label{app:functions-gamma}}

The time-dependent Keldysh correlation function $\gamma(s)$ [\Eq{eq:fermi-contraction}] with the relative time-argument $s=t-t' > 0$
is the basis of all these functions and appears explicitly in the time-nonlocal memory kernel \eq{eq:kineq-kernel}.
It involves an energy integration which can be explicitly carried out by inserting
$\tanh(z) = \sum_{n=-\infty}^{\infty}[z+i\pi(n+1/2)]^{-1}$
and closing an integration contour in the upper (lower) complex plane to pick up half of its poles in the residue theorem for the $\eta=+$ $(\eta=-)$ summand:
\begin{align}
\gamma(s)
&=  \sum_{n=-\infty}^\infty \sum_{\eta=\pm} \frac{1}{2\pi} \int d\omega \frac{e^{i\eta(\omega-\epsilon)s}}{\omega/(2T)+i\pi(n+1/2)}
\label{eq:gamma-solution} \\
&= i \, 2T \sum_{n=0}^\infty e^{-\pi T (2n+1)s} \sum_\eta \eta e^{-i\eta \epsilon s}
= 2T \frac{\sin(\epsilon s)}{\sinh ( \pi T s)}
\notag
.
\end{align}
From this representation we obtain the zeros
\begin{align}
	\gamma(t^\e)=0
	\quad
	\text{for } t^\e = \ell \frac{\pi}{\epsilon}
	\quad \ell=1,2,\ldots
	\label{app:te}
\end{align}
that lead to observable effects [\Eq{eq:tstar} ff.].
Its sign in the first interval, $\sign \, \gamma (s) = \sign \, \epsilon$ for all $s \in (0,\pi/\epsilon]$, determines the sign of $h(t)$ and $g(t)$ for \emph{all} $t> 0$.
We will restrict attention to the case $\epsilon \geq 0$ unless stated otherwise.

\subsection{$h(t)$ -- Time-local generator\label{app:functions-h}}

The function $h(t):=\int_0^t ds e^{-\frac{\Gamma}{2}s} \gamma(s)$ [\Eq{eq:h-def}]
appears in the time-local EOM \eq{eq:eom-averaged-parity} and the QME \eq{eq:tcl} of the main text.	
By integrating the decaying oscillations of $\gamma$ with a nonnegative, decaying envelope,
$h(t)$ has a stationary limit $h(\infty)=g(\infty)$ [\Eq{eq:g-infty2} and \eq{eq:g-infty1}].
For the same reason its sign is determined for all $t \geq 0$ by the sign of $\gamma(s)$ for $s \in (0,\pi/\epsilon]$: $\sign \, h(t) = \sign \, \epsilon$ for $t\geq 0$.	

The function $h(t)$ is always nonmonotic with extrema at every zero of the correlation function $\gamma(t)$ since $dh(t)/dt = e^{-\frac{\Gamma}{2}t} \gamma(t)$.
These are either local maxima at $t=(2\ell-1)\pi/\epsilon $ or local minima at $t=2\ell\pi/\epsilon$ for $\ell=1,2,\ldots$.
The global maximum always exceeds the stationary value $h(\infty)$ since in the sum
$h(\pi/\epsilon)-h(\infty)
=
\sum_{\ell=1}^\infty \int_{(2\ell-1)\pi/\epsilon}^{(2\ell+1)\pi/\epsilon} ds \, e^{-\Gamma s/2} \gamma(s) > 0
$ each integral is positive.
Moreover, the value
$h(\pi/\epsilon)=
\int_0^{\pi/\epsilon} ds \, e^{-\Gamma s/2} \gamma(s)$
can significantly exceed 1 for $\epsilon \neq 0$ and reaches its maximal value in the limit
\begin{align}
	\lim_{T, \Gamma \to 0}
	h \big(\tfrac{\pi}{\epsilon}\big)
	=
	\tfrac{2}{\pi} \, \text{Si}(\pi)
	\approx 1.179 > 1 = \lim_{T, \Gamma \to 0} h(\infty)
	,
	\label{eq:overshoot}
\end{align}
which can be seen from the final representation in \Eq{eq:gamma-solution} and the fact that the sign of $\gamma(s)$ does not change in the first interval.
Depending on parameters, several subsequent local maxima may exceed 1 as seen in \Fig{fig:functions}.
This leads to loss of CP-divisibility and the physically observable effects discussed in \Sec{sec:modes}.

\subsection{$g(t)$ -- Dynamical map\label{app:functions-g}}

The function $g(t)$ directly determines the parity evolution \eq{eq:eom-parity},
the exponent of the dynamical map \eq{eq:pi-superfermion-b} and the Kraus operators \eq{eq:krausfinal}.
By convoluting $h(t)$ with a nonnegative decaying envelope, the oscillations from the correlation function $\gamma(s)$ are further smoothed out
and the sign is as before $\sign \, g(t) = \sign \,  h(t) =\sign \, \epsilon$ for $t\geq 0$.
The key steps to understanding the properties of $g(t)$ are first the conversion to a \emph{single} time-integral over $\gamma(s)$,
\begin{align}
g(t)
&= \int_0^t ds' \frac{\sinh\left[\frac{\Gamma}{2}(t-s') \right]}{\sinh \left[\frac{\Gamma}{2}t \right]} \,\gamma(s')
,
\label{eq:g-bound}
\end{align}
by factorizing the double-integral obtained when inserting $h(\tau)$ into the definition \eq{eq:g-h}:
This is achieved by changing to relative $s'=\tau-s$ and cumulative time-variables $\theta =\tau + s$ the latter of which can be directly evaluated.
Then performing the relative-time integration ($s'$) after interchanging it
with the energy-integration ($\omega$) \emph{inside} $\gamma$ [\Eq{eq:fermi-contraction-a}]
we obtain a sum of contributions from all reservoir modes:
\begin{widetext}
\begin{subequations}
\begin{align}
g(t)
&= \frac{1}{\pi} \int d\omega \tanh\left(\frac{\omega+\epsilon}{2T} \right) \frac{\Gamma/2}{(\Gamma/2)^2 + \omega^2} \frac{ \cosh ({\Gamma}t/{2})-\cos (\omega t)}{\sinh ({\Gamma t}/{2})}
\label{eq:g-super-a}\\
&= g(\infty) \left(1-\frac{1}{e^{\Gamma t/2}+1} \right) +
 \frac{\Gamma t /2}{\sinh(\Gamma t /2)}
 \, 
 \frac{1}{\pi}\int d\omega \tanh\left(\frac{\omega+\epsilon}{2T} \right) \frac{\omega^2}{(\Gamma/2)^2 + \omega^2} \frac{t}{2}
\left(\frac{\sin(\omega t/2)}{\omega t/2}\right)^2
.
\label{eq:g-super-b}
\end{align}%
\end{subequations}%
\end{widetext}%
Inspection of the second line shows that the stationary value extracted from the expression \eq{eq:g-super-a},
\begin{align}
g(\infty)
 = \frac{1}{\pi} \int d\omega \tanh\left(\frac{\omega+\epsilon}{2T} \right) \frac{\Gamma/2}{(\Gamma/2)^2 + \omega^2}
\label{eq:ginfty-app}
,
\end{align}
is in fact the upper bound of $g(t)$: In \Eq{eq:g-super-b} the prefactor of $g(\infty)$ is upper bounded by 1
and the second term is always negative by symmetry of the integrand.
The magnitude of $g(\infty)$ is in turn upper bounded by its value in the limit $\epsilon \to \infty$:
\begin{align}
g(\infty) \leq \lim_{\epsilon\to\infty} g(\infty) = \frac{1}{\pi} \int d\omega \frac{\Gamma/2}{(\Gamma/2)^2 + \omega^2} = 1.
\label{eq:app-saturate}
\end{align} 
Including the signs, we obtain \Eq{eq:g-cp} of the main text:
\begin{align}
|g(t)|
\, \leq  \,
|g(\infty) |
\, \leq \,
 \lim_{|\epsilon|\to\infty} |g(\infty)|
\, \leq \,
1.
\end{align}
Evaluating the integral in \Eq{eq:ginfty-app} analogous to the calculation \eq{eq:Sigma-z-int} gives the representation  
\begin{align}
	g(\infty)
	= \frac{2}{\pi} \, \Im \, \psi \left( \frac{1}{2} + \frac{\Gamma/2+i\epsilon}{2\pi T} \right)
	\label{eq:g-infty2}
\end{align}
given in the main text [\Eq{eq:g-infty}] in terms of the Digamma function $\psi$
which reduces to the odd part $\tanh[\epsilon/(2T)]$ of the Fermi-distribution function in the weak-coupling limit $\Gamma \to 0$.
Finite $\Gamma$ broadens this step around $\epsilon=0$ as reflected by the linearization
\begin{align}
		g(\infty)
		& \approx \frac{\epsilon \,  \psi_1 \big( \frac{1}{2} + \frac{\Gamma / 2}{2\pi T} \big)}{\pi^2 T}
		=
		\begin{cases}
			\tfrac{4}{\pi} \tfrac{\epsilon}{\Gamma} & |\epsilon| \ll T      \ll \Gamma
			\\
			\tfrac{1}{2}\tfrac{\epsilon}{T}         & |\epsilon| \ll \Gamma \ll T
		\end{cases}
		,
\end{align}
using $\psi_1(z):=d\psi(z)/dz$ and $\psi_1(1/2)=\pi^2/2$.
The representation \eq{eq:g-infty2} furthermore shows that the stationary value approaches the upperbound \eq{eq:app-saturate}
as function of $\epsilon$ with a powerlaw due to the asymptotic behavior $\psi(z) \sim \ln z$:
\begin{align}
	|g(\infty)| \approx 1-\frac{\Gamma}{|\epsilon|}
	\qquad
	|\epsilon| \gg \Gamma, T
	.
\end{align}

\subsection{$dg(t)/dt$ -- Conditions for extrema \label{app:functions-extra}}

Using \Eq{eq:g-h} one finds a similar expression for the derivative of $g(t)$ in terms of the correlation function:
\begin{subequations}
	\begin{align}
\tfrac{d}{dt}g(t)
&= \frac{\Gamma}{2} \int_0^t ds \frac{\sinh \frac{\Gamma}{2}s}{\left(\sinh \frac{\Gamma}{2}t\right)^2} \gamma(s)
\\
&= \frac{\Gamma T}{\left(\sinh \frac{\Gamma}{2}t\right)^2} \int_0^t ds \frac{\sinh \frac{\Gamma}{2}s}{\sinh \pi T s} \sin \epsilon s 
\label{eq:dgdt-magic}\\
&= \frac{\Gamma}{1-e^{-\Gamma t}} \left[h(t)-g(t) \right]
.
\label{eq:dgdt-stationary}
\end{align}%
\label{eq:dgdt-general}%
\end{subequations}%
The form \eq{eq:dgdt-magic} gives direct access to the short-time behavior \eq{eq:hg-linear} of both functions determined by the Keldysh correlation function,
$h(t)=2 g(t) \approx \gamma(0^+) t = 2\epsilon/\pi t$.
Instead, \Eq{eq:dgdt-stationary} gives a simple proof that $g(t)$ and $h(t)$ have the same stationary limit as claimed in \Eq{eq:g-infty}:
\begin{align}
	g(\infty) &=
	h(\infty)
	=
	\int_0^\infty ds e^{-\frac{\Gamma}{2}s} \gamma(s)
	.
	\label{eq:g-infty1}
\end{align}
The latter integral agrees with result \eq{eq:ginfty-app} but is simpler to evaluate than the $t\to \infty$ limit of \Eq{eq:g-bound}.
This form also shows that $g(t)$ is non-monotonous if and only if $h(t)$ and $g(t)$ cross, a fact used in \Sec{sec:fixedpoint-tcl}.
In particular, an inspection of the form \eq{eq:dgdt-magic} allows one to investigate in which parameter regimes this can happen:

(i)
If ${\Gamma}/{2}=\pi T$, the remaining $s$-integration is simple and results in
\begin{align}
\tfrac{d}{dt}g(t)=
\frac{\Gamma T}{\epsilon} \, \frac{1-\cos(\epsilon t)}{[\sinh(\Gamma t/2)]^2}
,
\end{align}
showing that crossings of $g(t)$ and $h(t)$ occur at times $t^\c = 2 \ell \, \frac{\pi}{\epsilon}$ for $\ell =0,1,2,\ldots$ which coincide with the minima of $h(t)$.
Consequently, $h(t) \geq g(t)$ and $g(t)$ develops a plateau at times $t^\c$ where it touches $h(t)$ at its minimum
before continuing towards its stationary value which is in this case asymptotically reached as $dg(t)/dt \propto e^{-\Gamma t}$.

(ii)
For ${\Gamma}/{2}<\pi T$, the monotonously decreasing exponential envelope in the integrand \Eq{eq:dgdt-magic} ensures the integral is strictly positive
and thus $h(t)>g(t)$ for all times. Here, $dg(t)/dt$ decays faster than $e^{-\Gamma t}$.

(iii)
For ${\Gamma}/{2}>\pi T$ the envelope is monotonously increasing,
allowing positive contributions to the integral \eq{eq:dgdt-magic} to be overcompensated by later negative ones.
As a consequence, $h(t)<g(t)$ is possible and crossings occur in the vicinity of the minima of $h(t)$.
Here, $dg(t)/dt$ decays slower than $e^{-\Gamma t}$.

\begin{figure}[t]
	\includegraphics[width=1.0\linewidth]{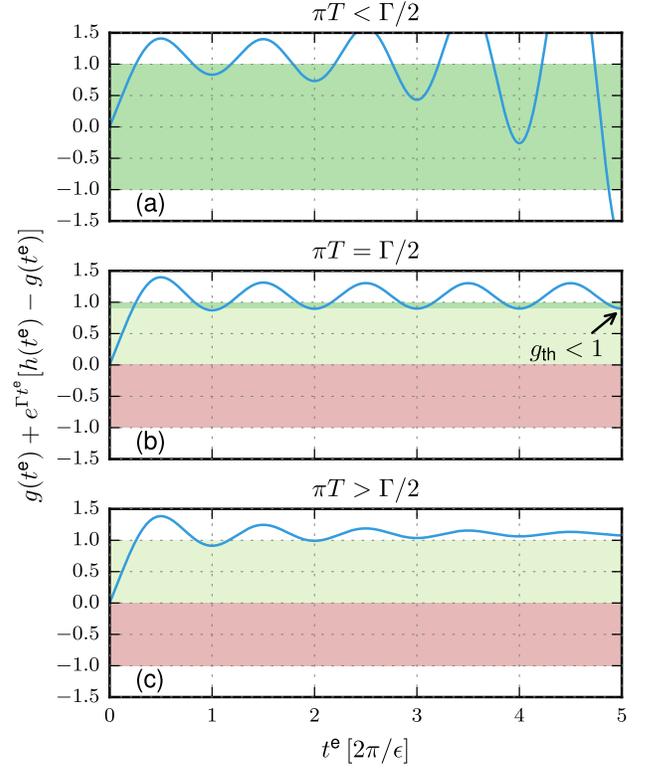}
	\caption{
		Right hand side of condition \eq{eq:app-initialcondition}
		for $\epsilon=\pi \Gamma$ and
		(a) $\pi T=0.01 \cdot \Gamma/2$,
		(b) $\pi T=\Gamma/2$ and
		(c) $\pi T=1.5 \cdot \Gamma/2$.
	}
	\label{fig:halting-condition}
\end{figure}

\subsection{Conditions for parity extrema\label{app:functions-parity}}

The condition $\brkt{(-\one)^n(t^\e)}=h(t^\e)$ for the appearance of a parity extremum at $t^\e$, \Eq{eq:extremum} of the main text,
can also be analyzed using \Eq{eq:dgdt-general}.
For this condition to hold, the initial parity must satisfy
\begin{align}
	\brkt{(-\one)^n(0)} & = g(t^\e) +  e^{\Gamma t^\e} [h(t^\e)-g(t^\e)]
	\label{eq:app-initialcondition}
	\\
	& =
	g(t^\e) + 
	\frac{4T}{1-e^{-\Gamma t^\e}}
	\int_0^{t^\e} ds \frac{\sinh \frac{\Gamma}{2}s}{\sinh \pi T s} \sin \epsilon s 
	\notag
	.
\end{align}
The right hand side is plotted in \Fig{fig:halting-condition} and qualitatively changes with the temperature / coupling ratio:

(a)
For $\pi T < \Gamma/2$, the second term in \Eq{eq:app-initialcondition} is an oscillating function of $t^\e$ whose envelope exponentially diverges after an initial decay.
Consequently, \emph{any} fixed initial condition $\brkt{(-\one)^n(0)} \in [-1,1]$ satisfies \Eq{eq:app-initialcondition} for infinitely many times $t^\e$
indicated by dark green in \Fig{fig:halting-condition}.
The converging parity thus keeps on oscillating.

(b)
For $\pi T = \Gamma/2$, this is possible only when the initial parity exceeds a threshold $g_\text{th}$ just below $1$ as indicated in \Fig{fig:halting-condition}.
Below this threshold there are either a few $t^\e$ solutions for $[0,g_\text{th})$ (at least one due to parity reentrance) as indicated in light green,
or the condition is \emph{never} satisfied for $[-1,0]$ (no parity extrema at all) as indicated in red.

(c)
For $\pi T > \Gamma/2$, the right-hand side of \Eq{eq:app-initialcondition} converges to a value exceeding 1 in an oscillating fashion:
The condition is satisfied only a finite number of times for an initial parity in $[0,1]$ and never for $[-1,0)$.

\subsection{$g(t)$ -- Special functions}

Finally, we express $g(t)$ in terms of special functions.
In \Eq{eq:gamma-solution} we insert the series for $\sinh$ in the denominator and express the numerator in terms of exponentials:
\begin{align}
& g(t) = \sum_{\eta, \chi = \pm} \eta \chi \frac{iT e^{\eta\frac{\Gamma}{2}t}}{\sinh\left[\frac{\Gamma}{2}t \right]} \sum_{n=0}^\infty \int_0^t ds \, e^{-[\pi T(2n+1)+\eta\frac{\Gamma}{2}+i\chi \epsilon]s} \notag
\\
	 &= - \sum_{\eta, \chi = \pm} \eta \chi \frac{iT e^{\eta\frac{\Gamma}{2}t}}{\sinh\left[\frac{\Gamma}{2}t \right]} \sum_{n=0}^\infty \frac{e^{-[\pi T (2n+1) +\eta\frac{\Gamma}{2}+i \chi \epsilon]t}-1}{\pi T(2n+1) + \eta\frac{\Gamma}{2}+i\chi \epsilon} \notag
\\
	&= \sum_{\eta = \pm} \frac{\eta}{\pi \sinh\left[\frac{\Gamma}{2}t \right]}
	 \text{Im}
	 \sum_{n=0}^\infty \frac{e^{-[\pi T +i\epsilon]t} e^{-2n\pi T t} - e^{\eta\frac{\Gamma}{2}t}}{n+\alpha_{\eta\Gamma}(T, \epsilon)}
	 \notag
\\
	&=
	 \sum_{\eta=\pm}
	 \eta \, 
	 \text{Im} \frac{ e^{-[\pi T + i\epsilon]t} \Phi\left(e^{-2\pi T t};1;\alpha_{\eta\Gamma} \right) +  e^{\eta \frac{\Gamma}{2}t}\psi \left( \alpha_{\eta\Gamma} \right)}{\pi \, \sinh\left[\frac{\Gamma}{2}t \right]}
	 .
\label{eq:g-explicit}
\end{align}
In the final result we identified the imaginary parts of the
Lerch function $\Im \, \Phi(z;s;\alpha) := \Im \sum_{n=0}^\infty \frac{z^n}{(n+\alpha)^s}$
and the Digamma function $\Im \, \psi (\alpha) := - \text{Im} \sum_{n=0}^\infty \frac{1}{n+\alpha}$
of the complex variable $\alpha_{\eta\Gamma} = \frac{1}{2} + \frac{\eta\Gamma/2+i\epsilon}{2\pi T}$.
In this expression the poles
of the Lerch and Digamma functions for $\alpha_{-\Gamma}$ cancel exactly~\cite{Saptsov14}.

\section{Real-time approach [\Sec{sec:superfermion}]\label{app:real-time}}

\subsection{Summing the two-stage expansion\label{app:superfermion}}

We first discuss the calculation of the exact dynamical map by direct summation of the perturbation series
\begin{align}
\Pi(t) & = \brkt{\hat{T} e^{-i\int_0^tL^V(s)}}^\E
= \sum_{k=0}^\infty \, \brkt{ \ones \, [ \ast (-iL^V) \ast \ones ]^k }^\E (t)
\label{eq:app-pi-expansion}
\end{align}
in two stages as explained in the main text.
Here, time-convolutions are denoted by $A \ast B(t) = \int_0^t ds A(t-s)B(s)$
and $\brkt{ \circ }^\E = \Tr_\E \circ \,  \rho^\E$ is evaluated by Wick contractions.
The identity superoperator $\ones$ denotes the free evolution in the interaction picture $L^V(\tau) := e^{iL^0 \tau} L^{V} e^{-iL^0\tau}$
which we only indicate by time-arguments throughout this appendix.
For this model, the coupling Liouvillian $L^V = \sum_{q=\pm} L^{Vq}$ acquires a mere phase factor from the interaction picture,
\begin{align}
L^{Vq}(\tau)
&= \sum_{\eta}\sqrt{\frac{\Gamma}{2\pi}} \int d\omega e^{i\eta(\omega-\varepsilon)\tau}J^q_{\eta\omega} G^{\bar{q}}_{\bar{\eta}}
,
\end{align}
and the Wick contractions \eq{eq:lv-contraction} can be written as:
\begin{align}
	&\brkt{ L^{V+}(\tau) L^{Vq}(\tau')}^\E
	\label{eq:app-lv-contraction}
	\\&
	= \tfrac{\Gamma}{2} \sum_\eta \left[
	\deltah(\tau-\tau') \delta_{q,-} + \eta \gamma(\tau-\tau') \delta_{q,+}
	\right]
	G^{+}_\eta G^{q}_{\bar{\eta}}
	\notag
	.
\end{align}

\emph{Stage one.}
For $T\to\infty$ only contractions between pairs of \emph{different} coupling components
$L^{V+} \propto G^+$ and $L^{V-} \propto G^-$ --in this order-- are nonzero in \Eq{eq:app-pi-expansion}.
Since these are time-local, all time-convolutions trivialize to products, leaving an ordinary power series
\begin{subequations}
		\begin{align}
	& \Pi_\infty(t)
	:=
	\lim_{T\to \infty} \Pi(t)
	\\
	&\quad= \mathcal{I} 
	- \underset{t \geq t_2 \geq t_1 \geq 0}{\int dt_2 dt_1}
	\brkt{L^{V+} L^{V-}}^\E \, \deltah(t_2-t_1)
	+ \ldots
	\label{eq:app-infty-series}
	\\
	&\quad  = \ones
	+  \hat{\Sigma}_\infty t + \tfrac{1}{2!} ( \hat{\Sigma}_\infty t)^2 + \ldots
	\,
	=
	\, 
	 e^{ \hat{\Sigma}_\infty t }
	 \label{eq:app-infty}
\end{align}
\end{subequations}
with the \emph{equal-time} correlator as the basic building block:
	\begin{align}
	\hat{\Sigma}_\infty
	:=
	-
	\brkt{ L^{V+} L^{V-}}^\E
	&=
	-
	\tfrac{\Gamma}{2}
	\sum_\eta G^+_\eta G^-_{\bar{\eta}}
	.
	\label{eq:app-Sigma-infinite}
\end{align}
It useful to note that $\sop{N}=\sum_\eta G^+_\eta G^-_{\bar{\eta}}$ counts the number of superfermions on the right
which gives rise to the relations
\begin{subequations}
\begin{align}
	G^+_\eta \Pi_\infty(t) G^+_{\bar{\eta}} & = e^{-\frac{1}{2}\Gamma t} G^+_\eta G^+_{\bar{\eta}} \Pi_\infty(t)
	\label{eq:app-GG-middle}
	\\
	\Pi_\infty(t) G^+_\eta G^+_{\bar{\eta}}  &= e^{-\Gamma t} G^+_\eta G^+_{\bar{\eta}}
	\label{eq:app-GG-left}
	.
\end{align}
The Pauli principle \eq{eq:pauli} furthermore implies
\begin{align}
	G^+_\eta G^+_{\bar{\eta}} \Pi_\infty(t)  &= G^+_\eta G^+_{\bar{\eta}}
	.
	\label{eq:app-GG-right}
\end{align}
\label{eq:app-GG}
\end{subequations}

\emph{Stage two.}
For finite $T$ the expansion \eq{eq:app-pi-expansion} contains additional \emph{time-nonlocal} contractions
between pairs of couplings of the same type $L^{V+} \propto G^+$.
In between these pairs, the time-local infinite-temperature contributions already summed up in $\Pi_\infty(t)$ may still appear
but due to the relations \eq{eq:app-GG} they give a simple decaying term.
More importantly, the Pauli principle \eq{eq:pauli} causes all contributions beyond first order in the time-nonlocal contraction to vanish identically:
\begin{widetext}%
\begin{subequations}%
\begin{align}%
	\hspace{-0.25cm}
	\Pi(t)
	&= \Pi_\infty(t)
	-
	\underset{t \geq t_2 \geq t_1 \geq 0}{\int dt_2 dt_1}
	\Pi_\infty(t-t_2) \, \brkt{L^{V+}(t_2) \Pi_\infty(t_2-t_1) L^{V+}(t_1)}^\E \, \Pi_\infty(t_1)
	+ \ldots
	\label{eq:app-pi-full}
	\\
	& = \Pi_\infty(t)
	-
	\underset{t \geq t_2 \geq t_1 \geq 0}{\int dt_2 dt_1}
	e^{-\Gamma (t-t_2)}  \brkt{L^{V+}(t_2) e^{-\tfrac{1}{2}\Gamma(t_2-t_1)} L^{V+}(t_1)}^\E
	\label{eq:app-trunc}
	\\
	&=
	\Pi_\infty(t)
	-
	\underset{t \geq t_2 \geq t_1 \geq 0}{\int dt_2 dt_1}
	\Gamma e^{-\Gamma (t-t_2)}  e^{-\tfrac{1}{2}\Gamma(t_2-t_1)} \gamma(t_2-t_1)
	\sum_{\eta} \eta G^+_\eta G^+_{\bar{\eta}}
	\,
	=
	\,
	\Pi_\infty(t)
	-
	\tfrac{1}{2}(1-e^{-\Gamma t})g(t) \sum_{\eta} \eta G^+_\eta G^+_{\bar{\eta}}
	.
	\label{eq:app-pi-superfermion-a}
\end{align}%
\end{subequations}%
\end{widetext}%
The renormalized series thus \emph{terminates} as claimed in the main text.
Reverting to the Schr\"odinger picture we obtain \Eq{eq:pi-superfermion-a}.	
To obtain the final form \eq{eq:pi-superfermion-b} we use \Eq{eq:app-GG-right}
to write the finite-temperature term in \Eq{eq:app-pi-superfermion-a} as a \emph{factor}:
\begin{subequations}
\begin{align}
	\Pi(t)
	&= \Big[ \mathcal{I} - \tfrac{1}{2} (1-e^{-\Gamma t})  g(t) \sum_\eta \eta G^+_\eta G^+_{\bar{\eta}}
	\Big] \Pi_\infty(t)
	\notag
	\\
	& = e^{- \frac{1}{2}(1-e^{-\Gamma t})  g(t) \sum_\eta \eta \mathcal{L}_\eta  }
	\, e^{\hat{\Sigma}_\infty t}
	\label{eq:full-pi-derivation-a}
	\\
	&= e^{\frac{\Gamma}{2}t \sum_\eta \left[1 - \eta g(t) \right] \mathcal{L}_\eta}.
	\label{eq:full-pi-derivation-b}
\end{align}
\end{subequations}
Written as an exponential it can be merged with $\Pi_\infty(t)=e^{\hat{\Sigma}_\infty t}$
using the Baker-Campbell-Hausdorff identity $e^X e^Y = e^{Y+\phi/(1-e^{-\phi})X}$ for operators obeying $[X,Y]_- = \phi X$
which results in \Eq{eq:full-pi-derivation-b}.
Transformation to Schr\"odinger picture $\Pi(t) \to e^{-iLt} \Pi(t)$ gives \Eq{eq:pi-superfermion-b} of the main text
as the identity $[L,G^+_\eta G^{\pm}_{\bar{\eta}}]=0$ allows one to simply add $L$ to the exponent.

\subsection{Time-nonlocal quantum master equation\label{app:superfermion-kernel}}

To derive the time-nonlocal QME \eq{eq:kineq2} in the main text within the real-time approach,
we note that the two stages can be written as self-consistent equations.
First, the infinite-temperature limit \eq{eq:app-infty-series} can be represented
in terms of the self-energy $\Sigma_\infty(t) := \hat{\Sigma}_\infty \deltah(t-t')$:
\begin{align}
	\Pi_\infty(t)
	\label{eq:app-dyson-infty}
	&= \mathcal{I} + \mathcal{I} \ast \Sigma_\infty \ast \Pi_\infty (t)
	.
\end{align}
This infinite-temperature limit feeds into the self-consistent form of \Eq{eq:app-pi-full},
\begin{align}
	\Pi(t)
	&= \Pi_\infty(t) + \Pi_\infty \ast \Delta \Sigma \ast \Pi (t)
	,
	\label{eq:app-dyson}
\end{align}
where the basic block is time-nonlocal:
\begin{subequations}
	\begin{align}
		&\Delta \Sigma(t_2-t_1) \notag \\
		 & \quad= - \brkt{L^{V+}(t_2) \Pi_\infty(t_2-t_1) L^{V+}(t_1)}^\E
		\\
		& \quad = - \Gamma \gamma(t_2-t_1) e^{-\frac{\Gamma}{2}(t_2-t_1)} \sum_\eta \eta G^+_\eta G^+_{\bar{\eta}}
		.
		\label{eq:app-Sigma-finite}
	\end{align}%
\end{subequations}%
To get to \Eq{eq:app-dyson} we used that in our model higher orders $[\Delta \Sigma \ast \Pi_\infty\ast]^k =0$ for $k =2,3,\ldots$
are zero by the Pauli principle \eq{eq:pauli} and may thus be added on the right hand side to complete $\Pi_\infty(t)$ to $\Pi(t)$.
Repeating this argument to also simplify the time-derivative, we obtain two coupled QMEs
\begin{subequations}
\begin{align}
	\tfrac{d}{dt}
	\Pi_\infty(t)
	=&  \Sigma_\infty \ast \Pi_\infty (t)
	\label{eq:app-kinetic-infty}
	\\
	\tfrac{d}{dt}
	\Pi(t)
	=&  \big( \Sigma_\infty  +\Delta \Sigma \big)  \ast \Pi (t)
	,
	\label{eq:app-kinetic-1}
\end{align}
which can be combined into the time-nonlocal QME
\begin{align}
	\tfrac{d}{dt}
	\Pi(t)
	=  \Sigma \ast \Pi (t)
	,
	\qquad
	\Sigma(t)=\hat{\Sigma}_\infty \deltah(t) +\Delta \Sigma(t)
	\label{eq:app-kinetic}
\end{align}
\end{subequations}
with the memory kernel $\Sigma(t)$ given by the sum of the time-local infinite-temperature kernel \eq{eq:app-Sigma-infinite}
and the time-nonlocal finite-temperature kernel \eq{eq:app-Sigma-finite}.
In the Schr\"odinger picture this is \Eq{eq:kineq2} of the main text.

Its Laplace transform $\Sigma(z)=\int_0^\infty dt e^{i z t} \Sigma(t)$ discussed in \Eq{eq:sigma-z} of the main text,
\begin{align}
	\label{eq:app-laplace-kernel}
	\Sigma(z)
		&=
	\frac{\Gamma}{2} \sum_\eta \int_0^\infty ds e^{i z s}
	\Big[ \deltah(s) - \eta e^{-\frac{\Gamma}{2}s} \gamma(s)  \Big]
	\mathcal{L}_\eta \\
	&=
	\frac{\Gamma}{2} \sum_\eta
	\Big[1 + i \frac{\eta}{\pi} \sum_{\chi=\pm} \chi \psi\left(\frac{1}{2} + \frac{\frac{\Gamma}{2}-i(z-\chi\epsilon)}{2\pi T}\right) \Big]
	\mathcal{L}_\eta \notag
	,
\end{align}
is obtained by an integration analogous to \Eq{eq:g-explicit}:
\begin{align}
	& \int_0^\infty ds  e^{i z s} e^{-\frac{\Gamma}{2}s} \gamma(s)
	\label{eq:Sigma-z-int}
	\\	&\quad=
	i  2T \sum_{\chi=\pm} \sum_{n=0}^\infty \chi \int_0^\infty ds e^{-[\pi T (2n+1)+\frac{\Gamma}{2} +i(\chi\epsilon-z)]s} 
	\notag
	\\
	&\quad=
	- i \frac{1}{\pi} \sum_{\chi=\pm} \chi \sum_{n=0}^\infty \frac{1}{n + \tfrac{1}{2} + \frac{\frac{\Gamma}{2}+i(\chi\epsilon-z)}{2\pi T}}
	\notag
	.
\end{align}

\subsection{Relation time-local and nonlocal generators\label{app:superfermion-tcl}}

\begin{table*}[t]
	\caption{\label{tab:kraus-rel}
		Sum-rules satisfied by the Kraus operators \eq{eq:krausfinal}.}.
	\begin{ruledtabular}
		\begin{tabular}{rcl rcl}
			$\sum\limits_{\eta=\pm} \Big[
			K_{\eta}^{0}(t)^\dag K_{\eta}^{0}(t) + K_{\eta}^{1}(t)^\dag K_{\eta}^{1}(t)
			\Big] $ 
			& $=$ & $\one$
			&
			$\sum\limits_{\eta=\pm}
			K_{\eta}^{0}(t)^\dag d_{\eta'}^\dag K_{\eta}^{0}(t)$ 		
			& $=$ & $ e^{(i\eta \varepsilon -\Gamma/2)t} d_{\eta'}^\dag$
			\\
			$\sum\limits_{\eta=\pm} \Big[
			K_{\eta}^{0}(t) K_{\eta}^{0}(t)^\dag + K_{\eta}^{1}(t) K_{\eta}^{1}(t)^\dag
			\Big] $ 
			& $=$ & $\one + (1-e^{-\Gamma t})g(t) (-\one)^n$
			&
			$\sum\limits_{\eta=\pm}
			K_{\eta}^{0}(t) d_{\eta'}^\dag K_{\eta}^{0}(t)^\dag$ 		
			& $=$ & $ e^{(-i\eta \varepsilon -\Gamma/2)t} d_{\eta'}^\dag$
			\\			
			$\sum\limits_{\eta=\pm} \Big[
			K_{\eta}^{0}(t) K_{\eta}^{0}(t)^\dag - K_{\eta}^{0}(t)^\dag K_{\eta}^{0}(t)
			\Big] $
			& $=$ & 0
			&
			$\sum\limits_{\eta=\pm}
			K_{\eta}^{1}(t) d_{\eta'}^\dag K_{\eta}^{1}(t)^\dag$ 		
			& $=$ & $0$				
			\\			
			$\sum\limits_{\eta=\pm} \Big[
			K_{\eta}^{1}(t) K_{\eta}^{1}(t)^\dag + K_{\eta}^{1}(t)^\dag K_{\eta}^{1}(t)
			\Big] $
			& $=$ &
			$(1-e^{-\Gamma t}) \one $
			&
			$\sum\limits_{\eta=\pm}
			K_{\eta}^{1}(t)^\dag d_{\eta'}^\dag K_{\eta}^{1}(t)$ 		
			& $=$ & $0$				
		\end{tabular}
	\end{ruledtabular}
\end{table*}

We next convert \Eq{eq:app-kinetic} to the time-\emph{local} QME \eq{eq:tcl} in the main text,
avoiding the inversion of $\Pi(t)$ that is required in the EOM approach [\App{app:eom-tcl}]:
\begin{align}
	\tfrac{d}{dt}
	\Pi(t)
	& =  \int_0^t dt \Sigma (t-t') \Pi (t')
	\label{eq:app-kinetic-tcl}
	\\
	&\overset{!}{=}
	\Big[ \int_0^t dt' \Sigma (t-t') \Big] \Pi (t)
	 = \Sigma^\TCL(t) \Pi (t)
	\notag
	.
\end{align}
As stressed in the main text [\Eq{eq:miracle} ff.] this is an \emph{exact} relation for our model as
$\Sigma(t-t')\Pi(t')=\Sigma(t-t')\Pi(t)$.
We can directly see this by splitting off in propagator $\Pi(t')=\Pi_\infty(t')+\Delta \Pi(t')$
and the memory kernel $\Sigma(t)=\hat{\Sigma}_\infty \deltah(t) +\Delta \Sigma(t)$
the corrections $\Delta$ to the infinite-temperature limit which are both proportional to $\sum_{\eta} \eta G^+_\eta G^+_\eta$.
Then by \Eq{eq:app-GG-left} and the Pauli principle \eq{eq:pauli} we have
\begin{subequations}
\begin{align}
	\Delta \Sigma(t-t') \Pi_\infty(t') & =\Delta \Sigma(t-t')
	\\
	\Delta \Sigma(t-t') \Delta \Pi(t') & =0
	,
\end{align}
\end{subequations}
causing the $t'$-dependence of $\Pi(t')$ to drop out.

\subsection{Divisor dynamics\label{app:superfermion-divisor}}

Finally, we show how one can factorize the evolution by applying the above considerations directly to the divisor $\Pi(t,t')=\Pi(t) \Pi(t')^{-1}$.
To this end, we write
\begin{subequations}
\begin{align}
	\Pi(t) & = \Pi_\infty(t) + \alpha(t) \sum_{\eta} \eta G^+_\eta G^+_\eta
	\\
	\Pi(t,t') & = \Pi_\infty(t-t') + \alpha(t,t') \sum_{\eta} \eta G^+_\eta G^+_\eta,
	\label{eq:app-divisor-superfermion}	
\end{align}%
\end{subequations}%
where $\alpha(t):=(1-e^{-\Gamma t})g(t)=\int_0^t ds e^{- \Gamma(t-s) } h(s)$ [\Eq{eq:g-h}]
and $\alpha(t,t')$ is to be determined.
Inserting this into the divisor equation $\Pi(t)=\Pi(t,t')\Pi(t')$
and using \Eq{eq:app-GG}, one finds a unique solution
by comparing coefficients, thereby confirming the ansatz:
\begin{align}
	\alpha(t,t') & =
	\alpha(t) - e^{-\Gamma (t-t')} \alpha(t')
	\label{eq:Pi-alpha}
	\\
	& =	\int_{t'}^t ds e^{- \Gamma(t-s) } h(s)
	=: (1-e^{-\Gamma (t-t)})g(t,t')
	\notag
	.
\end{align}
This recovers the EOM result \eq{eq:div-subst-final} for $g(t,t')$
and justifies the explicit construction \eq{eq:pi-divisor} of the divisor
within the real-time approach.

\section{Operator-sum approach [\Sec{sec:kraus}] \label{app:kraus}}

\subsection{Kraus operator construction -- Choi operator \label{app:kraus-choi}}

To derive the Kraus operator sum \eq{eq:krausfinal} of the main text
it is instructive to start from the spectral decomposition \eq{eq:pi-diag} in super-braket notation:
\begin{subequations}%
\begin{align}
	\Pi(t)
	=&
	\sum_{\eta} e^{(i\eta \varepsilon - \tfrac{1}{2} \Gamma )t}
	\, 
	\Ket{d_\eta^\dag} \, \Bra{  d_\eta^\dag }
	\label{eq:app-pi-1}
	\\
	&
	+ e^{-\Gamma t}
	\tfrac{1}{2} \, \Ket{(-\one)^n} \Big[ \Bra{(-\one)^n} - g(t) \Bra{\one}\Big]
	\label{eq:app-pi-2}
	\\
	&
	+ \tfrac{1}{2} \Big[ \Ket{\one} + g(t) \Ket{(-\one)^n} \Big] \Bra{\one}
	.
	\label{eq:app-pi-3}
\end{align}%
\label{eq:app-pi-diag}%
\end{subequations}%
We convert each of the three terms to a left-right action
\begin{subequations}
\begin{align}
	\Pi(t)=&
	\,
	e^{-\frac{1}{2}\Gamma t} \sum_\eta \bigg[
	e^{-i \eta \varepsilon t}
	\ket{\eta}\bra{\eta} \bullet\ket{\bar{\eta}}\bra{\bar{\eta}}
	\phantom{\bigg]}
	\label{eq:pi-intermediate-1}
	\\
	&
	+ \left(
	\cosh \tfrac{\Gamma t}{2}-\eta g(t) \sinh \tfrac{\Gamma t}{2}\right)
	\ket{\eta}\bra{\eta}\bullet \ket{\eta}\bra{\eta} 
	\label{eq:pi-intermediate-2}	
	 \\
	\phantom{\bigg[}
	&
	+ \left(1-\eta g(t) \sinh \tfrac{\Gamma t}{2}\right)
	\ket{\eta}\bra{\bar{\eta}}\bullet \ket{\bar{\eta}}\bra{\eta}  \bigg]
	\label{eq:pi-intermediate-3}	
\end{align}
\end{subequations}
by representing the fields $d_\eta = \ket{\eta}\bra{\bar{\eta}}$ and $d_\eta d_{\bar{\eta}} =\ket{\eta}\bra{\eta}$
in terms of ordinary brakets and collecting these on the left and right.
Already this preparatory step mixes up the different spectral components of $\Pi(t)$.
In particular, \Eq{eq:app-pi-2}-\eq{eq:app-pi-3} together produce the third term \eq{eq:pi-intermediate-3}.

The Choi operator \eq{eq:choi-def} is constructed by replacing basis superoperators by corresponding basis operators,
$\text{choi}( \ket{\eta}\bra{\eta'} \bullet \ket{\chi}\bra{\chi'} )
=
\ket{\eta\eta'}\bra{\chi'\chi}
$,
which act on states $\ket{\eta\eta'} = \ket{\eta} \otimes \ket{\eta'}$ of a doubled Hilbert space:
\begin{subequations}
	\begin{align}
	& \text{choi}(\Pi(t))
	=
	e^{-\frac{1}{2}\Gamma t}
	\Big\{ \cosh \tfrac{\Gamma t}{2}
	\one \otimes \one
	\label{eq:app-choi-1}
	\\
	&
	+ \sum_\eta \Big[
	e^{-i \eta \varepsilon t}
	\ket{\eta\eta}\bra{\bar{\eta}\bar{\eta}}
	-\eta g(t) \sinh \tfrac{\Gamma t}{2}
	\ket{\eta\eta}\bra{\eta\eta}
	\Big]
	\label{eq:app-choi-2}
	\\
	&
	+ \sum_\eta \Big[ 1-\eta g(t) \sinh \tfrac{\Gamma t}{2} \Big]
	\ket{\eta\bar{\eta}}\bra{\eta\bar{\eta}}
	\, 	\Big\}
	.
	\label{eq:app-choi-3}
\end{align}
\end{subequations}
The term \eq{eq:app-choi-3} coming from \Eq{eq:pi-intermediate-3} requires no diagonalization and gives rise to the eigenvalues \eq{eq:lambda1} given in the main text.
In contrast, the diagonalization of terms \eq{eq:pi-intermediate-1}-\eq{eq:pi-intermediate-2} results in the eigenvalues \eq{eq:lambda0}
and further mixes the spectral components \eq{eq:app-pi-1}-\eq{eq:app-pi-3}.
From the eigenvectors
\begin{gather}
\ket{K_\eta^0(t)} = \sqrt{\lambda^0_\eta(t)}
\,
\frac{\eta \sqrt{r(t)}^{\eta} \ket{- -} + \frac{1}{\sqrt{r(t)}^{\eta}} e^{-i \varepsilon t} \ket{++} }
{\sqrt{r(t)+\frac{1}{r(t)}} }
\notag
\\
\ket{K_\eta^1(t)} = \sqrt{\lambda^1_\eta(t)} \, \ket{\eta\bar{\eta}}
\label{eq:choi1}
\end{gather}
we obtain the Kraus operators \eq{eq:kraus0}-\eq{eq:kraus1} using the identities $\ket{\eta\eta} = d_{\eta} d_{\bar{\eta}}  \otimes \one \ket{\one}$,
$\ket{\eta\bar{\eta}} = d_{\eta} \otimes \one \ket{\one}$ and $\text{choi}( K_m \bullet K_m^\dag )=\ket{K_m}\bra{K_m}$.

\label{app:kraus-sumrule}
As noted in \Sec{sec:open-system} and highlighted in the above derivation,
the spectral components \eq{eq:app-pi-diag} of $\Pi(t)$ are nontrivially encoded in the Kraus operator-sum.
For our model this can be made more precise: The eigenvectors of $\Pi(t)$ can be derived from fact that
the Kraus operators have definite fermion-parity $(-\one)^n K^k_\eta(t) (-\one)^n =(-1)^k K^k_\eta(t)$
together with the quadratic sum-rules listed in Table \ref{tab:kraus-rel}.

\subsection{Effective environment density matrix\label{app:kraus-env}}

Finally, we construct the density matrix \eq{eq:effective-environment-state} of the effective environment
$\rho^{\E'}(t)^{k k'}_{\eta \eta'}
:=
\Tr_\S K_{\eta}^k(t) \rho(0) K_{\eta'}^{k'}(t)^\dag$
and first focus on the odd- and even-parity diagonal blocks $k=k'$ that remain when the initial system state $\rho(0)$ obeys superselection.
The odd block ($k=1$) is already diagonal in this basis,
\begin{subequations}
\begin{gather}
\label{eq:app-state-env-matrix-1}
\rho^{\E'}(t)^{11}_{\eta\eta'} = \delta_{\eta\eta'}\Lambda^{\E'1}_\eta \\
\quad \Lambda^{\E'1}_\eta = \tfrac{1}{2} \lambda^1_{\eta}(t) \Big[ 1+ \eta \brkt{(-\one)^n(0)}\Big] 
,
\label{eq:app-eigenvalues-env1}
\end{gather}
\label{eq:app-state-env-block1}%
\end{subequations}
with the Choi eigenvalues \eq{eq:lambda1}.
The even block ($k=0$) is nondiagonal and has to be explicitly diagonalized.
Using Eqs.~\eq{eq:lambda0} and \eq{eq:r-t} we obtain for the matrix elements and eigenvalues:%
\begin{widetext}%
\begin{subequations}%
	\begin{gather}%
\rho^{\E'}(t)^{00}_{\eta \eta'} =
 \tfrac{1}{2}
\bigg\{
\lambda^0_{\eta}(t) \left[
1+ \eta \brkt{(-\one)^n(0)} \frac{r(t)-\frac{1}{r(t)}}{r(t)+\frac{1}{r(t)}}
\right] \delta_{\eta \eta'}
- \brkt{(-\one)^n(0)} \,  \frac{2\sqrt{\lambda^0_{+}(t)\lambda^0_{-}(t)}}{r(t)+\frac{1}{r(t)}}
\delta_{\bar{\eta},\eta'}
\bigg\}
\label{eq:app-state-env-matrix-0}
\\
\Lambda^{\E'0}_{\eta}(t)
=
\tfrac{1}{2} \sum_{\eta''} \lambda^0_{\eta''}(t)
\left[
1+ \eta'' \brkt{(-\one)^n(0)} \frac{r(t)-\tfrac{1}{r(t)}}{r(t)+\tfrac{1}{r(t)}}
\right] 
\tfrac{1}{2}
\left[
1+\eta
\sqrt{1-
	\frac{\lambda^0_{+}(t) \lambda^0_{-}(t)
	\big[ 1- \brkt{(-\one)^n(0)}^2 \big]}
	{
	\Big\{ \tfrac{1}{2} \sum_{\eta'} \lambda^0_{\eta'}(t) 
	\left[
	1+ \eta' \brkt{(-\one)^n(0)} \frac{r(t)-\frac{1}{r(t)}}{r(t)+\frac{1}{r(t)}}
	\right]
	\Big\}^2
	}
}
\right]
.
\label{eq:app-eigenvalues-env0}%
\end{gather}%
\label{eq:app-state-env-block0}%
\end{subequations}%
\end{widetext}%
The result \eq{eq:app-eigenvalues-env0} shows that the effective environment density matrix depends on both
the Choi eigenvalues $\lambda^0_\eta(t)$ \emph{and} the Choi eigen\emph{vectors} through $r(t)$,
in addition to the initial state $\rho(0)$ through $\brkt{(-\one)^n(0)}$.
In particular, each eigenvalue is positive if and only if the initial state is positive, $|\brkt{(-\one)^n(0) }| \leq 1$,
and the evolution $\Pi(t)$ is CP, $|g(t)| \leq 1$.

\subsection{Factorization -- Effective environment modes\label{app:kraus-factor}}

The eigenvalues \eq{eq:app-eigenvalues-env1} and \eq{eq:app-eigenvalues-env0} are already products of positive factors
but the dynamics of the effective environment is easier analyzed in terms of the factorization
\begin{subequations}
	\begin{align}
	\Lambda^{\E'0}_\eta(t) & = \Lambda^{\E'+}_\eta(t) \cdot \Lambda^{\E'-}_\eta(t)
	\\
	\Lambda^{\E'1}_\eta(t) &= \Lambda^{\E'+}_\eta(t) \cdot \Lambda^{\E'-}_{\bar{\eta}}(t)
\end{align}%
\label{eq:app-factorize}%
\end{subequations}%
used in the main text [\Eq{eq:factors}].
These factors $\Lambda^{\E'\lambda}_\eta \geq 0$ are additionally normalized as $\sum_{\eta} \Lambda^{\E'\lambda}_\eta(t) =1$ and therefore represent probabilities.
From this ansatz we find by taking different sums of \Eq{eq:app-factorize} the expression
\begin{align}
	\Lambda^{\E'\lambda}_\eta(t) & = \Lambda^{\E'0}_\eta(t) + \Lambda^{\E'1}_{\lambda \cdot \eta}(t)
	\quad \text{with} \quad 
	\eta, \lambda = \pm
	.
	\label{eq:app-factors}
\end{align}
Substituting back into \Eq{eq:app-factorize} and using the normalization $\sum_{k=0,1} \Lambda^{\E'k}_\eta =1$
leads to a nontrivial condition on the eigenvalues of the two diagonal blocks in \Eq{eq:effective-environment-state}:
$\Lambda^{\E'0}_{+}\Lambda^{\E'0}_{-} = \Lambda^{\E'1}_{+}\Lambda^{\E'1}_{-}$.
One verifies using \Eqs{eq:app-eigenvalues-env1} and \eq{eq:app-eigenvalues-env0} that this indeed holds true independent of the physical parameters.
Thus, the eigenvalues of the parity-blocks $k=0$ and $k=1$ in \Eq{eq:effective-environment-state} are not independent:
Even though the effective two-fermion environment is coupled to the system,
its state \eq{eq:effective-environment-state} \emph{always} factorizes into two \emph{uncorrelated} fermion modes.
To obtain the representation \Eq{eq:factors} given in the main text, we insert \Eqs{eq:app-eigenvalues-env1} and \eq{eq:app-eigenvalues-env0} into \Eq{eq:app-factors}
and use the shorthand notation $c(t):=\coth(\Gamma t/2)$:
\begin{widetext}
\begin{subequations}
	\begin{align}
		\Lambda^{\E'\lambda}_{\eta}
		& =
		\frac{1}{2}
		\Big[
		1 + \eta \frac{
			\sqrt{
				[c(t)+ \brkt{(-\one)^n(0)} g(t)]^2 - [1-g(t)^2] [1-\brkt{(-\one)^n(0)}^2]
			}
			+ \lambda [\brkt{(-\one)^n(0)} - g(t)]
		}{c(t)+1}
		\Big]
		\\
		&
		=
		\frac{1}{2}
		\Big[
		1 + \eta \frac{
			\sqrt{
				[c(t)\brkt{(-\one)^n(0)} +  g(t)]^2 + [1-c(t)^2] [1-\brkt{(-\one)^n(0)}^2]
			}
			+ \lambda [\brkt{(-\one)^n(0)} - g(t)]
		}{c(t)+1}
		\Big]
		.
	\end{align}
\label{eq:app-factors-full}
\end{subequations}
\end{widetext}

The $\eta$-index corresponds to the ordering $0\leq \Lambda^{\E'\lambda}_{-}(t) \leq \Lambda^{\E'\lambda}_{+}(t) \leq 1$
and the second line shows that system-state eigenvalues, written as
\begin{align}
	\Lambda_\eta(t) = \tfrac{1}{2} \Bigg[
	1 + \eta \frac{ [c(t)-1]\brkt{(-\one)^n(0)} + 2 g(t) }{c(t)+1}
	\Bigg]
	,
	\label{eq:app-sys-evals}
\end{align}
are bounded by the eigenvalues of \emph{both} modes in a 'crossed' way:
\begin{align}
	\Lambda^{\E'\lambda}_{-}(t)
	\leq
	\Lambda_\lambda
	\leq
	\Lambda^{\E'\bar{\lambda}}_{+}
	.
	\label{eq:app-no-explanation}
\end{align}
They are \emph{not} bounded by the eigenvalues of \emph{one} of the effective modes,

Based on the representations \Eq{eq:app-factors-full}, one can identify the two simple special cases discussed in \Sec{sec:kraus}:
For pure initial system states with $\brkt{(-\one)^n(0)}=\sigma = \pm$ the eigenvalues of the effective environment modes reduce to
\begin{gather}
	\Lambda^{\E'\sigma}_\eta(t)
	=
	\tfrac{1}{2}[1+\sigma\cdot\eta]
	=
	\Lambda_{\sigma \cdot \eta}(0)	
	\\
	\Lambda^{\E'\bar{\sigma}}_\eta(t)
	=
	\tfrac{1}{2} \Big[
	1 + \sigma\cdot\eta \frac{[c(t)-1]\sigma  + 2 g(t) }{c(t)+1}
	\Big]
	=
	\Lambda_{\sigma \cdot \eta}(t)
	\notag
	.
\end{gather}
Comparing with \Eq{eq:app-sys-evals}, one sees that in this case one of the effective modes is locked to the spectrum of the pure initial system state $\rho(0)$
whereas the other mode has exactly the same spectrum as the system state.
Consequently, the factorization \eq{eq:entropy-env-stat} is trivially recovered in the stationary limit as $\rho(t) \to \rho(\infty)$.
Pure stationary system states are only reached in the off-resonant semigroup limit \eq{eq:eps-infty}
where $g(t) = \theta(t) g(\infty)$ instantly jumps to its stationary value $g(\infty)=\sigma=\pm$ while $\rho(t)$ still evolves in time.
In this case, the modes then reduce to
\begin{gather}
	\Lambda^{\E'\bar{\sigma}}_\eta(t)
	=
	\tfrac{1}{2}[1+\sigma\cdot\eta] 
	=
	\Lambda_{\sigma \cdot \eta}(\infty)
	\\
	\Lambda^{\E' \sigma }_\eta(t)
	=
	\tfrac{1}{2} \Big[
	1 + \eta \frac{[c(t)-1]  + 2 \sigma \brkt{(-\one)^n(0)} }{c(t)+1}
	\Big]
	\neq 
	\Lambda_{\sigma \cdot \eta}(t)
	\notag
	.
\end{gather}
Correspondingly, one of the effective environment modes instantly attains the \emph{stationary} system spectrum of $\rho(\infty)$
while the spectrum of the other mode evolves \emph{different} from $\rho(t)$, converging to the \emph{initial} system spectrum:
$\Lambda^{\E' \sigma }_\eta(\infty)=\tfrac{1}{2} [1 + \sigma\cdot\eta \brkt{(-\one)^n(0)} ]= \Lambda_{\sigma \cdot \eta}(0)$.
Finally, we note that the stationary limit \eq{eq:entropy-env-stat} of the effective environment is reached \emph{irrespective} of these special cases, as it should.
Which mode takes on the role of the initial, respectively stationary system spectrum
depends on both the initial $\brkt{(-\one)^n(0)}$ and stationary state $\brkt{(-\one)^n(\infty)}=g(\infty)$
through the sign $\sigma = \sign [ \brkt{(-\one)^n(0)}+ \brkt{(-\one)^n(\infty)}] $:
\begin{align}
	\Lambda^{\E'\sigma}_\eta(\infty)
	& =
	\tfrac{1}{2} \big[ 1 + \sigma\cdot\eta  \brkt{(-\one)^n(0)} \big]
	=
	\Lambda_{\sigma\cdot\eta}(0)	
	\\
	\Lambda^{\E'\bar{\sigma}}_\eta(\infty)
	&=
	\tfrac{1}{2} \big[ 1 + \sigma\cdot\eta \brkt{(-\one)^n(\infty)} \big]
	=
	\Lambda_{\sigma\cdot\eta}(\infty)	
	.
	\notag
\end{align}

\subsection{Pseudo-spin model without superselection\label{app:kraus-spin}}

Above and in the main text [\Eq{eq:effective-environment-state}]
we claimed that the off-diagonal blocks of $\rho^{\E'}(t)$ are zero for all $t \geq 0$ if $\rho(0)$ obeys superselection, i.e., $\brkt{d(0)}=0$.
An explicit calculation of the off-diagonal blocks
\begin{align}
&\rho^{\E'}(t)^{01}_{\eta \eta'}
=
\brkt{d_\eta}(0)
\,
\sqrt{
	\lambda^1_{\eta} \lambda^0_{\eta'}
}
\frac{ [\sqrt{r(t)}]^{-\eta \eta'} }{r(t)+1/r(t)}
[\eta' e^{-i \varepsilon t}]^{\frac{1+\eta}{2}}
\label{eq:app-state-env-blocks-offdiag}
\end{align}
indeed confirms that the dynamics preserves superselection and furthermore shows that the diagonal blocks remain unaltered.

As mentioned in \Sec{sec:model}, the model may also be considered as an unconventional but valid spin problem, cf. \Eq{eq:full-hamiltonian-spin}.
In this case there is no superselection constraint such that the effective environment state has no special block structure:
\begin{align}
\rho^{\E'}(t) =
\begin{bmatrix}
\rho^{\E'}(t)^{00} & \rho^{\E'}(t)^{01} \\
\rho^{\E'}(t)^{10} & \rho^{\E'}(t)^{11}
\end{bmatrix}
.
\label{eq:app-effective-environment-state}
\end{align}
Consequently, it can \emph{not} be considered as two fermions but only as two pseudo-spins.
Allowing for transverse initial pseudo-spin, $\brkt{d(0)} \neq 0$ the nonzero parity-off-diagonal blocks
make it harder to explicitly see the positivity of the state  \eq{eq:app-rho-t}.
It is equivalent to the positivity of any one of its diagonal blocks, say $\rho^{\E'}_{11}(t) \geq 0$, and its Schur complement~\cite{Schur-book},
\begin{align}
	\rho^{\E'}(t)^{00}
	- \rho^{\E'}(t)^{01} \frac{1}{\rho^{\E'}(t)^{11}} \rho^{\E'}(t)^{01}
	\geq 0
	\label{eq:app-stronger}
	.
\end{align}
Thus, without superselection it is \emph{not} sufficient that the other block $\rho^{\E'}(t)^{00} \geq 0$ is positive.
Explicitly verifying this and the other general facts discussed above
from \Eqs{eq:app-state-env-block1}, \eq{eq:app-state-env-block0} and \eq{eq:app-state-env-blocks-offdiag} is possible but cumbersome.
\bibliographystyle{aipnum4-1}
\end{document}